\documentclass{aastex631}
\usepackage{amsmath}
\usepackage{multirow}
\renewcommand{\vec}{\boldsymbol}

\begin{document} 

\title{Estimating microlensing parameters from observables and stellar isochrones with pyLIMASS.}
\author[0000-0002-6578-5078]{E. Bachelet}
\affiliation{IPAC, Mail Code 100-22, Caltech, 1200 E. California Blvd., Pasadena, CA 91125, USA;}

\author[0000-0003-0961-5231]{M. Hundertmark }
\affiliation{Zentrum f{\"u}r Astronomie der Universit{\"a}t Heidelberg, Astronomisches Rechen-Institut, M{\"o}nchhofstr. 12-14, 69120 Heidelberg, Germany; \label{heidelberg}}

\author[0000-0002-7669-1069]{S. Calchi Novati}
\affiliation{IPAC, Mail Code 100-22, Caltech, 1200 E. California Blvd., Pasadena, CA 91125, USA;}
    
\begin{abstract}
We present pyLIMASS, a novel algorithm for estimating the physical properties of the lensing system in microlensing events. The main idea of pyLIMASS is to combine all available information regarding the microlensing event, defined as \textit{observables}, and to estimate the parameter distributions of the system, such as the lens mass and distance. The algorithm is based on isochrones for the stars model and combine the observables using a Gaussian Mixtures approach. After describing the mathematical formalism and its implementation, we discuss the algorithm's performance on simulated and published events. Generally, the pyLIMASS estimations are in good agreement (i.e., within 1-$\sigma$) with the results of the selected published events, making it an effective tool to estimate the lens properties and their distribution. The applicability of the method was tested by using a catalog of realistically simulated events that could be observed by the future Galactic Bulge Time Domain Survey of the Nancy Grace Roman Space Telescope. By solely using constraints from the Roman lightcurves and images, pyLIMASS estimates the masses of the lens of the Roman catalog with a median precision of 20\% with almost no bias. 
\end{abstract}

\keywords{??}    
\date{Received ??; accepted ??}
    
   
%
\section{Introduction}



Measuring the mass of astrophysical objects is crucial to understanding their distribution and abundance in the Milky Way. However, this is often challenging due to their large distances and the technical limitations. Gravitational microlensing \citep{Einstein1936, Paczynski1986} is ideally suited to probe the mass function of objects in the entire Milky Way, because it observes the change in brightness of a background star, and the relatively faint lens object can be characterized solely by studying the light curve of the background star. Originally, the technique has been used to study lens populations of the the Milky Way halo by surveying the source located in the Magellanic Clouds \citep{Paczynski1986,Alcock2000,Tisserand2007,Wyrzykowski2011}. More recently, microlensing surveys have been focused towards the Galactic Bulge \citep{Bond2001,Udalski2003,Kim2016}, where the event rate is highest \citep{Sumi2013,Mroz2019}, but microlensing events have also been detected in the entire Galactic Disk \citep{Fukui2019,Wyrzykowski2023}. Based on these observations, more than two hundred planets have been detected \footnote{\url{https://exoplanetarchive.ipac.caltech.edu}}, as well as brown dwarfs \citep{Zhu2016,Chung2017,Shvartzvald2019,Bachelet2019}, free-floating planets \citep{Mroz2017, Mroz2018,Sumi2023}, stellar binaries (see for example  \citet{Street2019,Tsapras2019}) and stellar remnants \citep{Blackman2021,Sahu2022}. Moreover, the galactic structure can be inferred from the distribution of microlensing event parameters \citep{CalchiNovati2015, Awiphan2016,Mroz2017, Mroz2019}. Given its potential, a microlensing survey is one of the core missions of the upcoming Nancy Grace Roman Space Telescope \citep{Spergel2015,Penny2019,Johnson2020}.    

One of the difficulties of the method is to reconstruct the physical properties of the microlens, especially its mass and distance. Generally, the lightcurve of the microlensing event is the main phenomenon observed. From the modeling of the lightcurve, the Einstein ring crossing time $t_E$ is usually measured with great precision \citep{Gould2000}:
\begin{equation}
    t_E={{\theta_E}\over{\mu_{rel}}}
    \label{eq:tE}
\end{equation}
where $\vec{\mu_{rel}}=\vec{\mu_{L}}-\vec{\mu_{S}}$ is the relative proper motion vector between the lens and the source. It is measured in the observer frame (geocentric frame for example), and can be transformed to the heliocentric frame by \citep{Skowron2011}:
\begin{equation}
\vec{\mu_{rel,helio}} = \vec{\mu_{rel,geo}} +\pi_{rel}\vec{V_{\oplus,\perp}}     
\end{equation}
Here $\vec{V_{\oplus,\perp}}$ is the projected observer velocity in the plane of sky at the time of the event maximum and $\pi_{rel}=\pi_L-\pi_S$ is the relative parallax \citep{Gould2000}. $\theta_E$ is the angular Einstein ring radius that is a fundamental characteristic of the lensing phenomenon and defined as \citep{Gould2000}:
\begin{equation}
    \theta_E = \sqrt{\kappa M_L \pi_{rel}}
\end{equation}
with $\kappa$ a constant and $M_L$ the lens mass in solar units. $\theta_E$ can be measured if finite-source effects are seen in the lightcurve, commonly parameterized with $\rho_*={{\theta_S}\over{\theta_E}}$, where $\theta_S$ is the angular source radius \citep{Witt1994,Yoo2004}. We note that this parameter is also sometimes parameterized by $t_S=\rho_* t_E$, which is a measure of the duration of finite-source effects. The observations of the microlensing astrometric shifts \citep{Dominik2000,Sahu2022} or interferometric measurements during the peak of the event \citep{Dong2019,Cassan2022} also lead to a direct measurement of $\theta_E$, but these observations are generally challenging and expensive in telescope time. The measurement of $\theta_E$ can also be made by observing the lens and source several years after the event peak with high-resolution imagers \citep{Batista2015,Bennett2015,Bhattacharya2018,Vandorou2020}. Indeed, the observed lens and source angular separation $\delta$ at a time offset $\Delta t$ is directly related to $\theta_E$ via:
\begin{equation}
\pi_{rel} = \pi_L-\pi_S = \pi_E\theta_E={{\theta_E^2}\over{\kappa M_L}}= {{\mu_{rel}^2t_E^2}\over{\kappa M_L}}  = \Bigg({{t_E \delta\over{\Delta t}}}\Bigg)^2{{1}\over{\kappa M_L}}    
\label{eq:MLDL}
\end{equation}
where $\pi_E$ is the norm of the microlensing parallax vector \citep{Gould2004}. The microlensing parallax vector $\vec{\pi_E}$ is crucial to measure because it contains information about the lens mass and distance as well as the direction of the relative proper motion, for example in the equatorial system \citep{Gould2004}:
\begin{equation}
\vec{\pi_E} = (\pi_{EN},\pi_{EE})=  {{\pi_{E}}\over{\mu_{rel,geo}}}(\mu_{rel,geo,N},\mu_{rel,geo,E})   
\end{equation}
The microlensing parallax can be measured from a single location if the event is long enough, e.g. the annual parallax effect \citep{Gould2004}, or if the event is observed by several observatories with significant separations, e.g. the space based parallax \citep{Refsdal1966}. This has be done many times, especially with the Spitzer telescope, see for example \citet{CalchiNovati2015} and \citet{Zang2020}, and Gaia \citep{Wyrzykowski2020}. 

The measurements of the lens brightness $m_{L,\lambda}$, preferably in several bands or with spectroscopy, also provide strong constraints on the lens system \citep{Beaulieu2018}:
\begin{equation}
m_{L,\lambda}=5\log(D_L)-5+A_{L,\lambda}+M_0(\lambda;M_L;R_L;T_{eff,L}; Age_L;[Fe/H]_L)
\label{eq:mag_lens}
\end{equation}
where $A_{L,\lambda}$ is the absorption towards the lens and $M_0$ is the absolute magnitude of the lens in a given band $\lambda$. The latter can be estimated via spectral templates \citep{Kurucz1993,Allard1995}, stellar isochrones libraries \citep{Bressan2012,Choi2016} or empirical relation \citep{Pecaut2013, Benedict2016, Mann2019,Rabus2019}. 

Another source of constraints can be obtained by using models of the Milky Way. This has been done on many occasions and for different models \citep{Han1995, Han2003, Dominik2006,Bennett2014, Mroz2020, Specht2020, Koshimoto2021}. In particular, by comparing the timescale, optical depth and event rate distributions of microlensing events with Galactic Models predictions, constraints can be derived on the various population of objects as well as the kinematics and architecture of the Galaxy \citep{Mroz2020,Specht2020,Koshimoto2021}. These models can also be used to study individual event, by generating priors for the modeling of the lightcurves or to estimate the event physical parameters after the modeling. The downside is that the results are inevitably biased because of the different models assumptions. For example, \citet{Bachelet2022b} found a good agreement with the prediction from the Besan\c{c}on \citep{Robin2003}, the GalMod \citep{Pasetto2018}, \citet{Dominik2006} and \citet{Koshimoto2021} models, but small discrepancies exist.

Another difficulty is to combine the different measurements to reconstruct the properties of the lensing system. A natural way to do is to include any available exterior information with the form of prior during the modeling process. This is routinely done during the reanalysis of event with high-resolution imaging constraints \citep{Bennett2015,Bhattacharya2018,Vandorou2020}. However, the modeling of lightcurve can be non-trivial and time consuming, while the inclusion of all available information could also be difficult to add into modeling codes in practice. 

In recent years, the microlensing community has made great efforts to publish public codes for the modeling of microlensing events. The first package to be released was \texttt{pyLIMA} \citep{Bachelet2017}, a python-based software that includes multiple tools for analyzing microlensing events and is now widely used by the community. \texttt{MulensModel} by \citet{Poleski2019} is another public package for modeling microlensing events. Both packages make use of the \texttt{VBBinaryLensing} C++ library \citep{Bozza2010,Bozza2018}, a contour-integration code for the estimation of the microlenisng magnifcation. This library is also in use in the RTModel infrastructure, which provides real-time modeling for microlensing events since 2013\footnote{\url{http://www.fisica.unisa.it/gravitationastrophysics/RTModel.htm}} and has been recently released to the public\footnote{\url{https://github.com/valboz/RTModel}}. Similarly, Cheongho Han provides access to the community to real-time models made by his team \footnote{\url{http://astroph.chungbuk.ac.kr/~cheongho/modelling/2024/model_2024.php}}. More recently, the software \texttt{eesunhong} have been made public\footnote{\url{https://github.com/golmschenk/eesunhong}}. This code has been used for many years to analyze plenty of microlensing events \citep{Bennett1996,Bennett2010,Bennett2023}. 

As far as we know, none of these software provides tools to estimate the physical properties of the lensing system. While this is sometimes (relatively) straightforward to do so, when the microlensing parallax and finite-source effects are measured in the lightcurve for example, the combination of all the available information to accurately reconstruct the physical properties of the lenses is non trivial. Therefore, we propose in this work a new tool to achieve this goal. The algorithm is solely based on the information in observables, as defined in the next section and provided by the user with, by default, the additional prior of the stellar isochrones. The purpose of this tool is to combine the provided measurements (and stellar isochrones) to estimate the lensing system physical parameters in a timely manner. We note that this estimate is done by default without using any prior from Galactic Models. The algorithm is described in Section~\ref{sec:obstoparams} as well as its python implementation in pyLIMASS. Section~\ref{sec:simulations} and Section~\ref{sec:publishedevents} present the algorithm performance on simulated data and published analysis and we present our conclusions and potential upgrades in Section~\ref{sec:conclusion}.

\section{From observables to physical parameters}\label{sec:obstoparams}
\subsection{Definition of the observables}
In the following, we define an \textit{observable} $x$ as a measurable physical quantity, and its associated uncertainty, related to a microlensing event. The list of default observables defined in pyLIMASS can be found in the Appendix, but as an example the source magnitudes in different bands or the relative proper motion vector can be considered as observables. In general, an observable can be any kind of measurable quantity related to the microlensing event and does not have to be directly related to the modeling of the microlensing event data (photometric lightcurve, astrometric time series and/or high resolution imaging). For example, if the parallax of the lens is known directly from the Gaia satellite measurements \citep{Prusti2016, Vallenari2023}, then this is valuable information that is relevant to include as an observable.

\subsection{Inverse problem}
As discussed in the introduction, the parameters of most interest $P$, such as the lens mass and distance, are generally not measured directly and need to be reconstructed from a set of observables $X=(x_1,x_2,....x_i)$. This is a classic inverse problem and commonly expressed through Bayes' theorem in terms of probabilities:
\begin{equation}
    p(P|X)={{p(X|P)p(P)}\over{p(X)}}
    \label{eq:Bayes}
\end{equation}
$p(X|P)$ is the likelihood, $p(P)$ is the prior probability and $p(X)$ is a constant often referred as model evidence. The goal is to sample the posterior probabilities $p(P|X)\propto p(X|P)p(P)$, i.e. estimate the more plausible parameters P that replicates the set of observables X. The choice of priors is also important to sample the posterior probabilities. This is further discussed in the Appendix~\ref{app:priors}.

\subsection{Gaussian Mixture}

 Equation~\ref{eq:tE} shows that measuring only $t_E$ in a microlensing source event, yields weak, but not null, constraints on the lens mass. In fact, studying the parameters distributions of multiple events provides valuable information about the architecture of the Milky Way \citep{Sumi2011,CalchiNovati2015, Mroz2019,Sumi2023}. But the primary goal of our approach is to gather as much information as possible, expressed in terms of the observables as defined previously, about one specific microlensing event to reconstruct the physical parameters of the source and lens system. Ideally, it is desirable that the algorithm could handle constraints from multiple sources and of various forms. For example, $t_E$ is generally well measured from the lightcurve and leads to a constraint of the form $\mathcal{N}(t_E,\sigma_{t_E})$. Similarly, the blend flux is also generally well constrained from the lightcurve only. However, the hypothesis that the additional flux comes solely from the lens often turns to be wrong, especially towards the Galactic Bulge where the crowding is extreme. In fact, many high-resolution follow-up studies revealed unrelated companions at $\le 1$" from the target, see for example \citet{Blackman2021, Bhattacharya2021,Terry2021}. In that case, a safer assumption is to assume a constraint (for example in the V band) on the form $V_L\ge V_{blend}$. 
 
Therefore, we seek for a mechanism that supports such various distribution of constraints and also provide the likelihood $P(X|P)$ (or at least an approximation) of a set of parameters P given the set of observables X. To combine these observables, we model them using a multidimensional Gaussian Mixture (GM). GMs are frequently used for this purpose \citep{Kelly2007,Hao2010,Alsing2018,Fruhwirthschnatter2019,Ansari2021}, and the empirical distribution of the observables $X$ is then represented by G Gaussian components. The joint probability distribution of a single observation $y$ is then given by \citep{Fruhwirthschnatter2019}:

\begin{equation}
    \label{eq:GMpdf}
    p(y|\theta)= \sum_{k=1}^{G} \phi_k \mathcal{N}(\mu_k,\,\Sigma_k,y)	
\end{equation}

where $\theta=(\phi_k,\mu_k,\Sigma_k)$ is a set of G hyperparameters. $\mu_k$ and $\Sigma_k$ are the mean vector and covariance matrix associated with the multidimensional Gaussian component k. The parameter $\phi_k$ is the weight of the Gaussian components and by definition $\sum_{k=1}^{G}\phi_k=1$. 

The goal of the GM model is to approximate the underlying probability distribution of a sample of N observables $X=(x_1,x_2,....,x_N)$ or, in other words, find the parameters $\theta_f$ that maximize the likelihood:
\begin{equation}
    \label{eq:likelihoodGM}
    p(X|\theta) = \prod_{i=1}^{N} \sum_{k=1}^{G} \phi_k \mathcal{N}(\mu_k,\,\Sigma_k,x_i)	
\end{equation}
This is a non-trivial problem that is an area of active research \citep{Jin2016,Lucic2018}, but it can be solved by using the Expectation-Maximization (EM) algorithm \citep{Dempster1977} for example. Once the best hyperparameters $\theta_f$ are found, the likelihood of the observables serie $X$  for the parameters $P$ can be approximated via the GM model:
\begin{equation}
    \label{eq:GMpdf}
    p(X|P) \approx p(y|\theta_f)=\sum_{k=1}^{G} \phi_k \mathcal{N}(\mu_k,\,\Sigma_k,y)	
\end{equation}

It is worth noting that assuming the simplest case with no correlation between the parameters, Equation~\ref{eq:GMpdf} simplifies to:
\begin{equation}
    p(X|P)\approx \sum_{k=1}^{G} \phi_k \prod_{j=1}^{d}{{1}\over{\sqrt{2\pi}\sigma_{j,k}}}e^{-0.5{{(y-\mu_{j,k})^2}\over{\sigma_{j,k}^2}}}.
\end{equation}
In this equation $\mu_{j,k}$ and $\sigma_{j,k}$ represent the mean and the standard deviation of the Gaussian component k and the observable j, while d represents the total number of observables (i.e. the number of dimension). If the generated observable is close to the most probable region of a specific Gaussian k (i.e. $y\sim\mu_{k}$), the log-likelihood becomes:
\begin{equation}
    ln(p(X|P))\approx ln\biggl(\prod_{j=1}^{n}{{1}\over\sqrt{2\pi}\sigma_{j,k}}\biggr) \approx -\sum_{j=1}^{n} ln (\sqrt{2\pi}\sigma_{j,k})
\end{equation}
Here, it is assumed for simplicity that all weights $\phi$ are zero except for the value of the k component. Although this is an oversimplified model, it effectively demonstrates the general behaviour of the log-likelihood implemented in pyLIMASS.

\begin{figure}
    \centering
    \includegraphics[width=0.9\textwidth]{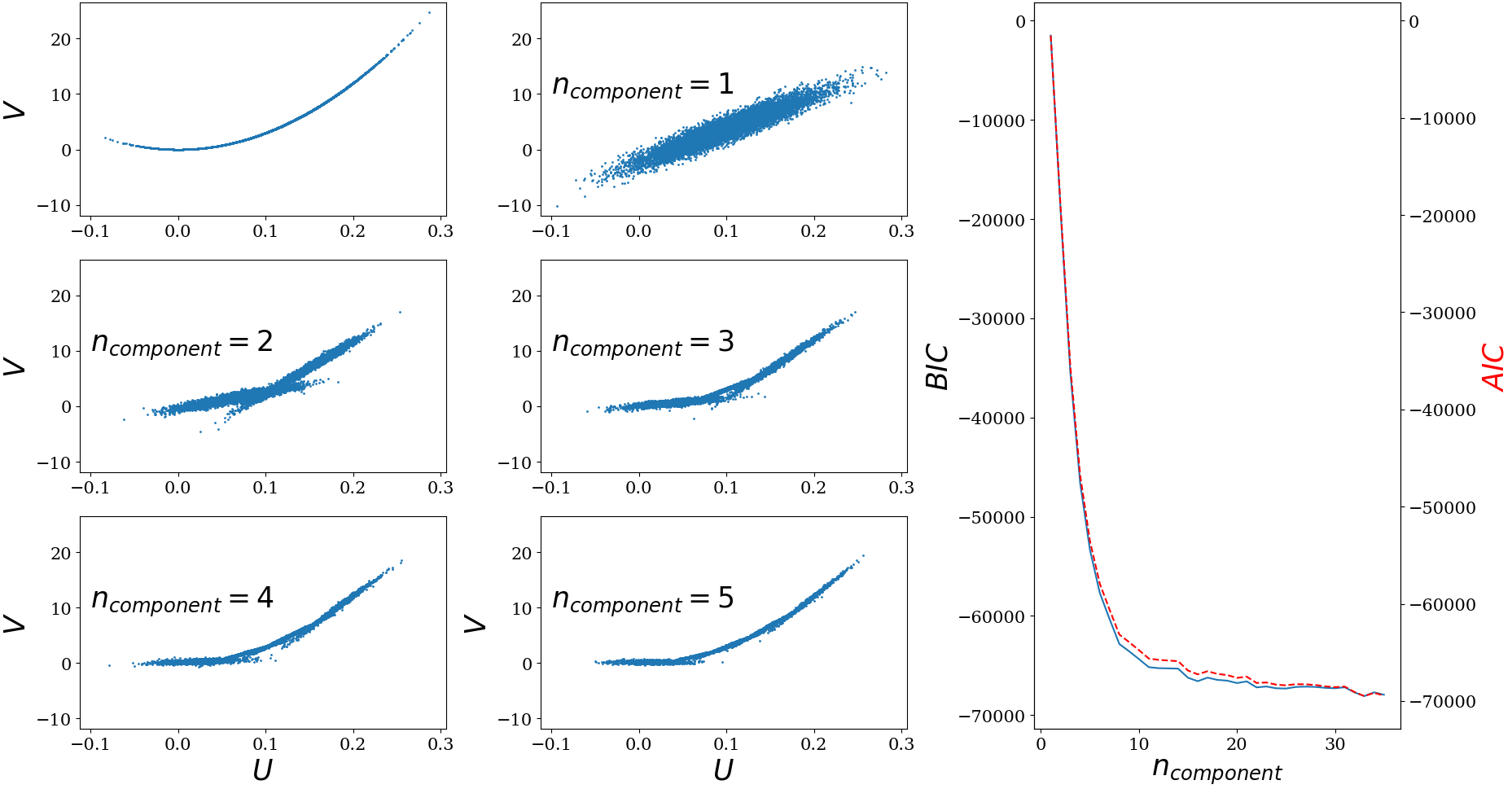}
    \caption{(Left) Simulated $V$ versus $U$ distribution (top left) and the different reconstructed GM distributions with different number of Gaussian components. As more components are added to the model, the simulated distribution is reconstructed with more fidelity. (Right) BIC (plain blue) and AIC (dashed red) from a validation sample versus the number of Gaussian component used in the GM model. Both metrics reaches a plateau after $\sim20$ components, indicating potential overfitting afterwards.}
    \label{fig:adhoc}
\end{figure}

The great advantage of GM is its simplicity as well as its potential to 
approximate multi-modal distributions with complex shapes \citep{Fruhwirthschnatter2019}. Such distributions are quite common in microlensing, where parameters degeneracies or competitive models can be found. To illustrate this, we realized an ad hoc simulation, where two random variables $U=0.1\pm0.05$ and $V=300 U^2$ are somehow measured. The resulting distribution is visible in Figure~\ref{fig:adhoc}. We then modeled different GM models using G components. It is clear that the original distribution is better replicate as the number of Gaussian is increased. However, there is a risk of overfitting by increasing arbitrarily the number of components. While the optimal number of components is problem dependent, statistical tools like the Bayesian information criterion (BIC, \citet{Schwarz1978}) or the Akaike Information Criterion (AIC, \citet{Akaike1974}) can be used to assess the significance of additional components. For the example presented, $\sim20$ components seems sufficient according to both metrics.

\subsection{Implementation in pyLIMASS}
We have implemented this approach in pyLIMASS, a new module of the pyLIMA package \citep{Bachelet2017}. First, it is worth noting that most of the relations presented earlier contain products, ratios, and power-laws that are much better represented in logarithmic space. Moreover some of the observables, such as $\rho_*$ or $t_E$, are strictly positive. The GM algorithm also performs better in logarithmic space because it represents the actual distribution more accurately and is not affected by floor and ceiling effects. Furthermore, the previously described distributions of parameters, such as $\rho_*$, are transformed into a sum of distributions, and consequently behave much better in the logarithmic space. As a result, the algorithm utilizes logarithmic quantities whenever possible. In pyLIMASS, GM models are generated with the scikit-learn implementation \citep{scikit-learn}, and the number of Gaussian functions used can be adjusted by the user, although it is set by default to the number of observable distributions ingested. Eight parameters $P$ are defined for the source and lens, namely the mass $M$ in solar mass units, the distance $D$ in kpc, the effective temperature $T_{eff}$ in Kelvin, the metallicity $Fe$ in dex, the logarithmic surface gravity $logg$, the geocentric North and East proper motions $\mu_N$ and $\mu_E$ in mas/yr and the V-band absorption $A_{V}$. For the lens body, it is more efficient to sample $\epsilon_D=D_L/D_S\le1$ and $\epsilon_{A_V}=A_{V,L}/A_{V,S}\le1$, these quantities are therefore used in pyLIMASS instead of $D_L$ and $A_{V,L}$. Most of the observables can be reconstructed from these parameters and compared with the GM distribution model.

The reconstruction of the observed magnitudes of the source and, possibly, the lens is the main notable exception. To estimate these observables, pyLIMASS uses an interpolation of stellar isochrones. In particular, the models described in \citet{Bressan2012}, available online\footnote{\url{http://stev.oapd.inaf.it/cgi-bin/cmd}}, that includes a broad range of age and metallicity values as well as a large catalog of passbands, have been implemented. By default (and for the rest of this study), pyLIMASS considers only stars with an age $\ge$ 1 Gyr, which can be adjusted to meet the user's needs. We would like to point out that pyLIMASS also supports the option to compute the magnitudes directly from spectral templates using the Spyctres library \footnote{\url{https://github.com/ebachelet/Spyctres}}. In short, Spyctres relies on stsynphot and synphot to simulate spectra of stars with the Kurucz \citep{Kurucz1993} or Phoenix \citep{Allard1995} libraries (depending on the user`s choice). However, this option is significantly slower and is not used in the following. The blackbody approximation option is also available, as it is much faster and still provides coherent results, except for the cooler stars. To estimate the absorption in the needed bands, pyLIMASS uses the absorption law of \citet{Wang2019}.

The observables $x$ are ingested in pyLIMASS as samples. If one has only knowledge of the mean $\mu$ and standard deviation $\sigma$ of given observables, then a sample of $\mathcal{N}(\mu,\sigma)$ is generated independently for each observable. However, if one has some prior knowledge about the correlation between observables, joint distributions can be used instead. For example, if the correlation between the parameters $\pi_{EN}$ and $\pi_{EE}$ is known from MCMC modeling of a specific event, then the corresponding samples should be included. 

To explore the parameter space, two algorithms are available by default in pyLIMASS. The \texttt{scipy} implementation \citep{Pauli2020} of the differential evolution algorithm \citep{Storn1997} is implemented to locate the region of maximum likelihood and the \texttt{emcee} package \citep{ForemanMackey2013} is available to run Monte-Carlo Marko Chain explorations. The description of the default parameters priors can be found in the Appendix. It is however straightforward for users to use their preferred sampler.

It is useful to know if a given model is plausible. One can compare the model likelihood $L$ to the GM model probability when $y=\mu_k$. Indeed, models centered close to the GM means will have a high likelihood (but potentially not the maximum). Noting $\hat{L}$ the likelihood at $y=\mu_k$, the quantity $-2(L-\hat{L})$ is chisquare distributed with d degrees of freedom, where d is the number of observables used.  Therefore, every model likelihood can be compared to $\hat{L}$ using a significance level $\alpha$. We have implemented such afunctionality in pyLIMASS and set a default of $\alpha=0.05$.

\begin{figure}
    \centering
    \includegraphics[width=0.99\textwidth]{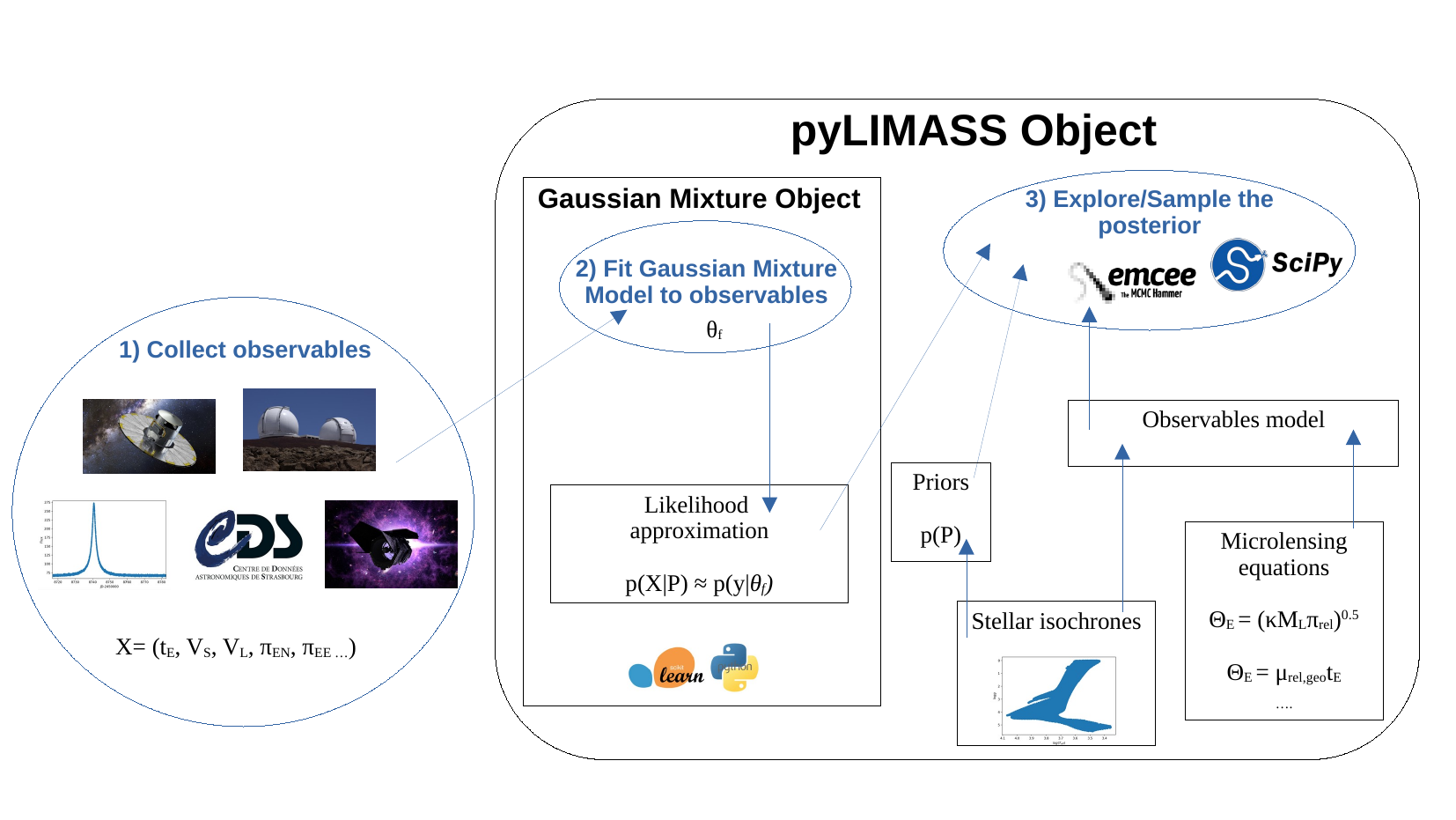} 
    \caption{Summary of the proposed algorithm. The first step is to collect the observables and inject them in the GM object. Then, the observables distribution are fitted by the GM model to found the best hyperparameters $\theta_f$. Finally, the likelihood function of the GM model is used in conjunction with user defined priors, stellar isochrones and microlensing parameters to explore the poterior probability $p(P|X)$.}
    
    \label{fig:pyLIMASS_chart}
\end{figure}

To conclude, the design of the algorithm can be found in the Figure~\ref{fig:pyLIMASS_chart} and can summarized in three steps as follow:
\begin{itemize}

    \item Collect the observables x and pass them to the algorithm in the form of distributions. It can be on the form of Normal distribution or MCMC chains for example
    \item Fit the observed distribution with a GM model to estimate $\theta_f$, via the EM algoritm for example. 
    \item Use the trained GM model to estimate the posterior probabilities of parameters P given the observables and priors via Equation~\ref{eq:Bayes} and Equation~\ref{eq:GMpdf}
   
\end{itemize}

\section{Tests on simulated examples}
\label{sec:simulations}

\subsection{A typical single lens event}
\label{sec:simpleexample}
In this section, we have simulated a "standard" microlensing event as it could be seen by the Roman microlensing survey. It is anticipated that Roman would observe the microlensing fields with at least two filters \citep{Penny2019}. Most of the images would be taken with the F146 wide filter but complementary observations with a $\sim$ daily cadence will be obtained with the narrow band F087 filter\footnote{\url{https://roman.gsfc.nasa.gov/science/WFI_technical.html}}. However, we note that the exact strategy is still not decided at the time of writing\footnote{\url{https://roman.gsfc.nasa.gov/science/ccs_community_input.html}} and multiple scenarii exist, that potentially include extra narrow band filters. We set the source to be a Sun like star with $M_S=1M_\odot$, $R_S=1 R_\odot$, $T_{eff}=5778$ K, $Fe_S=0$ dex, $logg_S=4.437$ cgs at $D_S=8$ kpc and with a visual absorption $A_V=4$ mag, with a (geocentric) proper motion $\mu_{S,N}=\mu_{S,E}=-3$ mas/yr. The lens is assumed to be an M-dwarf of $M_L=0.5M_\odot$, $R_L=0.5 R_\odot$, $T_{eff}=3600$, $Fe_L=0$ dex and $logg_L=4.74$ cgs at $D_L=4$ kpc and a visual absorption $A_V=2$ mag, with a (geocentric) proper motion  $\mu_{L,N}=2$ mas/yr and $\mu_{L,E}=-2$ mas/yr. Using these information, it is straightforward to generate any observables such as $t_E$, $\pi_E$, the lens and source magnitudes in the F087 and F146 bands and so on. Based on this simulation, several reconstruction fits have been performed using different set of observables. This intends to reproduce real case events where some quantities or effects are measured or not. The uncertainties of all observables have been arbitrary set to 10\% and without correlation. Results are summarized in Table~\ref{tab:simulatedevent}. The first experiment (Fit 1) used constraints from $\theta_E$, $D_S$, $\pi_{EN}$ and $\pi_{EE}$. This scenario could occur if Roman data measures the source parallax $\pi_S$ as well as the astrometric shift induced by the microlensing event \citep{Fardeen2023}, and therefore measuring $\theta_E$ directly, in addition of the measurement of the microlensing parallax vector $\vec{\pi_E}$ from the lightcurve modeling. In this case, the lens mass and distance are reconstructed with high fidelity, as one should expect according to Equation~\ref{eq:MLDL}.  However, the mass of the lens cannot be reconstructed to better than 14.6\% (i.e. $\pm0.073 M_\odot$, via error propagation) using only constraints from $\theta_E$, $\pi_{EN}$ and $\pi_{EE}$, and this is exactly what is recovered in Table~\ref{tab:simulatedevent}. Similarly, the lens distance is expected to be measure with a precision of 9 \% (i.e. $\pm0.36$ kpc) because it was assumed that the source distance is (somehow) known with a 10 \% precision. Fit 2 used $\theta_S$, $\rho_*$, $\pi_{EN}$ and $\pi_{EE}$ constraints and the reported precision on the lens mass measurement decreased as expected (the anticipated precision is 17\% in this case). But in this case the lens distance is not known because the source distance is only poorly constrained by $\theta_S$ and $\rho_*$. This scenario is realistic, as it is expected that $\rho_*$ should be measured on a significant number of events with lightcurves observed by the Roman space telescope, while $\theta_S$ should be routinely derived from the study of the color-magnitudes diagrams of Roman fields \citep{Yoo2004,Nataf2013}. The last scenario (Fit 3) corresponds to a case where few measurements can be made from the lightcurve itself (a single lens single source event for example), but the relative proper motion vector $\vec{\mu_{rel}}$ as well as the magnitudes of the source and lens in two bands have been measured via the modeling of the Roman images at the location of the microlensing event. This technique, often referred as "lens-flux analysis", is expected to provide constraints for a large fraction of events.  In this case, constraints from $t_E$, $F087_S$, $F146_S$, $F087_L$, $F146_L$, $\mu_{rel,N}$ and $\mu_{rel,E}$ have been used. In this scenario, while the precision of the lens mass decreases significantly, stellar parameters of the source and the lens, as well as the absorption, are known with better precision. The main exception is the metallicity, which is not constrained for all runs. Indeed, the reconstructed distribution of the metallicity is uniform and match the distribution of the isochrones (e.g. $Fe=-0.8\pm0.7$). This is expected, not only because the constraints on this parameter are the narrow $F087$ and wide $F146$ bands, but also because this parameter is highly degenerate with the age \citep{Worthey1994}. Similarly, the proper motion of the source and lens is always poorly constrained (but not their difference $\mu_{rel}$). Again, this is due to the lack of information in the provided observables.

\begin{table*}
\label{tab:simulatedevent}
    \centering
    \begin{tabular}{lcccc}
    \hline
    \hline
    
         Parameters [unit]& Model & Fit 1 & Fit 2 & Fit 3   \\[0.1cm]
         \hline 
          $M_S ~[M_\odot]$ & 1 &$1.1_{-0.4}^{+0.7}$ 
          &$1.1_{-0.4}^{+0.6}$ 
          &$1.1_{-0.4}^{+0.5}$  \\[0.1cm]
          $D_S ~[kpc]$ &8& $8.0_{-0.8}^{+0.8}$  
          &$34_{-22}^{+40}$ &
          $19_{-12}^{+23}$  \\[0.1cm]
          $T_{eff,S}~[K]$ & 5778 &$4800_{-1100}^{+1700}$ 
          & $5600_{-1100}^{+1900}$
          & $6100_{-1100}^{+1400}$ \\[0.1cm]
          $Fe_S~[dex]$ &0 &$-0.8_{-0.8}^{+0.9}$ 
          &$-0.8_{-0.8}^{+0.9}$ 
          & $-0.9_{-0.8}^{+0.9}$ \\[0.1cm]
          $logg_S~[cgs]$ &4.437 &$2_{-2}^{+2}$ 
          &$3.3_{-0.7}^{+0.9}$ 
          &$3.9_{-0.7}^{+0.7}$  \\[0.1cm]
          $\mu_{S,N}~[mas/yr]$ &-3& $-11_{-6}^{+10}$
          &$-11_{-7}^{+12}$ 
          &$-2_{-12}^{+12}$  \\[0.1cm]
          $\mu_{S,E}~[mas/yr]$ & -3&$-2_{-13}^{+12}$
          &$-2_{-12}^{+12}$  
          &$2_{-14}^{+13}$  \\[0.1cm]
          $A_{V,S}~[mag]$ &4&$1_{-1}^{+24}$ 
          &$1_{-1}^{+25}$ 
          &$4.0_{-0.8}^{+0.9}$  \\[0.1cm]
          $M_L ~[M_\odot]$ & 0.5 &$0.51_{-0.06}^{+0.07}$ 
          &$0.51_{-0.08}^{+0.09}$ 
          & $0.4_{-0.3}^{+0.3}$  \\[0.1cm]
          $D_L ~[kpc]$ & 4 &$4.0_{-0.3}^{+0.4}$ 
          &$6_{-2}^{+1}$ 
          &$4_{-2}^{+4}$  \\[0.1cm]
          $T_{eff,L}~[K]$ & 3660 &$3100_{-1700}^{+3700}$ 
          &$3100_{-1600}^{+3700}$ 
          &$3900_{-800}^{+1300}$  \\[0.1cm]
          $Fe_L~[dex]$ &0 &$-0.7_{-0.8}^{+0.9}$ 
          &$-0.8_{-0.8}^{+0.9}$ 
          &$-0.5_{-1.0}^{+0.7}$  \\[0.1cm]
          $logg_L~[cgs]$ &4.73 &$3_{-2}^{+2}$ 
          &$3_{-2}^{+2.0}$ 
          &$4.8_{-0.2}^{+0.3}$  \\
          $\mu_{L,N}~[mas/yr]$ & 2&$11_{-10}^{+6}$ 
          &$12_{-10}^{+6}$ 
          &$3_{-12}^{+12}$  \\[0.1cm]
          $\mu_{L,E}~[mas/yr]$ & -2&$3_{-13}^{+12}$ 
          &$2_{-12}^{+12}$ 
          &$3_{-14}^{+12}$ \\[0.1cm]
          $A_{V,L}~[mag]$ & 2&$0.1_{-0.1}^{+2.8}$ 
          &$0.1_{-0.1}^{+2.3}$ 
          &$2_{-2}^{+1}$  \\[0.1cm]

    \hline     
    \end{tabular}
    \caption{Estimated parameters of the source and the lens by using different set of observables. Fit 1 includes constraints from $\theta_E$, $D_S$, $\pi_{EN}$ and $\pi_{EE}$. Fit 2 uses $\rho_*$, $\theta_S$, $\pi_{EN}$ and $\pi_{EE}$ while Fit 3 uses $t_E$, $F087_S$, $F146_S$, $F087_L$, $F146_L$, $\mu_{rel,N}$ and $\mu_{rel,E}$.} 
    \label{tab:simulatedevent}
\end{table*}

\subsection{Simulation of a sample of lenses seen by Roman}
\label{sec:roman}
We generate a realistic sample of simulated events that can be expected from the Roman Galactic Bulge Time Domain Survey \citep{Penny2019}, with the Galactic model of \citet{Koshimoto2021}\footnote{\url{https://github.com/nkoshimoto/genstars}}. We generated 1000 single source and single lens events and simulated every lightcurve of the catalog using the pyLIMA software \citep{Bachelet2017}, including parallax \citep{Gould2004} and assuming Gaussian noise (with a zero point of $Z_p=27.4$ mag and an exposure time of 50 s). To ensure minimal signal in the lightcurves, $u_0$ was drawn from an uniform distribution $U(-1,1)$ while $t_0$ has been forced to be observed in the $\sim$ 72 day windows \citep{Penny2019}. We also consider that no other blends were present in the light of sight. Then, the gradient-like method included in pyLIMA was used to fit the lightcurve (and by starting from the true solution). The estimated model parameters and their corresponding uncertainties were ultimately used as a realistic set of observables for pyLIMASS. The parameters $t_E$, $\pi_{EN}$, $\pi_{EE}$ were used as observables, as well as constraints from $A_{V,S}$ and $\theta_S$ with a precision of 0.1 mag and 10\%, respectively, which are typical uncertainties reported in the literature. The apparent magnitudes of the source $m_S$ and lens $m_L$ in the F087 and F146 bands were also used, with a minimum precision of 0.001 mag. These measurements will be based on the lightcurve modeling as well as the modeling of the Roman images near the event location. A double-PSF fit provides strong constraints on the source and lens magnitudes as well as the relative proper motion $\mu_{rel}$ \citep{Bachelet2022}, which is included as the last observable. The typical astrometric precision can be approximated by $\Delta X={{FWHM}\over{\pi SNR}}$ \citep{Lindegren1978,Sozzetti2005}, where $FWHM$ is the Full-Width Half Maximum of the optical system and $SNR$ is the photometric signal-to-noise ratio. The anticipated Roman FWHM will be on the order of 1 pixel (i.e. 110 mas)\footnote{\url{https://roman.gsfc.nasa.gov/science/WFI_technical.html}} and therefore we assumed the relative precision of the relative proper motion to be equal to the less precise lens/source magnitudes components, i.e. $\Delta\mu_{rel}/\mu_{rel}\approx SNR=\Delta_{F146} = max(\Delta m_S,\Delta m_L)_{F146}$. The 1000 events were then analyzed with an MCMC exploratory run of the posterior using the 16 parameters P (after a search for the global maximum) and, for each event, we computed the mean and standard deviation of the chains for each parameters after the burn-in phase for convergence of the Markov chains.
\begin{figure}
    \centering
    \includegraphics[width=0.4\textwidth]{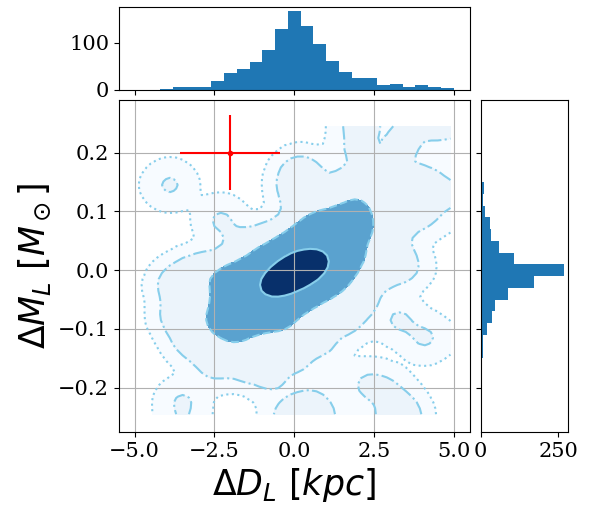}
    \includegraphics[width=0.4\textwidth]{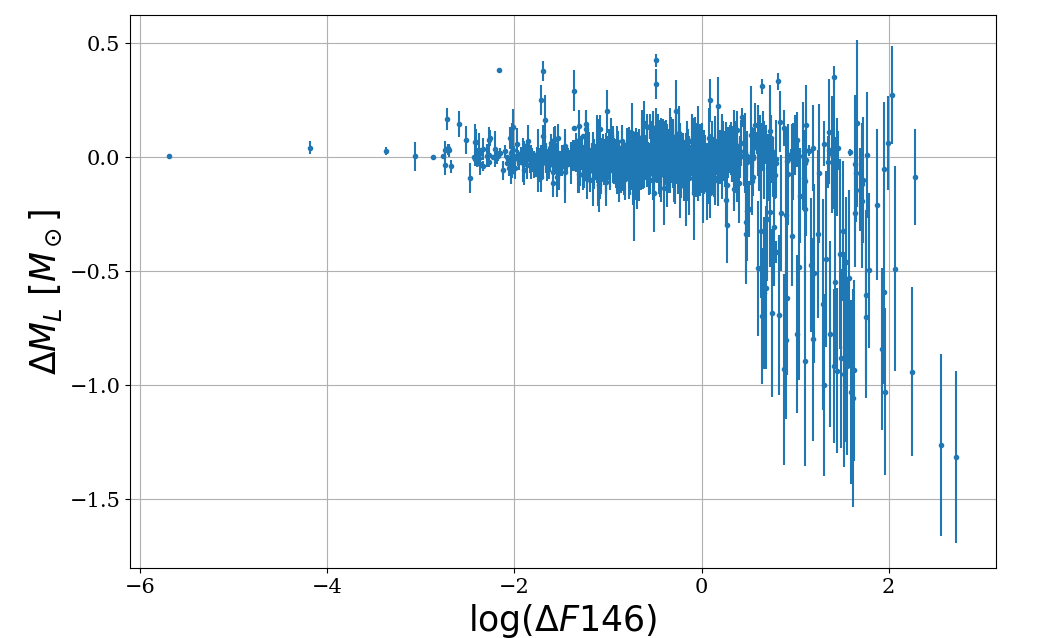}

    \caption{(Left)Lens mass accuracy distribution versus lens distance accuracy distribution for the sample of the Roman events presented in the text. The various levels of colours and lines indicates the 39\%, 86\%, 98.9\% and 99.9\% density contours from pyLIMASS while the 1-d distribution are also presented. The red cross indicated the median uncertainty, in both direction, reported by pyLIMASS. (Right) Distribution of the mass accuracy versus the maximum magnitude uncertainty between the lens and the source in the F146 band $\Delta F_{146}$.}
    \label{fig:Roman_MlDl}
\end{figure}

\begin{figure}
    \centering
    \includegraphics[width=0.99\textwidth]{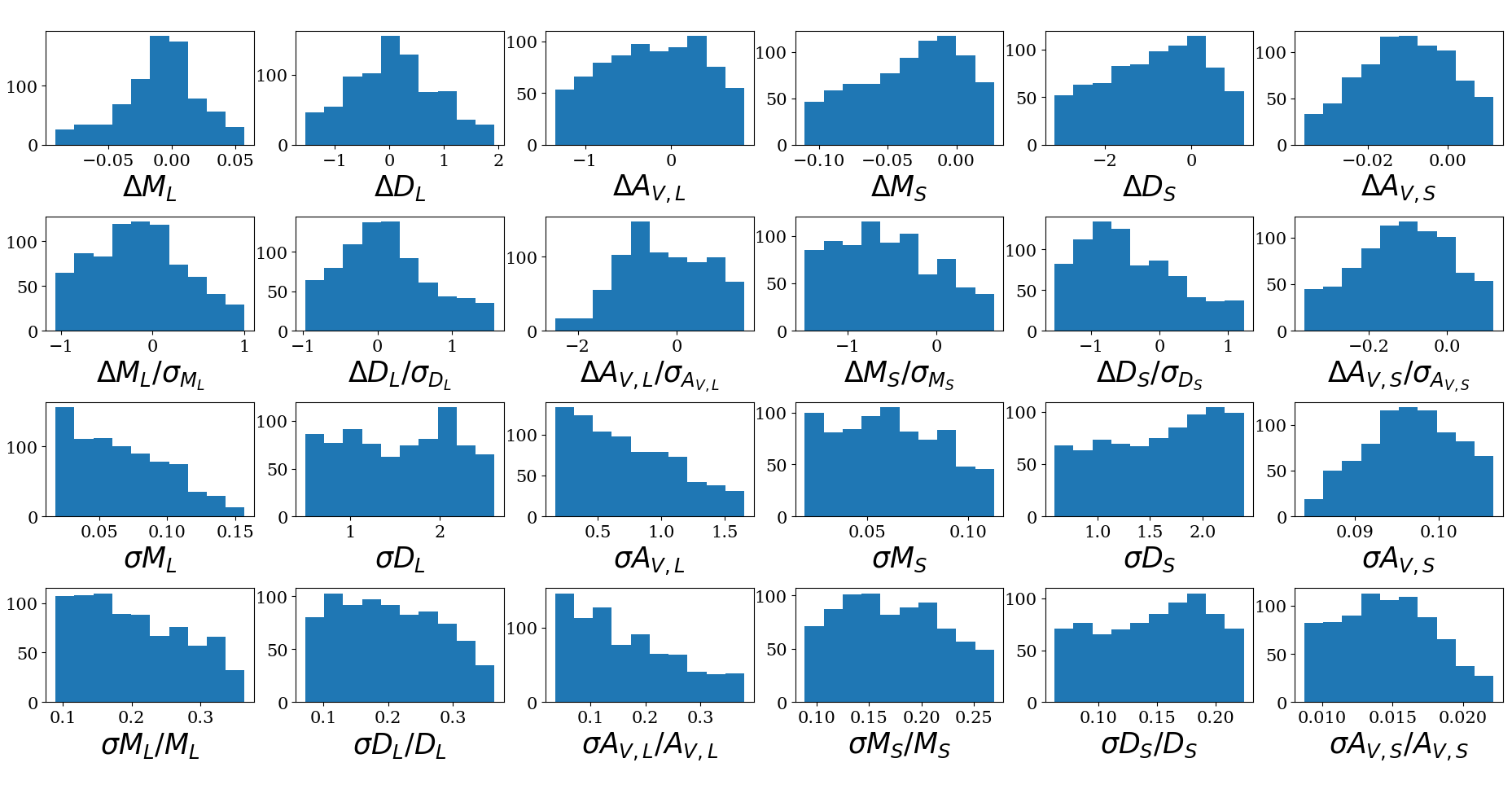} 
    \caption{Various distributions for six parameters of prime interest of the Roman sample. From top to bottom are displayed the distribution of the parameters accuracy, normalized accuracy, uncertainties and precision. The displayed range of parameters have been limited to the [10\%,90\%] percentiles for plotting purposes.}
    
    \label{fig:Roman_Summary}
\end{figure}

 In the Figure~\ref{fig:Roman_MlDl}, the reconstructed mass accuracy $\Delta M_L=M_{L,True}-M_{L,pyLIMASS}$ is presented versus the accuracy on the lens distance $\Delta D_L$. The overall accuracy is excellent with, a median $\Delta M_L=-0.007\pm0.0025 M_\odot$ and $\Delta D_L=0.08\pm0.75 $ kpc. However, a strong bias can be observed in the reconstructed lens mass for events where one of the components brightness (the source or the lens) is poorly known (i.e. very faint), that we quantified with  $\Delta_{F146}$. It is clear that lens masses are poorly reconstructed (both in terms of accuracy and precision) for $\Delta_{F146} \ge 1$ mag. This is expected because not only these magnitudes place constraints on the object masses via Equation~\ref{eq:mag_lens}, but also on $\mu_{rel}$ (and ultimately $\theta_E$). 
 Moreover, a clear correlation is visible in the Figure~\ref{fig:Roman_MlDl}. This is due to the fact that $\theta_E$ can be reconstructed from the injected observables (namely $t_E$ and $\mu_{rel}$) and that the lens mass can be approximated by $M_L\approx \theta_E^2  D_L/\kappa$ (Equation~\ref{eq:MLDL} ).  Figure~\ref{fig:Roman_Summary} shows the distribution of the accuracy, normalized accuracy, uncertainties and precision for six reconstructed parameters. The main result is that pyLIMASS can reconstruct the mass of the lenses with a median precision of ${{\sigma_{M_L}}\over {M_L}}=20 \%$ and almost no bias. The visual absorption values towards the sources are accurately reconstructed for most of the events, albeit with a small bias. This is to be expected, given the constraint on the absorption in the observables. However, less accurate and precise values are obtained for the distance and mass of the sources. This is because the algorithm can only put weak constraints on the source properties (via $\theta_S$ and the source magnitudes). And while the absorption is known accurately, the algorithm can adjust the stellar isochrones and the source distances to match the observed magnitudes. For example, if one selects isochrones within a specific color range $|F087-F146-0.5|\le0.1$ mag, the range of corresponding stellar parameters is wide, with $M\in[0.6,2]~M_\odot$, $T_{eff}\in [5000,6000]$ K, $[M/H]\in[-2,0.7]$ dex and $logg\in[4.8,1.9]$. Therefore, we argue that the accuracy and precision observed in the reconstructed parameters are fundamental issues rather than intrinsic problems of the algorithm, given the set of observables provided.

\section{Observed examples}
\label{sec:publishedevents}
In this section, we apply pyLIMASS on observed events published in the literature. Because all of these events, with the exception of Gaia16aye \citep{Wyrzykowski2023}, are located towards the Galactic Bulge,  we can only make a conservative assumption about the upper limit of the source distance $D_S<15$ kpc \citep{Penny2019}. A summary of the results can be found in the Table~\ref{tab:pyLIMASSvsPublications} but every events are discussed in more details thereafter.

\begin{table*}
    \centering
    \begin{tabular}{lcccc}
    \hline
    \hline
    
         Event & Reference  & $M_L~[M_\odot]$ & $D_L$ [kpc] & $D_S$ [kpc]   \\[0.2cm]
         \hline 
         \multirow{ 2}{*}{MOA-2009-BLG-266Lb} & \citet{Muraki2011}  & $0.56\pm0.09$ &$3.04\pm0.33$& $8.8$  \\
          &pyLIMASS & $0.56\pm0.12$ &$3.63\pm0.55$ & $11.4\pm2.2$\\ \hline
          
          \multirow{ 4}{*}{OGLE-2005-BLG-169Lb} &\citet{Gould2006} & $0.49\pm0.3$ &$2.7\pm1.6$&- \\
          &\citet{Batista2015}  & $0.65\pm0.05$ &$4.0\pm0.4$& $8.5\pm1.5$ \\
          &\citet{Bennett2015}  & $0.69\pm0.02$ &$4.1\pm0.4$& -\\
          &pyLIMASS & $0.7\pm0.1$ &$4.9\pm0.7$ & $11.7\pm1.9$ \\ \hline

          \multirow{ 2}{*}{OGLE-2017-BLG-1254L}
           & \citet{Zang2020}  & $0.60\pm0.03$ &$7.27\pm0.11$& $7.8\pm0.8$ \\
          &pyLIMASS & $0.62\pm0.07$ &$11.1\pm1.5$ & $12.4\pm1.9$ \\ 
          
           &pyLIMASS & $0.59\pm0.06$ &$7.5\pm0.7$ & $8.1\pm0.8$\\
           \hline

          \multirow{ 6}{*}{OGLE-2019-BLG-0960Lb} & \citet{Yee2021}  & $0.43\pm0.10$ &$0.83\pm0.17$& $7.56$\\
          &\citet{Yee2021} & $0.50\pm0.12$ &$0.98\pm0.21$& $7.56$\\
          &\citet{Yee2021}  & $0.42\pm0.09$ &$0.70\pm0.13$& $7.56$\\
           &\citet{Yee2021}  & $0.48\pm0.11$ &$0.81\pm0.17$& $7.56$ \\
          &pyLIMASS& $0.49\pm0.09$ &$1.3\pm0.3$ & $9.3\pm2.7$ \\
           &pyLIMASS& $0.5\pm0.1$ &$1.3\pm0.3$ & $9.4\pm2.7$  \\ \hline
  
          \multirow{ 3}{*}{MOA-2013-BLG-220Lb} & \citet{Vandorou2020}  & $0.88\pm0.05$ &$6.72\pm0.59$& $8.19\pm0.76$ \\
          &pyLIMASS & $0.99\pm0.16$ &$8.7\pm1.2$ & $11.3\pm1.8$\\ 
          &pyLIMASS & $0.624\pm0.058$ &$3.96\pm0.35$ & $4.72\pm0.47$\\ 
          \hline
          
          \multirow{ 2}{*}{Gaia16aye} & \citet{Wyrzykowski2020} & $0.93\pm0.07$ &$0.78\pm0.06$& $15.7\pm3.3$  \\
          &pyLIMASS& $0.97\pm0.13$ &$0.759\pm0.092$ & $15.3\pm3.3$ \\ 
          \hline
    \hline     
    \end{tabular}
    \caption{pyLIMASS estimated lenses mass $M_L$, lenses distance $D_L$ and sources distance for several events published in the literature. For the event MOA-2009-BLG-266Lb, $(V_S,I_S,H_S),H_{baseline},\pi_E,\phi_E,\rho_*,A_{V,S}$ have been used as observables. For OGLE-2005-BLG-169Lb, $t_E,\rho_*,(B_S,V_S,I_S,H_S),(B_L,V_L,I_L,H_L),\vec{\mu_{rel,helio}}$ have been used. Two runs have been made for the event OGLE-2017-BLG-1254L with $\pi_{EN},\pi_{EE},\rho_*,(I_S,H_S)$ and one run include $D_S$. Two runs have also been run in the case of OGLE-2019-BLG-0960Lb with $t_E,\rho_*,\pi_{EN},\pi_{EE},\rho_*,(V_S,I_S)$ and  $(I_L=18.65\pm0.1)$ mag or $(I_L\ge18.65)$ mag. $t_E,\rho_*,A_{V,S},\rho_*,(V_S,I_S),(K_{s,L})$ have been used for the event MOA-2013-BLG-220Lb and a second run also includes $I_L=21.10\pm0.1$ mag. For Gaia16aye, the observables used are $t_E,\rho_*,\vec{\pi_E},T_{eff,S},logg_S,Fe_S,(V_S,R_S,I_S)$.   } 
    \label{tab:pyLIMASSvsPublications}

\end{table*}

\subsection{MOA-2009-BLG-266Lb}
\begin{figure}[!h]
    \centering
    \includegraphics[width=0.5\textwidth]{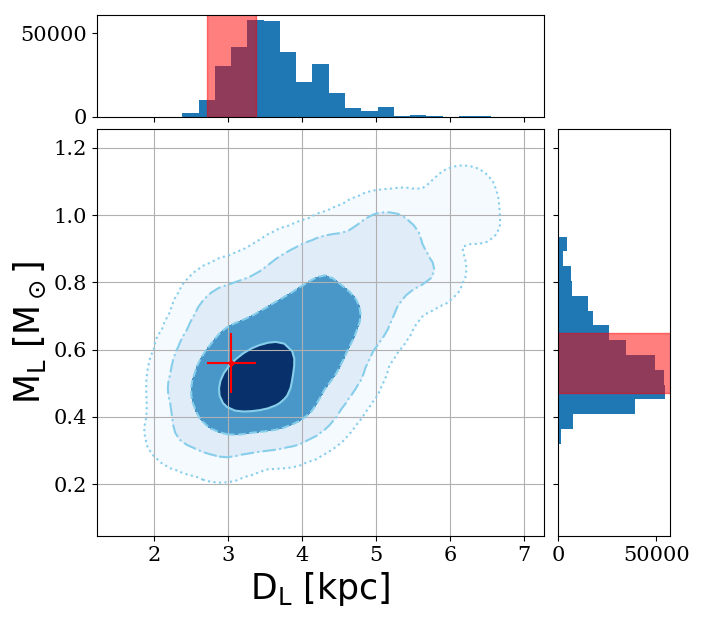}
    \caption{Lens mass versus lens distance for MOA-2009-266Lb \citep{Muraki2011}. The various levels of colours and lines indicates the 39\%, 86\%, 98.9\% and 99.9\% density contours from pyLIMASS while the measurement of \citet{Muraki2011} is displayed in red. The contours have been smoothed with a Gaussian kernel for plotting purposes.}
    \label{fig:kb09266}
\end{figure}

\citet{Muraki2011} reported the discovery of the anomalous microlensing event MOA-2009-BLG-266Lb. The lightcurve of the event exhibits finite-source effects and an annual parallax signal leading to the measurement of the Einstein ring radius $\theta_E=0.98\pm0.04$ mas and $\vec {{\pi_E}}=(\pi_{EN},\pi_{EE})=(0.1665\pm0.0503,0.1439\pm0.0035)$. Ultimately, the authors found the lens system to be composed of a super-Earth mass planet $m_p=10.4\pm1.7 M_\oplus$ orbiting a $M_L=0.56\pm0.09M_\odot$ host at $D_L=3.04+0.33$ kpc. They reported the source to be a red giant with $V_S=17.677\pm0.03$ mag, $I_S=15.856\pm0.030$ mag and $H_S=13.780\pm0.030$ mag. They also obtained high-resolution images with the NACO instrument on the Very Large Telescope (VLT) and the total flux at the target location is $H_{tot}=13.755\pm0.05$ mag. As displayed in the Figure~\ref{fig:kb09266}, by combining in the observables the source magnitudes (in V, I and H bands) as well as the baseline magnitude in H band (and assuming a conservative 0.1 mag errors for each measurements), the parallax vector in polar coordinates $\vec{{\pi_E}}=(\pi_E,\phi_E)$ and $\rho_*=0.00539\pm0.00011$, pyLIMASS estimates the lens system to be composed of a  $m_p=10.5\pm1.9 ~M_\oplus$ planet orbiting a $M_L=0.56\pm0.12~M_\odot$ host located at $D_L=3.63\pm0.55$ kpc from Earth, a solution in excellent agreement with \citet{Muraki2011}. However, it is relevant to point small discrepancies. First, pyLIMASS estimates the lens to be slightly more distant. This is mainly due to the fact that the source distance is not well constrained by the given observables with $D_S=11.4\pm2.2$ kpc, while \citet{Muraki2011} assumed $D_S=8.8$ kpc. pyLIMASS also estimates the angular source radius to be $\theta_S=4.94\pm0.36~ \mu as$, slightly smaller than the value reported by \citet{Muraki2011} with $\theta_S=5.2\pm0.2~\mu as$. We note however that using the color-angular radius relation of \citet{Adams2018}, the angular source radius is predicted to be $\theta_S=4.6 \mu as$ and $\theta_S=5.1\mu as$ (using the (V-I) and (V-H) relations of the "All Stars", respectively). Because $\rho_*$ is precisely measured, a small decrease in the angular source radius automatically leads to a smaller Einstein ring radius, that pyLIMASS estimates to be $\theta_E=0.919\pm0.069$ mas, and ultimately a smaller $\pi_{rel}$ since $M_L$ is precisely measured. Finally, we note that the pyLIMASS estimate of the source visual absorption $A_{V,S}=1.89\pm0.34$ mag is lower than the measurement of \citet{Muraki2011} $A_{V,S}=2.14$ mag, but in agreement with the OGLE Extinction Calculator\footnote{\url{https://ogle.astrouw.edu.pl/}}, based on the measurements of \citet{Nataf2013}, that reports $A_{V,S}=1.84$ mag. 

\subsection{OGLE-2005-BLG-169Lb}
Originally described in \citet{Gould2006} (G06), \citet{Batista2015} (Ba15) and \citet{Bennett2015} (Be15) have reanalyzed this event using high-resolution observations from HST and Keck and they were able to measure the relative proper motion as well as the source and lens magnitudes in the B, V, I and H bands. 

\begin{figure}[!h]
    \centering
    \includegraphics[width=0.5\textwidth]{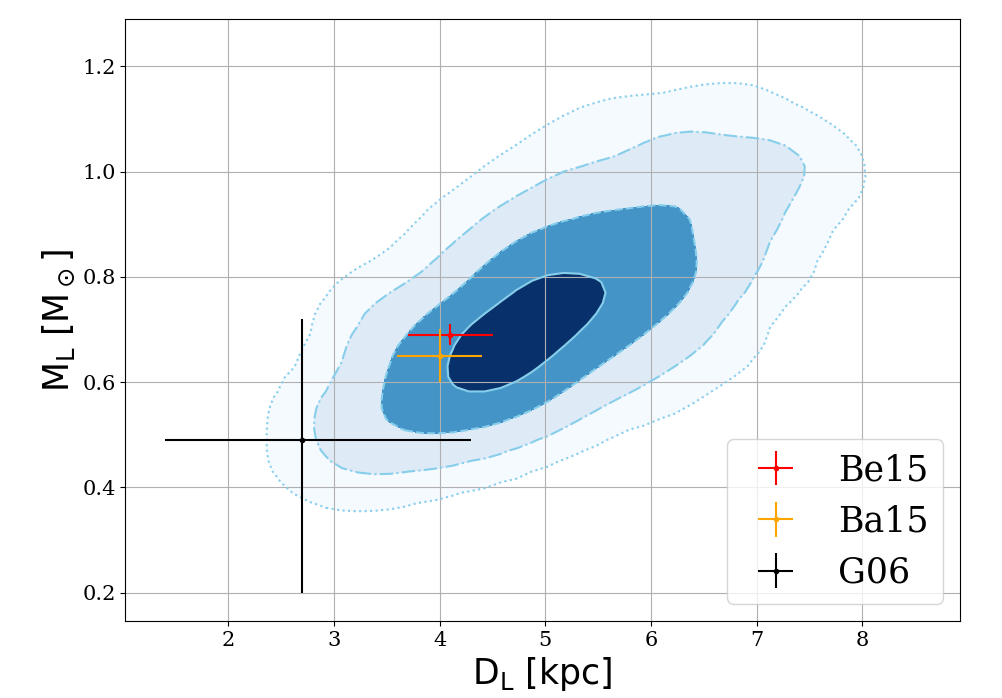}
    \caption{Same as Figure~\ref{fig:kb09266} but for the microlensing event OGLE-2005-169Lb. The measurements of G06, Ba15 and Be15 are displayed in black, orange and red respectively.}
    \label{fig:ob05169}
\end{figure}

As observables, we used $t_E=41.8\pm2.9$ days, $\rho_*=0.00048\pm0.00004$, the source magnitudes $B_S=23.382\pm0.06$ mag, $V_S=22.21\pm0.04$ mag, $I_S=20.555\pm0.05$ mag $H_S=18.81\pm0.08$ mag and the lens magnitudes $B_l=24.72\pm0.15$ mag, $V_L=22.783\pm0.07$ mag, $I_L=20.493\pm0.05$ mag and $H_L=18.20\pm0.10$ from Ba15 and Be15. We also used $\mu_{rel,hel,N}=4.87\pm0.12$ mas/yr and $\mu_{rel,hel,E}=5.63\pm0.12$ mas/yr from Ba15. pyLIMASS estimates are consistent with the three studies, as shown in the Figure~\ref{fig:ob05169}, with $M_L=0.72\pm0.10 M_\odot$ and $D_L=4.94\pm0.70$ kpc. pyLIMASS estimates of $\theta_S=0.401\pm0.023\mu as$ and $\theta_E=0.823\pm0.040$ mas are compatible but smaller compared to G06 measurements of $\theta_S=0.44\pm0.04\mu as$ and $\theta_E=1.00\pm0.22$. However, we note, that the angular source radius is precisely known from the measurement of the relative proper motion, since:
\begin{equation}
\theta_S = \theta_E\rho_* = \mu_{rel,geo} \rho_* t_E
\label{eq:theta_Smu_rel}
\end{equation}
Using the $\mu_{rel,geo}=7.0\pm0.2$ mas/yr estimate from Ba15 with the $\rho_*$ and $t_E$ values from Be15, we found $\theta_S=0.387\pm0.048\mu as$, close to the pyLIMASS estimate, but different from the values of G06 and Ba15 (Be15 does not report $\theta_S$ values in their Table 1). We note that pyLIMASS estimates of the Einstein ring are very similar to the one obtained by Be15, i.e. $\theta_E=0.848\pm0.027$ mas, when the constraints of the relative proper motion $\mu_{rel}$ are integrated in the modeling. Again, the discrepancy between the lens distance is due to the difference in the source distance estimated by pyLIMASS with $D_S=11.7\pm1.9$ kpc, while Ba15 assumed $D_S=8.5\pm1.5$ kpc. 

\subsection{OGLE-2017-BLG-1254L}
\begin{figure}[!h]
    \centering
    \includegraphics[width=0.5\textwidth]{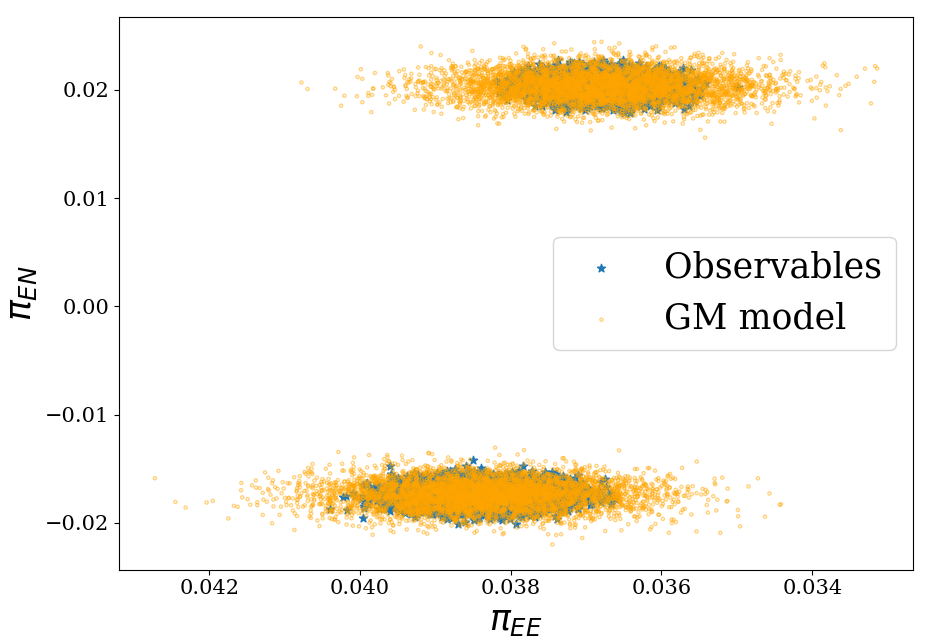}
    \caption{Distribution of the parallax components for the two models of OGLE-2017-1254L described in \citet{Zang2020} (blue stars) and samples generated by the GM model (orange dots).}
    \label{fig:ob171254_gm}
\end{figure}

\begin{figure}[!h]
    \centering
    \includegraphics[width=0.95\textwidth]{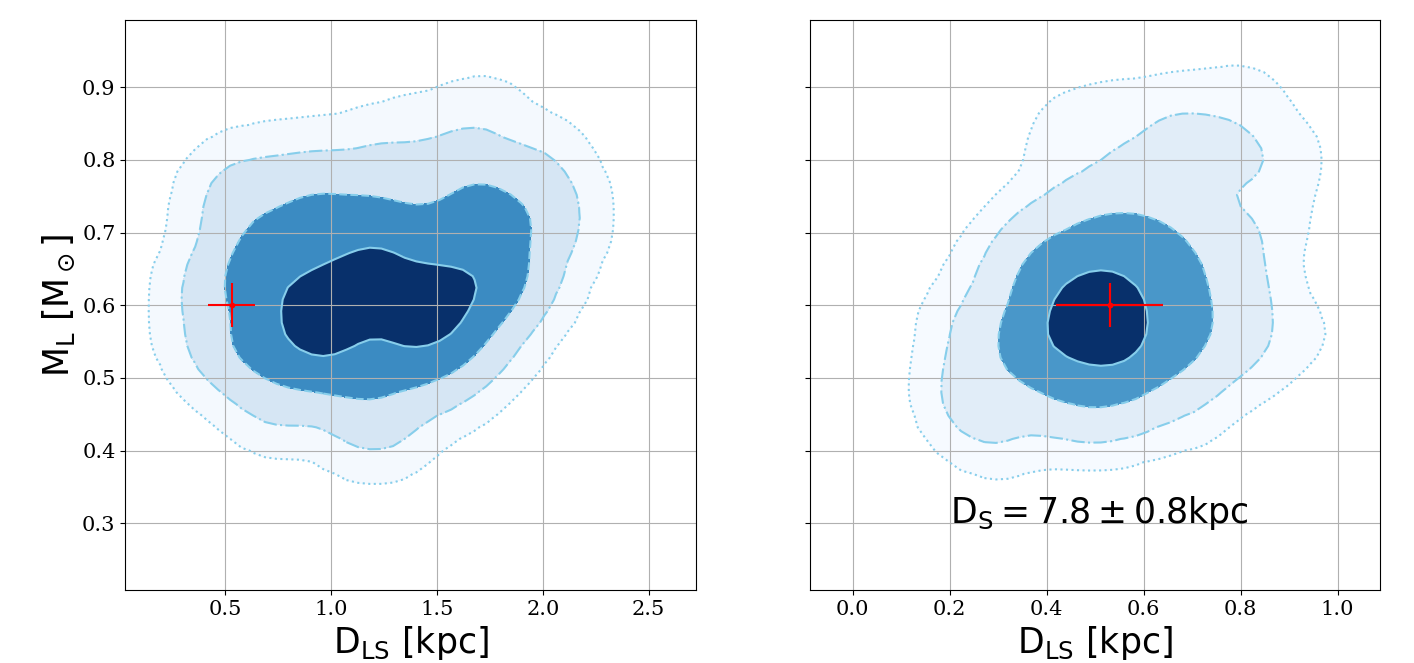}
    \caption{Distribution of the lens mass and the lens-source distance for the microlensing event OGLE-2017-1254L \citep{Zang2020}.(Left) The pyLIMASS distribution with weak constraints on the source distance (i.e. $2<D_S<15$kpc) and in red the measurements from \citet{Zang2020}. (Right) The same distribution but assuming the source distance $D_S=7.8\pm0.8$ kpc as an observable.}
    \label{fig:ob171254}
\end{figure}

As presented in \citet{Zang2020}, this event has been monitored by ground based telescopes as well as the Spitzer Space Telescope. The separation between the Earth and Spitzer at the time of the event allowed a precise measurement of the parallax vector. Despite a bimodal degeneracy, the impact on the lens mass and distance is not substantial \citep{Zang2020}. It is, however, a good test case for pyLIMASS as the two different modes can be injected simultaneously in the observables as the GM can handle multi-modal distributions, as can be seen in the Figure~\ref{fig:ob171254_gm}. We note that initial guesses on the position needs to be passed on to the GM to obtain good convergence of the model. Using $\pi_{EN}$, $\pi_{EE}$, $\rho_*$, $I_S$ and $H_S$ from \citet{Zang2020}, we obtain the lens-source distance $D_{LS}=1.21\pm0.35$ kpc and $M_L=0.62\pm0.07M_\odot$, in agreement with the published results $M_L=0.60\pm0.03$ and $D_{LS}=0.53\pm0.11$ kpc. Again, the relative low precision on $D_{LS}$ is mainly due to the low precision on the source distance estimate with $D_S=12.4\pm1.8$, to be compared with $D_S=7.8\pm0.8$ kpc from \citet{Zang2020}. Indeed, as pointed by \citet{Zang2020}, the distance between the
lens and the source $D_{LS}=D_S(\epsilon_{D}-1)$ is generally determined more precisely than $D_S$ and $D_L$ separately. As can be seen in the Figure~\ref{fig:ob171254}, if one assumes the source distance as an observable, the mass estimate from pyLIMASS stays unchanged with $M_L=0.59\pm0.06 M_\odot$ but the lens-source distances are almost identical to the published values with $D_{LS}=0.52\pm0.10$ kpc.

\subsection{OGLE-2019-BLG-0960Lb}
Presented by \citet{Yee2021} as the "smallest microlensing planet" (it was rather the smallest mass ratio microlensing planet at the time), the lens system is well characterized because both finite-source effets with $\rho\sim3\times10^{-5}$ and the parallax vector are well measured, despite the four-fold degeneracy \citep{Gould2004}. \citet{Yee2021} also secure three bands measurements of the source magnitudes that includes $V_s=21.34\pm0.1$ mag and $I_s=19.81\pm0.05$ mag. Coupled with their measurements of the extinction $A_{V,S}=1.64\pm0.1$ mag, they estimated $\theta_S=0.625\pm0.028 \mu as$ and then $\theta_E=1.9-2.1$ mas depending on the lightcurve modeling solutions. Ultimately, the four solutions are in agreement at the 1-$\sigma$ level with $M_L\sim0.45M_\odot$ at $D_L\sim0.8$ kpc, compatible with the measured blend light $I_b=18.65\pm0.1$ mag \citep{Yee2021}. We note that their source color $(V-I)_0=0.86$ mag is compatible with an early K-dwarf star \citet{Bessel1988, Pecaut2013}. However, this source stellar type is in tension with their adopted source distance of $D_S=7.56$ kpc. Indeed, at this distance, one can anticipate a significantly smaller source angular radius of $\theta_S\sim0.5\mu as$. 

\begin{figure}[!h]
    \centering
    \includegraphics[width=0.95\textwidth]{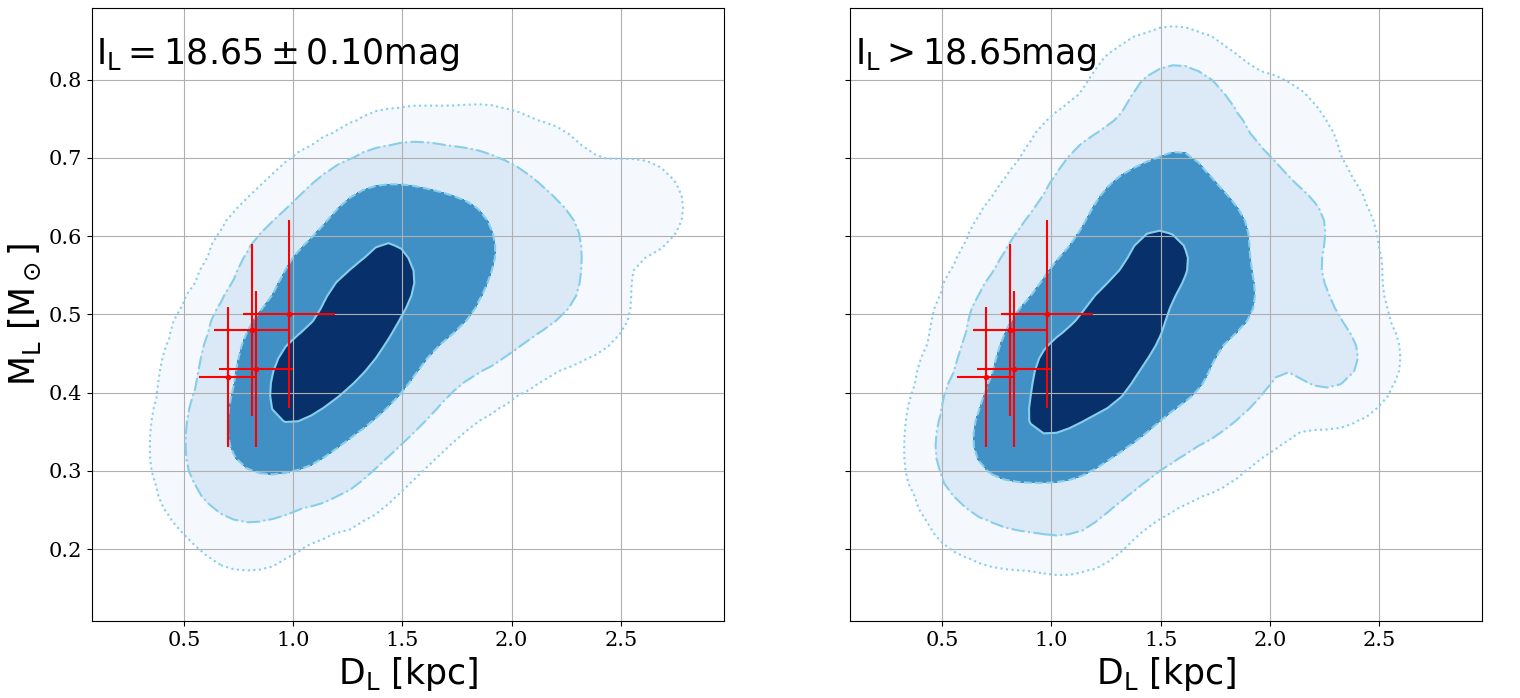}
    \caption{Distribution of the lens mass and the lens distance for the microlensing event OGLE-2019-0960Lb \citep{Yee2021}.(Left) The pyLIMASS distribution with strong constraint on the lens flux and in red the measurements from \citet{Yee2021}. (Right) The same distribution but assuming an lower limit on the lens brightness.}
    \label{fig:ob190960}
\end{figure}

We have injected the four different solutions of \citet{Yee2021} in pyLIMASS and used the observables $t_E$, $\rho_*$, $\pi_{EN}$, $\pi_{EE}$, the source magnitudes in the $V_S$ and $I_S$ bands as well as the lens magnitude in the $I$ band. For the latter, we assumed two different distributions (and so two runs of pyLIMASS). The first run is assuming that the entire blend light is due to the lens ($I_L=I_{blend}$) and the second run uses only a lower limit ($I_L>I_{blend}$). Results are presented in the Figure~\ref{fig:ob190960}. Both scenario are well compatible and found a slightly smaller Einstein ring radius $\theta_E=1.6\pm0.2$ mas, due to the smaller angular source radius  $\theta_S=0.51\pm0.05 \mu as$. Both runs agree at the 1-$\sigma$ level with \citet{Yee2021} that the lens is an M-dwarf $M_L=0.49\pm0.09 M_\odot$ at $D_L=1.3\pm0.3$ kpc (for $I_L=I_{blend}$) and $M_L=0.5\pm0.1 M_\odot$ at $D_L=1.3\pm0.30$ kpc (with $I_L\ge I_{blend}$), and a similar source distance of $D_S=9.3\pm2.7$ kpc.

\subsection{MOA-2013-BLG-220Lb}

\begin{figure}[!h]
    \centering
    \includegraphics[width=0.95\textwidth]{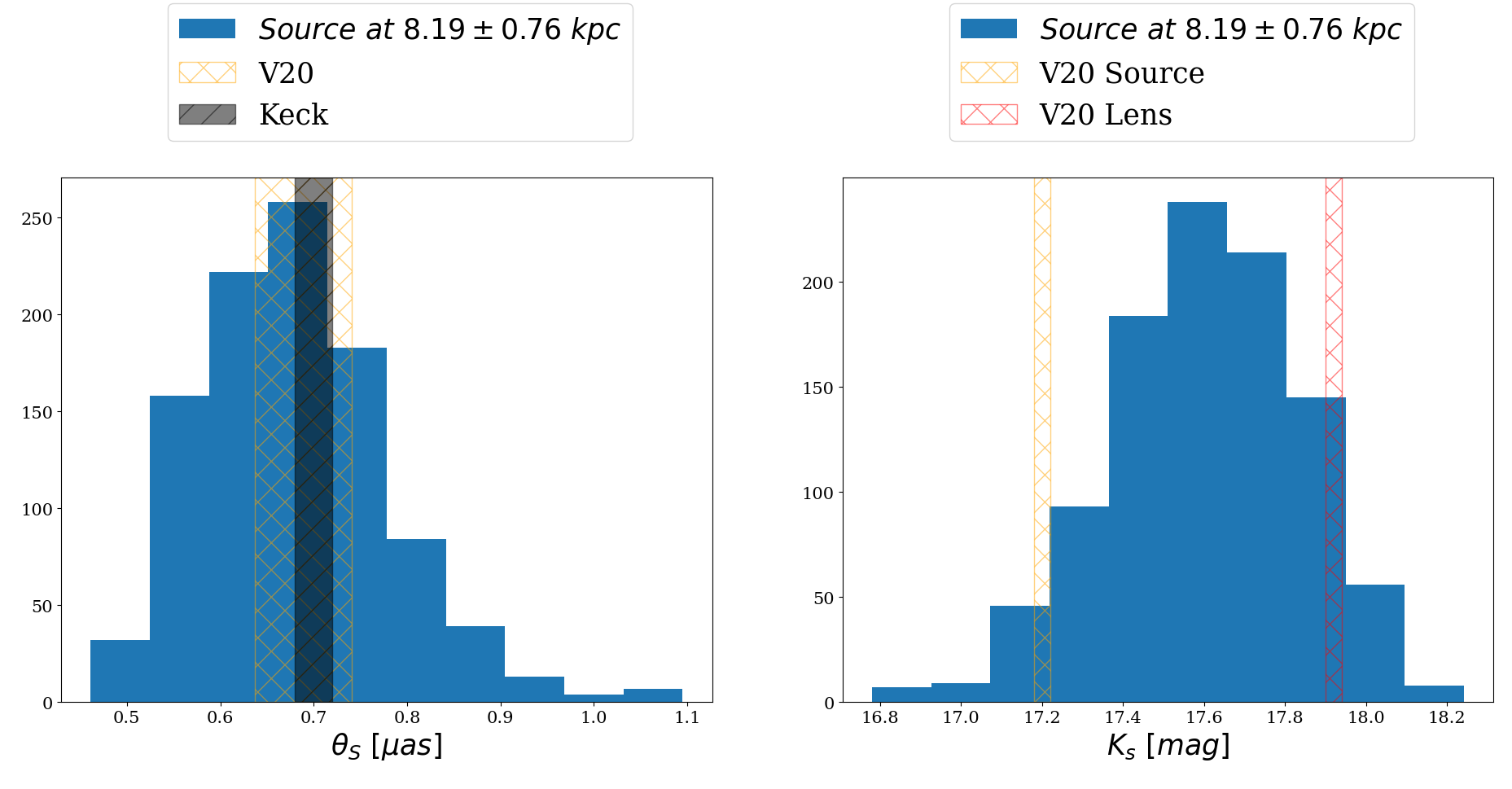}
    \caption{(Left) Reconstruced angular source radius of the source of MOA-2013-220Lb from the $(V,I)_S$ measurements of \citet{Vandorou2020} and the stellar isochrones of \citet{Bressan2012}, assuming the source at $D_S=8.19\pm0.76$ kpc. This is in good agreement with the estimation of V20 (yellow) as well as the measurement from Equation~\ref{eq:theta_Smu_rel}. (Right) The estimated magnitude of the source in the $K_s$ band from the isochrones is in tension in the measurement of the source (yellow) magnitude reported by V20. The lens magnitude is alsor reported in red.}
    \label{fig:KB13220_error}
\end{figure}

Originally published by \citet{Yee2014} (Y14 thereafter), MOA-2013-220Lb shows strong anomalies at its peaks that allows a precise measurement of the lens-planet mass ratio $q=0.003$ as well as $\rho_*$ and $t_E$. Unfortunately, only a single mass-distance relation can be extracted from the lightcurve ($\theta_E$) and Y14 could not make any firm conclusions on the nature of the host. However, they were able to place upper limits on the lens mass $M_L<0.77M_\odot$ and distance $D_L<6.5$ kpc, based on blend flux arguments. \citet{Vandorou2020} (V20 thereafter) published a follow-up study of this event where they observed the event location with high-resolution imagers with the Keck telescope in 2015 and 2019. Their results are in contradiction with Y14 limits as they conclude the lens to be $M_L=0.88\pm0.05 M_\odot$ at $D_L=6.72\pm0.59$ kpc. Using slightly different methods, both studies agree on the measurement of the angular source radius $\theta_S$, and finally on the measurement of $\theta_E$ since $\rho_*$ is well constrained by the lightcurve. For instance, V20 found $\theta_S=0.689\pm0.052 \,\mu as$ based on the source properties $((V-I),I)_{0,S} = (0.592\pm0.038,18.057\pm0.054)$. However, this source color is in tension with their Keck source measurement in the $K_s$ band. Indeed, V20 reported $K_s=17.20\pm0.02$ mag and therefore, assuming $A_{Ks}\sim0.2$ mag based on their measurement of $A_V=2.1$ mag, the measured color is $(I-K_s)_0=1.06$ mag, 0.3 mag redder that the expected infrared color derived from their optical color and the tabulations of \citet{Bessel1988} and \citet{Pecaut2013}. Similarly, the spectral type of such main sequence star implies $R_S\sim1.4~ R_\odot$ \citep{Pecaut2013}, leading to a significantly different angular radius $\theta_S=0.795 \mu as$ if the source is located at $D_S=8.19\pm0.76$ kpc (V20). In fact, similarly to OGLE-2005-BLG-169Lb, the source and Einstein ring angular radii are known with great precision from the measurements of $t_E$, $\rho_*$ and $\mu_{rel}$ (i.e. Equation~\ref{eq:theta_Smu_rel}), leading to $\theta_E=0.454\pm0.005$ mas and $\theta_S=0.70\pm0.01 \mu as$. As can be seen in the Figure~\ref{fig:KB13220_error}, the angular radius of the source measured from the proper motion accurately matches the one predicted by V20, strongly raising the confidence in the $(V-I)_S$ color and absorption measurements. We also present in the Figure~\ref{fig:KB13220_error} the result of a Monte-Carlo experiment designed to recover the angular radius of the source based on the closest match between 
 $(V,I)_S$ mag and the stellar isochrones of \citet{Bressan2012}, that is also in good agreement with V20. However, if one assumes $D_S=8.19\pm0.76$ kpc from V20, the predicted $K_s$ magnitude of the source from this sample is  $K_{s,S}=17.6\pm0.2$ mag, in tension with the $K_{s,S}=17.20\pm0.02$ mag from V20. The cause of these tension is not clear and it could be due to incorrect calibrations, a wrong estimation of the absorption  and distance to the source, a source/lens misendification or the presence of a companion to the source. The latter hypothesis seems plausible. Indeed, if one assumes the infrared magnitude of the primary to be $K_s=17.6$ mag, then the potential companion predicted magnitude would be $K_s=18.5$ mag, too faint to be detected in the $V$ and $I$ bands. The FWHM of 2019 Keck images presented in V20 were $\sim60$ mas, so any binary stars with a projected separation $a_\perp\le60~\rm{mas}\times8.19~\rm{kpc}\le500$ AU would be included in the FWHM of the images, which is realistic.

\begin{figure}[!h]
    \centering
    \includegraphics[width=0.9\textwidth]{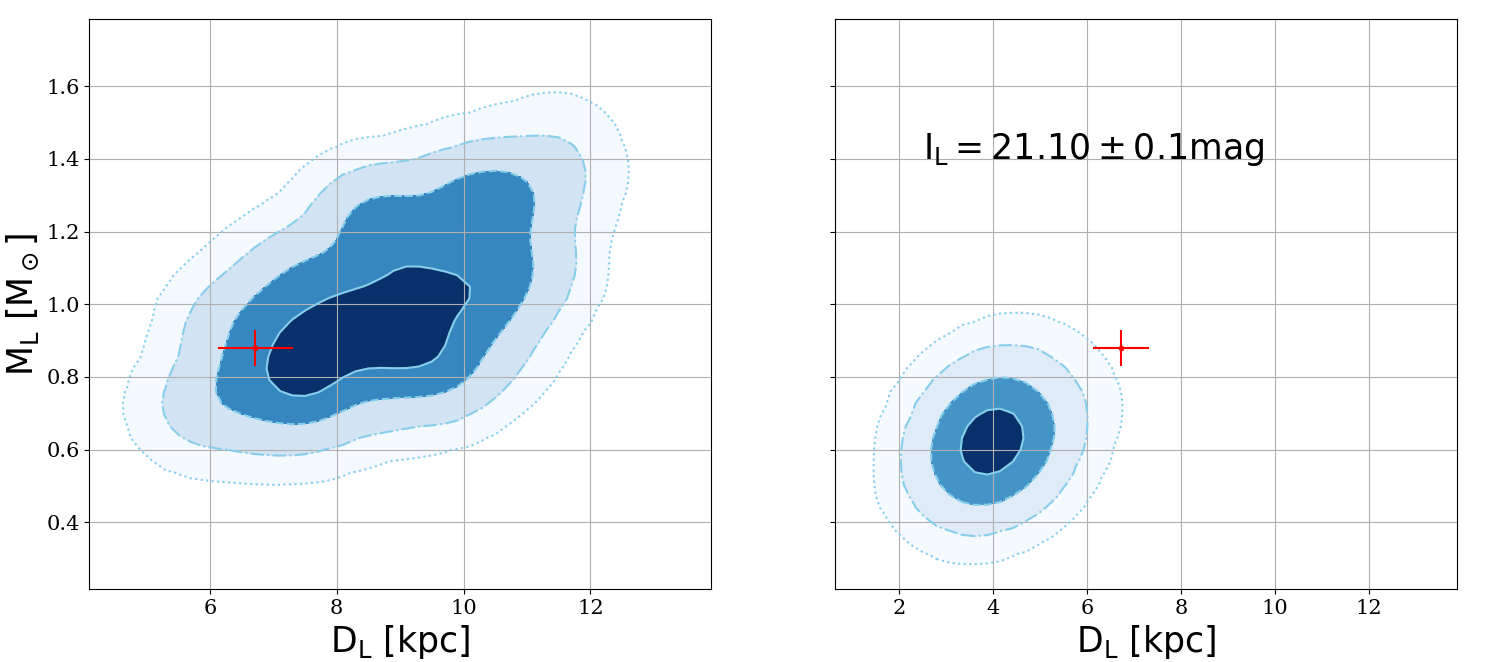}
    \caption{Distribution of the lens mass and the lens distance for the microlensing event MOA-2013-BLG-220Lb \citep{Yee2014,Vandorou2020}.(Left) The pyLIMASS distribution with the lens flux in $K_s$ bands and in red the measurements from V20. (Right) The same distribution but assuming the extra constraint on the lens flux $I_L=21.10\pm0.1$ mag.}
    \label{fig:kb13220}
\end{figure}

It is clear nevertheless from Figure~\ref{fig:KB13220_error} that the optical colors of the source are in good agreement with the derived relative proper motion from the Keck images. Therefore, we included these information in the pyLIMASS estimation, along with the measurements of $t_E$, $\rho$, $A_{V,S}$ and $K_{s,L}$ of V20, and results are presented in the Figure~\ref{fig:kb13220}. The estimated mass agrees with, but is less precise than, V20 with $M_L=0.99\pm0.16 M_\odot$, while lens distance is poorly constrained with $D_L=8.7\pm1.2$ kpc, mainly because the source distance is almost unconstrained by the observables with $D_S=11.3\pm1.8$kpc. The main reason is that, while the effective temperature of the source has been well measured with $T_{eff,S}=6200\pm200$ K, the absolute magnitudes of the source is not well constrained. Indeed, a wide range of metallicity ($Fe=-1.00\pm0.70$) and surface gravity values ($logg=3.95\pm0.16$) is possible at this location of the Hertzsprung-Russel diagram. The decrease in precision for the lens mass is also due to the fact that V20 used isochrones with a fixed age of 6.4\,Gyr, while pyLIMASS considers a wider range of ages ([1-10] Gyr), allowing for more flexibility in term of mass and luminosity. Moreover, the estimated absorption towards the lens is poorly measured with $A_{V,L}=1.16\pm0.65$ mag, mainly because only the $K_s$ band has been used, while V20 used a precise empirical relation for the lens absorption.  It is relevent to note that pyLIMASS estimation is compatible with the blend flux constraint highlighted by Y14. However, as underlined by V20, the blend brightness reported by Y14 is extremely faint with $I_{blend}=21.10\pm0.05$ mag and is potentially biased. It could be affected by an incorrect background estimation when performing crowded field photometry. While this does not affect the estimation of the source magnitudes, it could significantly bias the estimate of such a faint blend. For completness, a second run was conducted by including the additional constraint $I_L=I_{blend}$. From the Keck images presented in V20, it is clear that the field of view around the target is not overly crowded and only two components appear to be present and, therefore, it seems natural to assume $I_L=I_{blend}$. The solution leads to a closer and lighter lens with $M_L=0.624\pm0.058 M_\odot$ at $D_L=3.96\pm0.35$ kpc. We note that in this scenario, the source distance decreases significantly to $D_S=4.72\pm0.47$ kpc, and its predicted infrared magnitude is $K_{s,S}=17.507\pm0.065$ mag. A source at this distance is potentially possible, but given the already mentionned cautions about the faint blend, this scenario seems unlikely.

\subsection{Gaia16aye}
\begin{figure}[!h]
    \centering
    \includegraphics[width=0.45\textwidth]{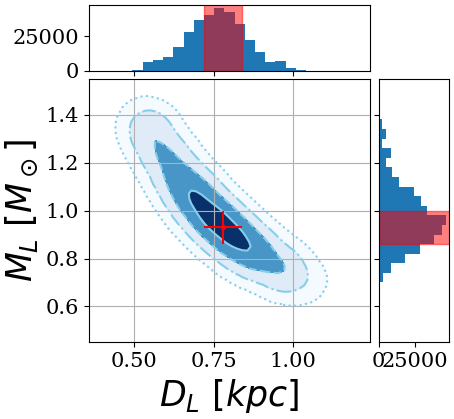}
    \caption{Distribution of the lens mass and the lens distance for the microlensing event Gaia16aye \citep{Wyrzykowski2020}. The measurements of W20 are represented in red.}
    \label{fig:gaiamldl}
\end{figure}

\begin{figure}[!h]
    \centering
    \includegraphics[width=0.9\textwidth]{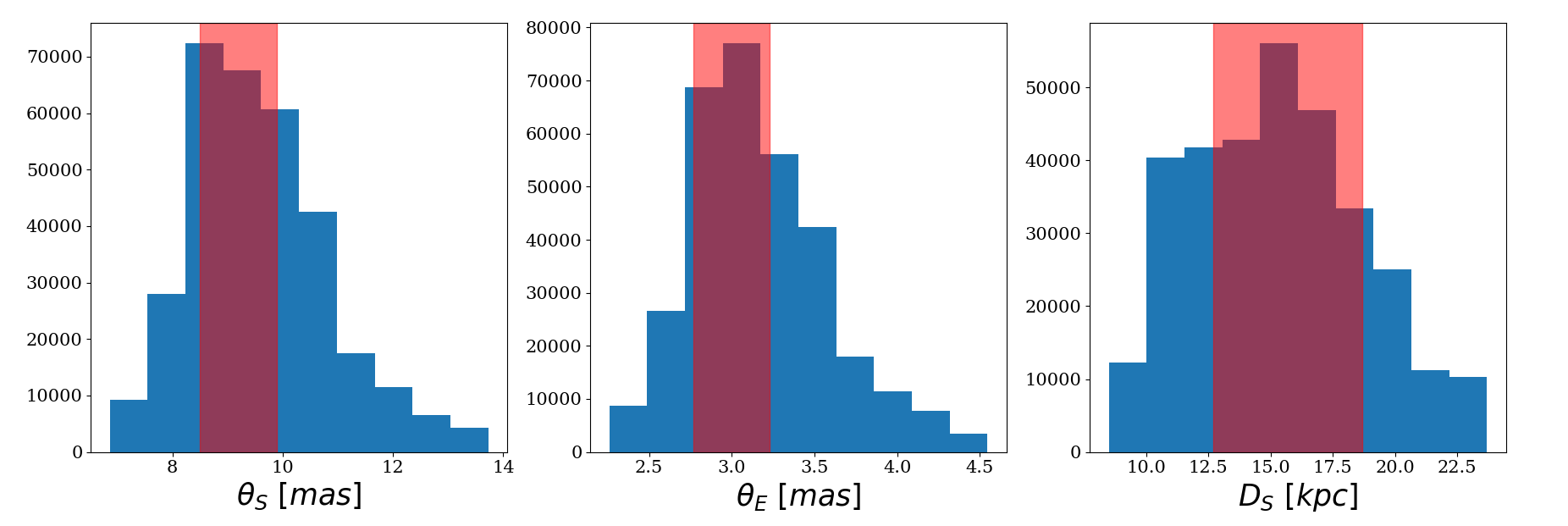}
    \caption{pyLIMASS posterior distribution of the angular source radius of the source (Left), Einstein angular radii (Middle) and source distance (Right) for the event Gaia16aye \citep{Wyrzykowski2020}. The $1\sigma$ measurements of W20 are represented in red.}
    \label{fig:gaiathetas}
\end{figure}

Gaia16aye is a long duration event detected by the Gaia telescope \citep{Wyrzykowski2020} (W20 thereafter). Thanks to a dense monitorig from the ground, this event has been precisely characterized. In particular, the lens orbital motion and the space parallax have been detected (W20), and the lens is a binary stars of $M_{L,1}=0.57\pm0.05$ and $M_{L,2}=0.36\pm0.03$ at $780\pm60$ pc. We used information from $t_E=111.09\pm0.41$ d, $\log_{10}(\rho)=-2.519\pm0.003$, $\pi_{EN}=-0.373\pm0.002$, $\pi_{EE}=-0.145\pm0.001$, $T_{eff,S}=3933\pm133$ K, $logg_S=2.2\pm1.4$, $Fe_s=-0.08\pm0.41$ as well as the magnitude of the source $(V,R,I)_S=(16.61\pm0.02, 15.62\pm0.02, 14.70\pm0.02)$ mag. Because the lens is a binary star, we set pyLIMASS to consider the lens as non stellar (so no constraints from the isochrones are imposed). Results can be found in Figure~\ref{fig:gaiamldl} and Figure~\ref{fig:gaiathetas}. With a total mass $M_L=0.97\pm0.13M_\odot$ at $D_L=0.759\pm0.092$ kpc, pyLIMASS results are fully consistent results with W20. Moreover, as can be seen in Figure~\ref{fig:gaiathetas}, both $\theta_S=10\pm1 \mu as$ and $\theta_E=3.2\pm0.5$ mas are in good agreement with W20 that reported $\theta_S=9.2\pm0.7 \mu as$ and $\theta_E=3.04\pm0.24$ mas. Similarly, the source distance $D_S=15.3\pm3.3$ kpc is in excellent agreement with W20 estimation $D_S=15.7\pm3.0$ kpc.

\section{Conclusions}
\label{sec:conclusion}
We present a new algorithm to estimate the most probable physical properties of microlensing events. Based on stellar isochrones and a Gaussian Mixture approach to combine the various information collected on events, the algorithm explores the entire parameter space and recovers the most plausible scenarii. With more than thirty possible observables currently available, the user can provide as much as information available to constrain the parameters of any events. The algorithm is solely based on these observables (e.g. measurements) and the stellar isochrones, and does not assume any distribution on the parameters. In particular, it does not use any models of the Milky Way but users can easily adapt the algorithm to include such priors information.

We have demonstrated the robustness and fidelity of the algorithm on a simulated sample of planetary hosts that can be observed by the future Galactic Bulge Time Domain Survey of the Nancy Grace Roman Space Telescope. Using a set of observables that should be routinely obtained from observations, we found the algorithm to accurately and precisely reconstruct the simulated parameters as well as returning realistic uncertainties. We also run the algorithm on several events published in the literature. The algorithm generally found an agreement at a 1-$\sigma$ level with the published results. The discrepancies are generally due to tensions between the measurements presented in the original studies, such ad for OGLE-2019-0960Lb or MOA-2013-BLG-220Lb, rather than the algorithm.

It is clear from Section~\ref{sec:roman} and  Section~\ref{sec:publishedevents} that pyLIMASS tends to systematically overestimate the source distances. This is mainly due for two reasons. First, the algorithm does not assume any models of the Milky Way. Therefore, there are no informative priors about the source distances provided (by default), which is a major difference with most of analysis presented in the literature. Indeed, most studies used galactic models priors or simply assume a source distance equal to the distance of the Red Giant Clump. A second explanation of this effect is detailed in the Section~\ref{sec:roman}. The combination of the absorption flexibility with the diversity of stars inside the isochrones allows a vast panel of models to accurately replicate the source apparent magnitudes injected in the observables. Fortunately, the imprecise measurement of the source distance moderately impact the estimation of the lens parameters, but explains for some part the lower precision measured by pyLIMASS versus the literature seen in Section~\ref{sec:publishedevents}.

We note that the algorithm is fast to run. For example, it takes about 300 seconds to run 20000 chains with 32 walkers (with the emcee package of \citet{ForemanMackey2013}) on a personal laptop equipped with a Intel(R) Xeon(R) W-10855M CPU @ 2.80GHz and uses about 400 Mo of memory. Therefore, it is an ideal tool to estimate the parameters of microlensing events at large scale. The presented results have been obtained with a python implementation of the algorithm, named pyLIMASS, that will be officially released in the upcoming release 2.0 of the pyLIMA software (Bachelet et al. in prep.) but that can already be found online\footnote{\url{https://github.com/ebachelet/pyLIMA}}.

In the future, several improvements could be made to the algorithm. In particular, the inclusion of a secondary body, both for the source and the lens, would bring extra constraints. Indeed, not only the observed magnitudes would be better replicated, but also the orbital elements (and therefore the mass) of the systems could be constraints via the observation of second order effects during the events, namely the orbital of the lens \citep{Skowron2011,Udalski2018} and the xallarap effect \citep{Rota2021}.

\appendix

\section{Implementation of the parameters priors}
\label{app:priors}
As usual with Bayesian approaches, the choice of the parameters priors p(P) is critical \citep{Tak2018}. While the priors distributions can be changed by the user, we described in this section the default options.

As discussed previously, the microlensing system is described by  sixteen parameters of interest $P$. The more trivial, but not innocent \citep{BailerJones2015,Eadie2023}, choice is to use a uniform prior. This is a valid approach, as long as lower and upper limits are finite and the range not too narrow \citep{Tak2018}, and are often referred as weakly informative prior. But in the present case, such priors on stellar parameters could create non-astrophysical objects and generate non-physical observables. To address this, pyLIMASS incorporates a prior probability distribution on the stellar parameters by comparing the stellar parameters $M$, $T_{eff}$, $Fe$ and $logg$ with the stellar isochrones. In practice, the squared Euclidean distance $\delta^2$ is computed between the drawn parameters and all of the isochrones catalog, and the minimum distance is selected. The distance is then turned into a probability $p(P)$ via:
\begin{equation}
    p(P)= e^{-\delta^2/\sigma^2}
\end{equation} 
where $\sigma$ is a relaxing term that is set to $\sigma\sim0.05$. We note that this needs to be systematically applied to the source, but not necessarily to the lens object. Indeed, the lens could be an object that is not present in the catalog of stellar isochrones, such as a free-floating planet, a brown dwarf or a stellar remnant. The isochrone constraints for the lens are not set by default, but the user can force the lens to be a stellar object by activating this constraint. If not, then uniform prior on the stellar parameters are used. Similarly, uniform priors for the distances $D$, proper motions $\mu$, visual absorption $A_{v,s}$ are used and a summary can be found in the Table~\ref{tab:priors}. This is a similar approach as \citet{Mroz2021} and \citet{Kruszynska2022}.

\begin{table*}
    \centering
    \begin{tabular}{lcc}
    \hline
    \hline
    
         Parameters&Unit&Priors  \\[0.1cm]
         \hline 
         $M_S$ & $M_\odot$ & Isochrones \\
         $D_S$ & $kpc$ & U(0,100) \\
         $\log_{10}(T_{eff,S})$ & $K$ & Isochrones \\
         $Fe_S$ & dex & Isochrones \\
         $logg_S$ &  & Isochrones \\
         $\mu_{N,S}$ & mas/yr & U(-20,20) \\
         $\mu_{E,S}$ & mas/yr & U(-20,20) \\
         $A_{V,S}$ & mag & U(-20,20) \\
         \hline
         $M_L$ & $M_\odot$ & Isochrones or U(0,100) \\
         $\epsilon_D$ &  & U(0,1)\\
         $\log_{10}(T_{eff,L})$& $K$ & Isochrones or U(3,5) \\
         $Fe_L$ & dex & Isochrones or U(-2,0.5) \\
         $logg_L$ &  & Isochrones or U(0,8) \\
         $\mu_{N,L}$ & mas/yr & U(-20,20) \\
         $\mu_{E,L}$ & mas/yr & U(-20,20) \\
         $\epsilon_{A_{V}}$ &  & U(0,1) \\ 
    \hline     
    \end{tabular}
    \caption{List of default priors used in pyLIMASS. There are tow possibilities for the lens, depending if the lens is forced to be stellar (see text).U(a,b) indicates a uniform distribution of the prior between a and b.} 
    \label{tab:priors}
\end{table*}

\section{List of pyLIMASS default observables}
The following list shows the default observables that can be included in pyLIMASS. There is no mandatory selection of observables, but at least one observable needs to be provided by the user. 
We note that in cases where the heliocentric proper motions are provided as observable, the reference time $t_{0,par}$ as well as the equatorial event coordinates ($\alpha$ and $\delta$) must be passed to the algorithm to calculate the geocentric to heliocentric transformation \citep{Skowron2011}.  

\begin{itemize}
\begin{minipage}{0.2\linewidth}
    \item $\log(D_S)$
    \item $\log(\pi_S)$
    \item $\log(M_S)$
    \item $\log(T_{eff,S})$
    \item $Fe_S$
    \item $logg_S$
    \item $\mu_{S,N}$
    \item $\mu_{S,E}$
    \item $\log(A_{V,S})$
    \item $\log(\theta_S)$
    \item $\log (R_S)$

\end{minipage}
\begin{minipage}{0.2\linewidth}

    \item $\log(D_L)$
    \item $\log(\epsilon_D)$
    \item $\log(\pi_L)$
    \item $\log(M_L)$
    \item $\log(T_{eff,L})$
    \item $Fe_L$
    \item $logg_L$
    \item $\mu_{L,N}$
    \item $\mu_{L,E}$
    \item $\log(A_{V,L})$
    \item $\log(\epsilon_{A_V})$

\end{minipage}
\begin{minipage}{0.2\linewidth}

    \item $\log(\theta_L)$
    \item $\log (R_L)$
    \item $\log(t_E)$
    \item $\log(\rho_*)$
    \item $\log(t_S)$
    \item $\log(\theta_E)$
    \item $\pi_{EN}$
    \item $\pi_{EE}$
    \item $\log(\pi_E)$
    \item $\phi_E$

\end{minipage}
\begin{minipage}{0.4\linewidth}
    \item $\mu_{rel,geo,N}$

    \item $\mu_{rel,geo,E}$
    \item $\log(\mu_{rel,geo})$
    \item $\mu_{rel,helio,N}$
    \item $\mu_{rel,helio,E}$
    \item $\log(\mu_{rel,helio})$
    \item $\log(\theta_E)$
    \item Source apparent magnitudes
    \item Lens apparent magnitudes
    \item Baseline apparent magnitudes
\end{minipage}

\end{itemize}

We thank J-P. Beaulieu and K. Vandorou for fruitful discussion regarding this work. This work was authored by employees of Caltech/IPAC under Contract No. 80GSFC21R0032 with the National Aeronautics and Space Administration. This research has made use of the NASA Exoplanet Archive, which is operated by the California Institute of Technology, under contract with the National Aeronautics and Space Administration under the Exoplanet Exploration Program.  This work has made use of data from the European Space Agency (ESA) mission
{\it Gaia} (\url{https://www.cosmos.esa.int/gaia}), processed by the {\it Gaia}
Data Processing and Analysis Consortium (DPAC,
\url{https://www.cosmos.esa.int/web/gaia/dpac/consortium}). Funding for the DPAC
has been provided by national institutions, in particular the institutions
participating in the {\it Gaia} Multilateral Agreement. This research has made use of the VizieR catalogue access tool, CDS, Strasbourg, France.

\bibliography{pyLIMASS.bib}

\begin{thebibliography}{}
\expandafter\ifx\csname natexlab\endcsname\relax\def\natexlab#1{#1}\fi
\providecommand{\url}[1]{\href{#1}{#1}}
\providecommand{\dodoi}[1]{doi:~\href{http://doi.org/#1}{\nolinkurl{#1}}}
\providecommand{\doeprint}[1]{\href{http://ascl.net/#1}{\nolinkurl{http://ascl.net/#1}}}
\providecommand{\doarXiv}[1]{\href{https://arxiv.org/abs/#1}{\nolinkurl{https://arxiv.org/abs/#1}}}

\bibitem[{{Adams} {et~al.}(2018){Adams}, {Boyajian}, \& {von
  Braun}}]{Adams2018}
{Adams}, A.~D., {Boyajian}, T.~S., \& {von Braun}, K. 2018, \mnras, 473, 3608,
  \dodoi{10.1093/mnras/stx2367}

\bibitem[{{Akaike}(1974)}]{Akaike1974}
{Akaike}, H. 1974, IEEE Transactions on Automatic Control, 19, 716

\bibitem[{{Alcock} {et~al.}(2000){Alcock}, {Allsman}, {Alves}, {Axelrod},
  {Becker}, {Bennett}, {Cook}, {Dalal}, {Drake}, {Freeman}, {Geha}, {Griest},
  {Lehner}, {Marshall}, {Minniti}, {Nelson}, {Peterson}, {Popowski}, {Pratt},
  {Quinn}, {Stubbs}, {Sutherland}, {Tomaney}, {Vandehei}, \&
  {Welch}}]{Alcock2000}
{Alcock}, C., {Allsman}, R.~A., {Alves}, D.~R., {et~al.} 2000, \apj, 542, 281,
  \dodoi{10.1086/309512}

\bibitem[{{Allard} \& {Hauschildt}(1995)}]{Allard1995}
{Allard}, F., \& {Hauschildt}, P.~H. 1995, \apj, 445, 433,
  \dodoi{10.1086/175708}

\bibitem[{{Alsing} {et~al.}(2018){Alsing}, {Silva}, \& {Berti}}]{Alsing2018}
{Alsing}, J., {Silva}, H.~O., \& {Berti}, E. 2018, \mnras, 478, 1377,
  \dodoi{10.1093/mnras/sty1065}

\bibitem[{{Ansari} {et~al.}(2021){Ansari}, {Agnello}, \& {Gall}}]{Ansari2021}
{Ansari}, Z., {Agnello}, A., \& {Gall}, C. 2021, \aap, 650, A90,
  \dodoi{10.1051/0004-6361/202039675}

\bibitem[{{Awiphan} {et~al.}(2016){Awiphan}, {Kerins}, \&
  {Robin}}]{Awiphan2016}
{Awiphan}, S., {Kerins}, E., \& {Robin}, A.~C. 2016, \mnras, 456, 1666,
  \dodoi{10.1093/mnras/stv2625}

\bibitem[{{Bachelet} {et~al.}(2017){Bachelet}, {Norbury}, {Bozza}, \&
  {Street}}]{Bachelet2017}
{Bachelet}, E., {Norbury}, M., {Bozza}, V., \& {Street}, R. 2017, \aj, 154,
  203, \dodoi{10.3847/1538-3881/aa911c}

\bibitem[{{Bachelet} {et~al.}(2019){Bachelet}, {Bozza}, {Han}, {Udalski},
  {Bond}, {Beaulieu}, {Street}, {Kim}, {Bramich}, {Cassan}, {Dominik}, {Figuera
  Jaimes}, {Horne}, {Hundertmark}, {Mao}, {Menzies}, {Ranc}, {Schmidt},
  {Snodgrass}, {Steele}, {Tsapras}, {Wambsganss}, {RoboNet Collaboration},
  {Mr{\'o}z}, {Soszy{\'n}ski}, {Szyma{\'n}ski}, {Skowron}, {Pietrukowicz},
  {Koz{\l}owski}, {Poleski}, {Ulaczyk}, {Pawlak}, {OGLE Collaboration}, {Abe},
  {Barry}, {Bennett}, {Bhattacharya}, {Donachie}, {Fukui}, {Hirao}, {Itow},
  {Kawasaki}, {Kondo}, {Koshimoto}, {Li}, {Matsubara}, {Muraki}, {Miyazaki},
  {Nagakane}, {Rattenbury}, {Suematsu}, {Sullivan}, {Sumi}, {Suzuki},
  {Tristram}, {Yonehara}, \& {MOA Collaboration}}]{Bachelet2019}
{Bachelet}, E., {Bozza}, V., {Han}, C., {et~al.} 2019, \apj, 870, 11,
  \dodoi{10.3847/1538-4357/aaedb9}

\bibitem[{{Bachelet} {et~al.}(2022{\natexlab{a}}){Bachelet}, {Tsapras},
  {Gould}, {Street}, {Bennett}, {Hundertmark}, {Bozza}, {Bramich}, {Cassan},
  {Dominik}, {Horne}, {Mao}, {Saha}, {Wambsganss}, {Zang}, {ROME/REA
  Collaboration}, {Abe}, {Barry}, {Bennett}, {Bhattacharya}, {Bond}, {Fukui},
  {Fujii}, {Hirao}, {Itow}, {Kirikawa}, {Kondo}, {Koshimoto}, {Matsubara},
  {Matsumoto}, {Miyazaki}, {Muraki}, {Olmschenk}, {Ranc}, {Okamura},
  {Rattenbury}, {Satoh}, {Sumi}, {Suzuki}, {Silva}, {Toda}, {Tristram},
  {Vandorou}, {Yama}, {MOA Collaboration}, {Albrow}, {Chung}, {Han}, {Hwang},
  {Jung}, {Ryu}, {Shin}, {Shvartzvald}, {Yee}, {Cha}, {Kim}, {Kim}, {Lee},
  {Lee}, {Lee}, {Park}, {Pogge}, {KMTNet Collaboration}, {Udalski}, {Mr{\'o}z},
  {Poleski}, {Skowron}, {Szyma{\'n}ski}, {Soszy{\'n}ski}, {Pietrukowicz},
  {Koz{\l}owski}, {Ulaczyk}, {Rybicki}, {Iwanek}, {Wrona}, {Gromadzki}, \&
  {OGLE Collaboration}}]{Bachelet2022b}
{Bachelet}, E., {Tsapras}, Y., {Gould}, A., {et~al.} 2022{\natexlab{a}}, \aj,
  164, 75, \dodoi{10.3847/1538-3881/ac78ed}

\bibitem[{{Bachelet} {et~al.}(2022{\natexlab{b}}){Bachelet}, {Specht}, {Penny},
  {Hundertmark}, {Awiphan}, {Beaulieu}, {Dominik}, {Kerins}, {Maoz}, {Meade},
  {Nucita}, {Poleski}, {Ranc}, {Rhodes}, \& {Robin}}]{Bachelet2022}
{Bachelet}, E., {Specht}, D., {Penny}, M., {et~al.} 2022{\natexlab{b}}, \aap,
  664, A136, \dodoi{10.1051/0004-6361/202140351}

\bibitem[{{Bailer-Jones}(2015)}]{BailerJones2015}
{Bailer-Jones}, C. A.~L. 2015, \pasp, 127, 994, \dodoi{10.1086/683116}

\bibitem[{{Batista} {et~al.}(2015){Batista}, {Beaulieu}, {Bennett}, {Gould},
  {Marquette}, {Fukui}, \& {Bhattacharya}}]{Batista2015}
{Batista}, V., {Beaulieu}, J.~P., {Bennett}, D.~P., {et~al.} 2015, \apj, 808,
  170, \dodoi{10.1088/0004-637X/808/2/170}

\bibitem[{{Beaulieu}(2018)}]{Beaulieu2018}
{Beaulieu}, J.-P. 2018, Universe, 4, 61, \dodoi{10.3390/universe4040061}

\bibitem[{{Benedict} {et~al.}(2016){Benedict}, {Henry}, {Franz}, {McArthur},
  {Wasserman}, {Jao}, {Cargile}, {Dieterich}, {Bradley}, {Nelan}, \&
  {Whipple}}]{Benedict2016}
{Benedict}, G.~F., {Henry}, T.~J., {Franz}, O.~G., {et~al.} 2016, \aj, 152,
  141, \dodoi{10.3847/0004-6256/152/5/141}

\bibitem[{{Bennett}(2010)}]{Bennett2010}
{Bennett}, D.~P. 2010, \apj, 716, 1408, \dodoi{10.1088/0004-637X/716/2/1408}

\bibitem[{{Bennett} \& {Rhie}(1996)}]{Bennett1996}
{Bennett}, D.~P., \& {Rhie}, S.~H. 1996, \apj, 472, 660, \dodoi{10.1086/178096}

\bibitem[{{Bennett} {et~al.}(2014){Bennett}, {Batista}, {Bond}, {Bennett},
  {Suzuki}, {Beaulieu}, {Udalski}, {Donatowicz}, {Bozza}, {Abe}, {Botzler},
  {Freeman}, {Fukunaga}, {Fukui}, {Itow}, {Koshimoto}, {Ling}, {Masuda},
  {Matsubara}, {Muraki}, {Namba}, {Ohnishi}, {Rattenbury}, {Saito}, {Sullivan},
  {Sumi}, {Sweatman}, {Tristram}, {Tsurumi}, {Wada}, {Yock}, {MOA
  Collaboration}, {Albrow}, {Bachelet}, {Brillant}, {Caldwell}, {Cassan},
  {Cole}, {Corrales}, {Coutures}, {Dieters}, {Dominis Prester}, {Fouqu{\'e}},
  {Greenhill}, {Horne}, {Koo}, {Kubas}, {Marquette}, {Martin}, {Menzies},
  {Sahu}, {Wambsganss}, {Williams}, {Zub}, {PLANET Collaboration}, {Choi},
  {DePoy}, {Dong}, {Gaudi}, {Gould}, {Han}, {Henderson}, {McGregor}, {Lee},
  {Pogge}, {Shin}, {Yee}, {{\ensuremath{\mu}}FUN Collaboration},
  {Szyma{\'n}ski}, {Skowron}, {Poleski}, {Koz{\l}owski}, {Wyrzykowski},
  {Kubiak}, {Pietrukowicz}, {Pietrzy{\'n}ski}, {Soszy{\'n}ski}, {Ulaczyk},
  {OGLE Collaboration}, {Tsapras}, {Street}, {Dominik}, {Bramich}, {Browne},
  {Hundertmark}, {Kains}, {Snodgrass}, {Steele}, {RoboNet Collaboration},
  {Dekany}, {Gonzalez}, {Heyrovsk{\'y}}, {Kandori}, {Kerins}, {Lucas},
  {Minniti}, {Nagayama}, {Rejkuba}, {Robin}, \& {Saito}}]{Bennett2014}
{Bennett}, D.~P., {Batista}, V., {Bond}, I.~A., {et~al.} 2014, \apj, 785, 155,
  \dodoi{10.1088/0004-637X/785/2/155}

\bibitem[{{Bennett} {et~al.}(2015){Bennett}, {Bhattacharya}, {Anderson},
  {Bond}, {Anderson}, {Barry}, {Batista}, {Beaulieu}, {DePoy}, {Dong}, {Gaudi},
  {Gilbert}, {Gould}, {Pfeifle}, {Pogge}, {Suzuki}, {Terry}, \&
  {Udalski}}]{Bennett2015}
{Bennett}, D.~P., {Bhattacharya}, A., {Anderson}, J., {et~al.} 2015, \apj, 808,
  169, \dodoi{10.1088/0004-637X/808/2/169}

\bibitem[{{Bennett} {et~al.}(2023){Bennett}, {Bhattacharya}, {Beaulieu},
  {Koshimoto}, {Blackman}, {Bond}, {Ranc}, {Rektsini}, {Terry}, {Vandorou},
  {Lu}, {Baptiste Marquette}, {Olmschenk}, \& {Suzuki}}]{Bennett2023}
{Bennett}, D.~P., {Bhattacharya}, A., {Beaulieu}, J.-P., {et~al.} 2023, arXiv
  e-prints, arXiv:2311.00627, \dodoi{10.48550/arXiv.2311.00627}

\bibitem[{{Bessell} \& {Brett}(1988)}]{Bessel1988}
{Bessell}, M.~S., \& {Brett}, J.~M. 1988, \pasp, 100, 1134,
  \dodoi{10.1086/132281}

\bibitem[{{Bhattacharya} {et~al.}(2018){Bhattacharya}, {Beaulieu}, {Bennett},
  {Anderson}, {Koshimoto}, {Lu}, {Batista}, {Blackman}, {Bond}, {Fukui},
  {Henderson}, {Hirao}, {Marquette}, {Mroz}, {Ranc}, \&
  {Udalski}}]{Bhattacharya2018}
{Bhattacharya}, A., {Beaulieu}, J.~P., {Bennett}, D.~P., {et~al.} 2018, \aj,
  156, 289, \dodoi{10.3847/1538-3881/aaed46}

\bibitem[{{Bhattacharya} {et~al.}(2021){Bhattacharya}, {Bennett}, {Beaulieu},
  {Bond}, {Koshimoto}, {Lu}, {Blackman}, {Vandorou}, {Terry}, {Batista},
  {Marquette}, {Cole}, {Fukui}, {Henderson}, \& {Ranc}}]{Bhattacharya2021}
{Bhattacharya}, A., {Bennett}, D.~P., {Beaulieu}, J.~P., {et~al.} 2021, \aj,
  162, 60, \dodoi{10.3847/1538-3881/abfec5}

\bibitem[{{Blackman} {et~al.}(2021){Blackman}, {Beaulieu}, {Bennett},
  {Danielski}, {Alard}, {Cole}, {Vandorou}, {Ranc}, {Terry}, {Bhattacharya},
  {Bond}, {Bachelet}, {Veras}, {Koshimoto}, {Batista}, \&
  {Marquette}}]{Blackman2021}
{Blackman}, J.~W., {Beaulieu}, J.~P., {Bennett}, D.~P., {et~al.} 2021, \nat,
  598, 272, \dodoi{10.1038/s41586-021-03869-6}

\bibitem[{{Bond} {et~al.}(2001){Bond}, {Abe}, {Dodd}, {Hearnshaw}, {Honda},
  {Jugaku}, {Kilmartin}, {Marles}, {Masuda}, {Matsubara}, {Muraki}, {Nakamura},
  {Nankivell}, {Noda}, {Noguchi}, {Ohnishi}, {Rattenbury}, {Reid}, {Saito},
  {Sato}, {Sekiguchi}, {Skuljan}, {Sullivan}, {Sumi}, {Takeuti}, {Watase},
  {Wilkinson}, {Yamada}, {Yanagisawa}, \& {Yock}}]{Bond2001}
{Bond}, I.~A., {Abe}, F., {Dodd}, R.~J., {et~al.} 2001, \mnras, 327, 868,
  \dodoi{10.1046/j.1365-8711.2001.04776.x}

\bibitem[{{Bozza}(2010)}]{Bozza2010}
{Bozza}, V. 2010, \mnras, 408, 2188, \dodoi{10.1111/j.1365-2966.2010.17265.x}

\bibitem[{{Bozza} {et~al.}(2018){Bozza}, {Bachelet}, {Bartoli{\'c}}, {Heintz},
  {Hoag}, \& {Hundertmark}}]{Bozza2018}
{Bozza}, V., {Bachelet}, E., {Bartoli{\'c}}, F., {et~al.} 2018, \mnras, 479,
  5157, \dodoi{10.1093/mnras/sty1791}

\bibitem[{{Bressan} {et~al.}(2012){Bressan}, {Marigo}, {Girardi}, {Salasnich},
  {Dal Cero}, {Rubele}, \& {Nanni}}]{Bressan2012}
{Bressan}, A., {Marigo}, P., {Girardi}, L., {et~al.} 2012, \mnras, 427, 127,
  \dodoi{10.1111/j.1365-2966.2012.21948.x}

\bibitem[{{Calchi Novati} {et~al.}(2015){Calchi Novati}, {Gould}, {Udalski},
  {Menzies}, {Bond}, {Shvartzvald}, {Street}, {Hundertmark}, {Beichman}, {Yee},
  {Carey}, {Poleski}, {Skowron}, {Koz{\l}owski}, {Mr{\'o}z}, {Pietrukowicz},
  {Pietrzy{\'n}ski}, {Szyma{\'n}ski}, {Soszy{\'n}ski}, {Ulaczyk},
  {Wyrzykowski}, {OGLE Collaboration}, {Albrow}, {Beaulieu}, {Caldwell},
  {Cassan}, {Coutures}, {Danielski}, {Dominis Prester}, {Donatowicz},
  {Lon{\v{c}}ari{\'c}}, {McDougall}, {Morales}, {Ranc}, {Zhu}, {PLANET
  Collaboration}, {Abe}, {Barry}, {Bennett}, {Bhattacharya}, {Fukunaga},
  {Inayama}, {Koshimoto}, {Namba}, {Sumi}, {Suzuki}, {Tristram}, {Wakiyama},
  {Yonehara}, {MOA Collaboration}, {Maoz}, {Kaspi}, {Friedmann}, {Wise Group},
  {Bachelet}, {Figuera Jaimes}, {Bramich}, {Tsapras}, {Horne}, {Snodgrass},
  {Wambsganss}, {Steele}, {Kains}, {RoboNet Collaboration}, {Bozza}, {Dominik},
  {J{\o}rgensen}, {Alsubai}, {Ciceri}, {D'Ago}, {Haugb{\o}lle}, {Hessman},
  {Hinse}, {Juncher}, {Korhonen}, {Mancini}, {Popovas}, {Rabus}, {Rahvar},
  {Scarpetta}, {Schmidt}, {Skottfelt}, {Southworth}, {Starkey}, {Surdej},
  {Wertz}, {Zarucki}, {MiNDSTEp Consortium}, {Gaudi}, {Pogge}, {DePoy}, \&
  {{\ensuremath{\mu}}FUN Collaboration}}]{CalchiNovati2015}
{Calchi Novati}, S., {Gould}, A., {Udalski}, A., {et~al.} 2015, \apj, 804, 20,
  \dodoi{10.1088/0004-637X/804/1/20}

\bibitem[{{Cassan} {et~al.}(2022){Cassan}, {Ranc}, {Absil}, {Wyrzykowski},
  {Rybicki}, {Bachelet}, {Le Bouquin}, {Hundertmark}, {Street}, {Surdej},
  {Tsapras}, {Wambsganss}, \& {Wertz}}]{Cassan2022}
{Cassan}, A., {Ranc}, C., {Absil}, O., {et~al.} 2022, Nature Astronomy, 6, 121,
  \dodoi{10.1038/s41550-021-01514-w}

\bibitem[{{Choi} {et~al.}(2016){Choi}, {Dotter}, {Conroy}, {Cantiello},
  {Paxton}, \& {Johnson}}]{Choi2016}
{Choi}, J., {Dotter}, A., {Conroy}, C., {et~al.} 2016, \apj, 823, 102,
  \dodoi{10.3847/0004-637X/823/2/102}

\bibitem[{{Chung} {et~al.}(2017){Chung}, {Zhu}, {Udalski}, {Lee}, {Ryu},
  {Jung}, {Shin}, {Yee}, {Hwang}, {Gould}, {and}, {Albrow}, {Cha}, {Han},
  {Kim}, {Kim}, {Kim}, {Kim}, {Lee}, {Park}, {Pogge}, {KMTNet Collaboration},
  {Poleski}, {Mr{\'o}z}, {Pietrukowicz}, {Skowron}, {Szyma{\'n}ski},
  {Soszy{\'n}ski}, {Koz{\l}owski}, {Ulaczyk}, {Pawlak}, {OGLE Collaboration},
  {Beichman}, {Bryden}, {Calchi Novati}, {Carey}, {Fausnaugh}, {Gaudi},
  {Henderson}, {Shvartzvald}, {Wibking}, \& {Spitzer Team}}]{Chung2017}
{Chung}, S.~J., {Zhu}, W., {Udalski}, A., {et~al.} 2017, \apj, 838, 154,
  \dodoi{10.3847/1538-4357/aa67fa}

\bibitem[{Dempster {et~al.}(1977)Dempster, Laird, \& Rubin}]{Dempster1977}
Dempster, A.~P., Laird, N.~M., \& Rubin, D.~B. 1977, Journal of the Royal
  Statistical Society: Series B (Methodological), 39, 1,
  \dodoi{https://doi.org/10.1111/j.2517-6161.1977.tb01600.x}

\bibitem[{{Dominik}(2006)}]{Dominik2006}
{Dominik}, M. 2006, \mnras, 367, 669, \dodoi{10.1111/j.1365-2966.2006.10004.x}

\bibitem[{{Dominik} \& {Sahu}(2000)}]{Dominik2000}
{Dominik}, M., \& {Sahu}, K.~C. 2000, \apj, 534, 213, \dodoi{10.1086/308716}

\bibitem[{{Dong} {et~al.}(2019){Dong}, {M{\'e}rand}, {Delplancke-Str{\"o}bele},
  {Gould}, {Chen}, {Post}, {Kochanek}, {Stanek}, {Christie}, {Mutel},
  {Natusch}, {Holoien}, {Prieto}, {Shappee}, \& {Thompson}}]{Dong2019}
{Dong}, S., {M{\'e}rand}, A., {Delplancke-Str{\"o}bele}, F., {et~al.} 2019,
  \apj, 871, 70, \dodoi{10.3847/1538-4357/aaeffb}

\bibitem[{{Eadie} {et~al.}(2023){Eadie}, {Speagle}, {Cisewski-Kehe},
  {Foreman-Mackey}, {Huppenkothen}, {Jones}, {Springford}, \&
  {Tak}}]{Eadie2023}
{Eadie}, G.~M., {Speagle}, J.~S., {Cisewski-Kehe}, J., {et~al.} 2023, arXiv
  e-prints, arXiv:2302.04703, \dodoi{10.48550/arXiv.2302.04703}

\bibitem[{{Einstein}(1936)}]{Einstein1936}
{Einstein}, A. 1936, Science, 84, 506, \dodoi{10.1126/science.84.2188.506}

\bibitem[{{Fardeen} {et~al.}(2023){Fardeen}, {McGill}, {Perkins}, {Dawson},
  {Abrams}, {Lu}, {Ho}, \& {Bird}}]{Fardeen2023}
{Fardeen}, J., {McGill}, P., {Perkins}, S.~E., {et~al.} 2023, arXiv e-prints,
  arXiv:2312.13249, \dodoi{10.48550/arXiv.2312.13249}

\bibitem[{{Foreman-Mackey} {et~al.}(2013){Foreman-Mackey}, {Hogg}, {Lang}, \&
  {Goodman}}]{ForemanMackey2013}
{Foreman-Mackey}, D., {Hogg}, D.~W., {Lang}, D., \& {Goodman}, J. 2013, \pasp,
  125, 306, \dodoi{10.1086/670067}

\bibitem[{Fr{\"u}hwirth-Schnatter {et~al.}(2019)Fr{\"u}hwirth-Schnatter,
  Celeux, \& Robert}]{Fruhwirthschnatter2019}
Fr{\"u}hwirth-Schnatter, S., Celeux, G., \& Robert, C.~P. 2019, {Handbook of
  Mixture Analysis} ({Taylor \& Francis}).
\newblock \url{https://hal.science/hal-03943314}

\bibitem[{{Fukui} {et~al.}(2019){Fukui}, {Suzuki}, {Koshimoto}, {Bachelet},
  {Vanmunster}, {Storey}, {Maehara}, {Yanagisawa}, {Yamada}, {Yonehara},
  {Hirano}, {Bennett}, {Bozza}, {Mawet}, {Penny}, {Awiphan}, {Oksanen},
  {Heintz}, {Oberst}, {B{\'e}jar}, {Casasayas-Barris}, {Chen}, {Crouzet},
  {Hidalgo}, {Klagyivik}, {Murgas}, {Narita}, {Palle}, {Parviainen},
  {Watanabe}, {Kusakabe}, {Mori}, {Terada}, {de Leon}, {Hernandez}, {Luque},
  {Monelli}, {Monta{\~n}es-Rodriguez}, {Prieto-Arranz}, {Murata}, {Shugarov},
  {Kubota}, {Otsuki}, {Shionoya}, {Nishiumi}, {Nishide}, {Fukagawa}, {Onodera},
  {Villanueva}, {Street}, {Tsapras}, {Hundertmark}, {Kuzuhara}, {Fujita},
  {Beichman}, {Beaulieu}, {Alonso}, {Reichart}, {Kawai}, \&
  {Tamura}}]{Fukui2019}
{Fukui}, A., {Suzuki}, D., {Koshimoto}, N., {et~al.} 2019, \aj, 158, 206,
  \dodoi{10.3847/1538-3881/ab487f}

\bibitem[{{Gaia Collaboration} {et~al.}(2016){Gaia Collaboration}, {Prusti},
  {de Bruijne}, {Brown}, {Vallenari}, {Babusiaux}, {Bailer-Jones}, {Bastian},
  {Biermann}, {Evans}, {Eyer}, {Jansen}, {Jordi}, {Klioner}, {Lammers},
  {Lindegren}, {Luri}, {Mignard}, {Milligan}, {Panem}, {Poinsignon},
  {Pourbaix}, {Randich}, {Sarri}, {Sartoretti}, {Siddiqui}, {Soubiran},
  {Valette}, {van Leeuwen}, {Walton}, {Aerts}, {Arenou}, {Cropper}, {Drimmel},
  {H{\o}g}, {Katz}, {Lattanzi}, {O'Mullane}, {Grebel}, {Holland}, {Huc},
  {Passot}, {Bramante}, {Cacciari}, {Casta{\~n}eda}, {Chaoul}, {Cheek}, {De
  Angeli}, {Fabricius}, {Guerra}, {Hern{\'a}ndez}, {Jean-Antoine-Piccolo},
  {Masana}, {Messineo}, {Mowlavi}, {Nienartowicz}, {Ord{\'o}{\~n}ez-Blanco},
  {Panuzzo}, {Portell}, {Richards}, {Riello}, {Seabroke}, {Tanga},
  {Th{\'e}venin}, {Torra}, {Els}, {Gracia-Abril}, {Comoretto},
  {Garcia-Reinaldos}, {Lock}, {Mercier}, {Altmann}, {Andrae}, {Astraatmadja},
  {Bellas-Velidis}, {Benson}, {Berthier}, {Blomme}, {Busso}, {Carry},
  {Cellino}, {Clementini}, {Cowell}, {Creevey}, {Cuypers}, {Davidson}, {De
  Ridder}, {de Torres}, {Delchambre}, {Dell'Oro}, {Ducourant}, {Fr{\'e}mat},
  {Garc{\'\i}a-Torres}, {Gosset}, {Halbwachs}, {Hambly}, {Harrison}, {Hauser},
  {Hestroffer}, {Hodgkin}, {Huckle}, {Hutton}, {Jasniewicz}, {Jordan},
  {Kontizas}, {Korn}, {Lanzafame}, {Manteiga}, {Moitinho}, {Muinonen},
  {Osinde}, {Pancino}, {Pauwels}, {Petit}, {Recio-Blanco}, {Robin}, {Sarro},
  {Siopis}, {Smith}, {Smith}, {Sozzetti}, {Thuillot}, {van Reeven}, {Viala},
  {Abbas}, {Abreu Aramburu}, {Accart}, {Aguado}, {Allan}, {Allasia},
  {Altavilla}, {{\'A}lvarez}, {Alves}, {Anderson}, {Andrei}, {Anglada Varela},
  {Antiche}, {Antoja}, {Ant{\'o}n}, {Arcay}, {Atzei}, {Ayache}, {Bach},
  {Baker}, {Balaguer-N{\'u}{\~n}ez}, {Barache}, {Barata}, {Barbier}, {Barblan},
  {Baroni}, {Barrado y Navascu{\'e}s}, {Barros}, {Barstow}, {Becciani},
  {Bellazzini}, {Bellei}, {Bello Garc{\'\i}a}, {Belokurov}, {Bendjoya},
  {Berihuete}, {Bianchi}, {Bienaym{\'e}}, {Billebaud}, {Blagorodnova},
  {Blanco-Cuaresma}, {Boch}, {Bombrun}, {Borrachero}, {Bouquillon}, {Bourda},
  {Bouy}, {Bragaglia}, {Breddels}, {Brouillet}, {Br{\"u}semeister},
  {Bucciarelli}, {Budnik}, {Burgess}, {Burgon}, {Burlacu}, {Busonero}, {Buzzi},
  {Caffau}, {Cambras}, {Campbell}, {Cancelliere}, {Cantat-Gaudin}, {Carlucci},
  {Carrasco}, {Castellani}, {Charlot}, {Charnas}, {Charvet}, {Chassat},
  {Chiavassa}, {Clotet}, {Cocozza}, {Collins}, {Collins}, {Costigan}, {Crifo},
  {Cross}, {Crosta}, {Crowley}, {Dafonte}, {Damerdji}, {Dapergolas}, {David},
  {David}, {De Cat}, {de Felice}, {de Laverny}, {De Luise}, {De March}, {de
  Martino}, {de Souza}, {Debosscher}, {del Pozo}, {Delbo}, {Delgado},
  {Delgado}, {di Marco}, {Di Matteo}, {Diakite}, {Distefano}, {Dolding}, {Dos
  Anjos}, {Drazinos}, {Dur{\'a}n}, {Dzigan}, {Ecale}, {Edvardsson}, {Enke},
  {Erdmann}, {Escolar}, {Espina}, {Evans}, {Eynard Bontemps}, {Fabre},
  {Fabrizio}, {Faigler}, {Falc{\~a}o}, {Farr{\`a}s Casas}, {Faye}, {Federici},
  {Fedorets}, {Fern{\'a}ndez-Hern{\'a}ndez}, {Fernique}, {Fienga}, {Figueras},
  {Filippi}, {Findeisen}, {Fonti}, {Fouesneau}, {Fraile}, {Fraser}, {Fuchs},
  {Furnell}, {Gai}, {Galleti}, {Galluccio}, {Garabato}, {Garc{\'\i}a-Sedano},
  {Gar{\'e}}, {Garofalo}, {Garralda}, {Gavras}, {Gerssen}, {Geyer}, {Gilmore},
  {Girona}, {Giuffrida}, {Gomes}, {Gonz{\'a}lez-Marcos},
  {Gonz{\'a}lez-N{\'u}{\~n}ez}, {Gonz{\'a}lez-Vidal}, {Granvik}, {Guerrier},
  {Guillout}, {Guiraud}, {G{\'u}rpide}, {Guti{\'e}rrez-S{\'a}nchez}, {Guy},
  {Haigron}, {Hatzidimitriou}, {Haywood}, {Heiter}, {Helmi}, {Hobbs},
  {Hofmann}, {Holl}, {Holland}, {Hunt}, {Hypki}, {Icardi}, {Irwin}, {Jevardat
  de Fombelle}, {Jofr{\'e}}, {Jonker}, {Jorissen}, {Julbe}, {Karampelas},
  {Kochoska}, {Kohley}, {Kolenberg}, {Kontizas}, {Koposov}, {Kordopatis},
  {Koubsky}, {Kowalczyk}, {Krone-Martins}, {Kudryashova}, {Kull}, {Bachchan},
  {Lacoste-Seris}, {Lanza}, {Lavigne}, {Le Poncin-Lafitte}, {Lebreton},
  {Lebzelter}, {Leccia}, {Leclerc}, {Lecoeur-Taibi}, {Lemaitre}, {Lenhardt},
  {Leroux}, {Liao}, {Licata}, {Lindstr{\o}m}, {Lister}, {Livanou}, {Lobel},
  {L{\"o}ffler}, {L{\'o}pez}, {Lopez-Lozano}, {Lorenz}, {Loureiro},
  {MacDonald}, {Magalh{\~a}es Fernandes}, {Managau}, {Mann}, {Mantelet},
  {Marchal}, {Marchant}, {Marconi}, {Marie}, {Marinoni}, {Marrese},
  {Marschalk{\'o}}, {Marshall}, {Mart{\'\i}n-Fleitas}, {Martino}, {Mary},
  {Matijevi{\v{c}}}, {Mazeh}, {McMillan}, {Messina}, {Mestre}, {Michalik},
  {Millar}, {Miranda}, {Molina}, {Molinaro}, {Molinaro}, {Moln{\'a}r},
  {Moniez}, {Montegriffo}, {Monteiro}, {Mor}, {Mora}, {Morbidelli}, {Morel},
  {Morgenthaler}, {Morley}, {Morris}, {Mulone}, {Muraveva}, {Musella},
  {Narbonne}, {Nelemans}, {Nicastro}, {Noval}, {Ord{\'e}novic},
  {Ordieres-Mer{\'e}}, {Osborne}, {Pagani}, {Pagano}, {Pailler}, {Palacin},
  {Palaversa}, {Parsons}, {Paulsen}, {Pecoraro}, {Pedrosa}, {Pentik{\"a}inen},
  {Pereira}, {Pichon}, {Piersimoni}, {Pineau}, {Plachy}, {Plum}, {Poujoulet},
  {Pr{\v{s}}a}, {Pulone}, {Ragaini}, {Rago}, {Rambaux}, {Ramos-Lerate},
  {Ranalli}, {Rauw}, {Read}, {Regibo}, {Renk}, {Reyl{\'e}}, {Ribeiro},
  {Rimoldini}, {Ripepi}, {Riva}, {Rixon}, {Roelens}, {Romero-G{\'o}mez},
  {Rowell}, {Royer}, {Rudolph}, {Ruiz-Dern}, {Sadowski}, {Sagrist{\`a}
  Sell{\'e}s}, {Sahlmann}, {Salgado}, {Salguero}, {Sarasso}, {Savietto},
  {Schnorhk}, {Schultheis}, {Sciacca}, {Segol}, {Segovia}, {Segransan},
  {Serpell}, {Shih}, {Smareglia}, {Smart}, {Smith}, {Solano}, {Solitro},
  {Sordo}, {Soria Nieto}, {Souchay}, {Spagna}, {Spoto}, {Stampa}, {Steele},
  {Steidelm{\"u}ller}, {Stephenson}, {Stoev}, {Suess}, {S{\"u}veges}, {Surdej},
  {Szabados}, {Szegedi-Elek}, {Tapiador}, {Taris}, {Tauran}, {Taylor},
  {Teixeira}, {Terrett}, {Tingley}, {Trager}, {Turon}, {Ulla}, {Utrilla},
  {Valentini}, {van Elteren}, {Van Hemelryck}, {van Leeuwen}, {Varadi},
  {Vecchiato}, {Veljanoski}, {Via}, {Vicente}, {Vogt}, {Voss}, {Votruba},
  {Voutsinas}, {Walmsley}, {Weiler}, {Weingrill}, {Werner}, {Wevers},
  {Whitehead}, {Wyrzykowski}, {Yoldas}, {{\v{Z}}erjal}, {Zucker}, {Zurbach},
  {Zwitter}, {Alecu}, {Allen}, {Allende Prieto}, {Amorim},
  {Anglada-Escud{\'e}}, {Arsenijevic}, {Azaz}, {Balm}, {Beck}, {Bernstein},
  {Bigot}, {Bijaoui}, {Blasco}, {Bonfigli}, {Bono}, {Boudreault}, {Bressan},
  {Brown}, {Brunet}, {Bunclark}, {Buonanno}, {Butkevich}, {Carret}, {Carrion},
  {Chemin}, {Ch{\'e}reau}, {Corcione}, {Darmigny}, {de Boer}, {de Teodoro}, {de
  Zeeuw}, {Delle Luche}, {Domingues}, {Dubath}, {Fodor}, {Fr{\'e}zouls},
  {Fries}, {Fustes}, {Fyfe}, {Gallardo}, {Gallegos}, {Gardiol}, {Gebran},
  {Gomboc}, {G{\'o}mez}, {Grux}, {Gueguen}, {Heyrovsky}, {Hoar}, {Iannicola},
  {Isasi Parache}, {Janotto}, {Joliet}, {Jonckheere}, {Keil}, {Kim},
  {Klagyivik}, {Klar}, {Knude}, {Kochukhov}, {Kolka}, {Kos}, {Kutka}, {Lainey},
  {LeBouquin}, {Liu}, {Loreggia}, {Makarov}, {Marseille}, {Martayan},
  {Martinez-Rubi}, {Massart}, {Meynadier}, {Mignot}, {Munari}, {Nguyen},
  {Nordlander}, {Ocvirk}, {O'Flaherty}, {Olias Sanz}, {Ortiz}, {Osorio},
  {Oszkiewicz}, {Ouzounis}, {Palmer}, {Park}, {Pasquato}, {Peltzer}, {Peralta},
  {P{\'e}turaud}, {Pieniluoma}, {Pigozzi}, {Poels}, {Prat}, {Prod'homme},
  {Raison}, {Rebordao}, {Risquez}, {Rocca-Volmerange}, {Rosen}, {Ruiz-Fuertes},
  {Russo}, {Sembay}, {Serraller Vizcaino}, {Short}, {Siebert}, {Silva},
  {Sinachopoulos}, {Slezak}, {Soffel}, {Sosnowska}, {Strai{\v{z}}ys}, {ter
  Linden}, {Terrell}, {Theil}, {Tiede}, {Troisi}, {Tsalmantza}, {Tur},
  {Vaccari}, {Vachier}, {Valles}, {Van Hamme}, {Veltz}, {Virtanen}, {Wallut},
  {Wichmann}, {Wilkinson}, {Ziaeepour}, \& {Zschocke}}]{Prusti2016}
{Gaia Collaboration}, {Prusti}, T., {de Bruijne}, J.~H.~J., {et~al.} 2016,
  \aap, 595, A1, \dodoi{10.1051/0004-6361/201629272}

\bibitem[{{Gaia Collaboration} {et~al.}(2023){Gaia Collaboration}, {Vallenari},
  {Brown}, {Prusti}, {de Bruijne}, {Arenou}, {Babusiaux}, {Biermann},
  {Creevey}, {Ducourant}, {Evans}, {Eyer}, {Guerra}, {Hutton}, {Jordi},
  {Klioner}, {Lammers}, {Lindegren}, {Luri}, {Mignard}, {Panem}, {Pourbaix},
  {Randich}, {Sartoretti}, {Soubiran}, {Tanga}, {Walton}, {Bailer-Jones},
  {Bastian}, {Drimmel}, {Jansen}, {Katz}, {Lattanzi}, {van Leeuwen}, {Bakker},
  {Cacciari}, {Casta{\~n}eda}, {De Angeli}, {Fabricius}, {Fouesneau},
  {Fr{\'e}mat}, {Galluccio}, {Guerrier}, {Heiter}, {Masana}, {Messineo},
  {Mowlavi}, {Nicolas}, {Nienartowicz}, {Pailler}, {Panuzzo}, {Riclet}, {Roux},
  {Seabroke}, {Sordo}, {Th{\'e}venin}, {Gracia-Abril}, {Portell}, {Teyssier},
  {Altmann}, {Andrae}, {Audard}, {Bellas-Velidis}, {Benson}, {Berthier},
  {Blomme}, {Burgess}, {Busonero}, {Busso}, {C{\'a}novas}, {Carry}, {Cellino},
  {Cheek}, {Clementini}, {Damerdji}, {Davidson}, {de Teodoro}, {Nu{\~n}ez
  Campos}, {Delchambre}, {Dell'Oro}, {Esquej}, {Fern{\'a}ndez-Hern{\'a}ndez},
  {Fraile}, {Garabato}, {Garc{\'\i}a-Lario}, {Gosset}, {Haigron}, {Halbwachs},
  {Hambly}, {Harrison}, {Hern{\'a}ndez}, {Hestroffer}, {Hodgkin}, {Holl},
  {Jan{\ss}en}, {Jevardat de Fombelle}, {Jordan}, {Krone-Martins}, {Lanzafame},
  {L{\"o}ffler}, {Marchal}, {Marrese}, {Moitinho}, {Muinonen}, {Osborne},
  {Pancino}, {Pauwels}, {Recio-Blanco}, {Reyl{\'e}}, {Riello}, {Rimoldini},
  {Roegiers}, {Rybizki}, {Sarro}, {Siopis}, {Smith}, {Sozzetti}, {Utrilla},
  {van Leeuwen}, {Abbas}, {{\'A}brah{\'a}m}, {Abreu Aramburu}, {Aerts},
  {Aguado}, {Ajaj}, {Aldea-Montero}, {Altavilla}, {{\'A}lvarez}, {Alves},
  {Anders}, {Anderson}, {Anglada Varela}, {Antoja}, {Baines}, {Baker},
  {Balaguer-N{\'u}{\~n}ez}, {Balbinot}, {Balog}, {Barache}, {Barbato},
  {Barros}, {Barstow}, {Bartolom{\'e}}, {Bassilana}, {Bauchet}, {Becciani},
  {Bellazzini}, {Berihuete}, {Bernet}, {Bertone}, {Bianchi}, {Binnenfeld},
  {Blanco-Cuaresma}, {Blazere}, {Boch}, {Bombrun}, {Bossini}, {Bouquillon},
  {Bragaglia}, {Bramante}, {Breedt}, {Bressan}, {Brouillet}, {Brugaletta},
  {Bucciarelli}, {Burlacu}, {Butkevich}, {Buzzi}, {Caffau}, {Cancelliere},
  {Cantat-Gaudin}, {Carballo}, {Carlucci}, {Carnerero}, {Carrasco},
  {Casamiquela}, {Castellani}, {Castro-Ginard}, {Chaoul}, {Charlot}, {Chemin},
  {Chiaramida}, {Chiavassa}, {Chornay}, {Comoretto}, {Contursi}, {Cooper},
  {Cornez}, {Cowell}, {Crifo}, {Cropper}, {Crosta}, {Crowley}, {Dafonte},
  {Dapergolas}, {David}, {David}, {de Laverny}, {De Luise}, {De March}, {De
  Ridder}, {de Souza}, {de Torres}, {del Peloso}, {del Pozo}, {Delbo},
  {Delgado}, {Delisle}, {Demouchy}, {Dharmawardena}, {Di Matteo}, {Diakite},
  {Diener}, {Distefano}, {Dolding}, {Edvardsson}, {Enke}, {Fabre}, {Fabrizio},
  {Faigler}, {Fedorets}, {Fernique}, {Fienga}, {Figueras}, {Fournier},
  {Fouron}, {Fragkoudi}, {Gai}, {Garcia-Gutierrez}, {Garcia-Reinaldos},
  {Garc{\'\i}a-Torres}, {Garofalo}, {Gavel}, {Gavras}, {Gerlach}, {Geyer},
  {Giacobbe}, {Gilmore}, {Girona}, {Giuffrida}, {Gomel}, {Gomez},
  {Gonz{\'a}lez-N{\'u}{\~n}ez}, {Gonz{\'a}lez-Santamar{\'\i}a},
  {Gonz{\'a}lez-Vidal}, {Granvik}, {Guillout}, {Guiraud},
  {Guti{\'e}rrez-S{\'a}nchez}, {Guy}, {Hatzidimitriou}, {Hauser}, {Haywood},
  {Helmer}, {Helmi}, {Sarmiento}, {Hidalgo}, {Hilger}, {H{\l}adczuk}, {Hobbs},
  {Holland}, {Huckle}, {Jardine}, {Jasniewicz}, {Jean-Antoine Piccolo},
  {Jim{\'e}nez-Arranz}, {Jorissen}, {Juaristi Campillo}, {Julbe}, {Karbevska},
  {Kervella}, {Khanna}, {Kontizas}, {Kordopatis}, {Korn}, {K{\'o}sp{\'a}l},
  {Kostrzewa-Rutkowska}, {Kruszy{\'n}ska}, {Kun}, {Laizeau}, {Lambert},
  {Lanza}, {Lasne}, {Le Campion}, {Lebreton}, {Lebzelter}, {Leccia}, {Leclerc},
  {Lecoeur-Taibi}, {Liao}, {Licata}, {Lindstr{\o}m}, {Lister}, {Livanou},
  {Lobel}, {Lorca}, {Loup}, {Madrero Pardo}, {Magdaleno Romeo}, {Managau},
  {Mann}, {Manteiga}, {Marchant}, {Marconi}, {Marcos}, {Marcos Santos},
  {Mar{\'\i}n Pina}, {Marinoni}, {Marocco}, {Marshall}, {Martin Polo},
  {Mart{\'\i}n-Fleitas}, {Marton}, {Mary}, {Masip}, {Massari},
  {Mastrobuono-Battisti}, {Mazeh}, {McMillan}, {Messina}, {Michalik}, {Millar},
  {Mints}, {Molina}, {Molinaro}, {Moln{\'a}r}, {Monari}, {Mongui{\'o}},
  {Montegriffo}, {Montero}, {Mor}, {Mora}, {Morbidelli}, {Morel}, {Morris},
  {Muraveva}, {Murphy}, {Musella}, {Nagy}, {Noval}, {Oca{\~n}a}, {Ogden},
  {Ordenovic}, {Osinde}, {Pagani}, {Pagano}, {Palaversa}, {Palicio},
  {Pallas-Quintela}, {Panahi}, {Payne-Wardenaar}, {Pe{\~n}alosa Esteller},
  {Penttil{\"a}}, {Pichon}, {Piersimoni}, {Pineau}, {Plachy}, {Plum}, {Poggio},
  {Pr{\v{s}}a}, {Pulone}, {Racero}, {Ragaini}, {Rainer}, {Raiteri}, {Rambaux},
  {Ramos}, {Ramos-Lerate}, {Re Fiorentin}, {Regibo}, {Richards}, {Rios Diaz},
  {Ripepi}, {Riva}, {Rix}, {Rixon}, {Robichon}, {Robin}, {Robin}, {Roelens},
  {Rogues}, {Rohrbasser}, {Romero-G{\'o}mez}, {Rowell}, {Royer}, {Ruz Mieres},
  {Rybicki}, {Sadowski}, {S{\'a}ez N{\'u}{\~n}ez}, {Sagrist{\`a} Sell{\'e}s},
  {Sahlmann}, {Salguero}, {Samaras}, {Sanchez Gimenez}, {Sanna},
  {Santove{\~n}a}, {Sarasso}, {Schultheis}, {Sciacca}, {Segol}, {Segovia},
  {S{\'e}gransan}, {Semeux}, {Shahaf}, {Siddiqui}, {Siebert}, {Siltala},
  {Silvelo}, {Slezak}, {Slezak}, {Smart}, {Snaith}, {Solano}, {Solitro},
  {Souami}, {Souchay}, {Spagna}, {Spina}, {Spoto}, {Steele},
  {Steidelm{\"u}ller}, {Stephenson}, {S{\"u}veges}, {Surdej}, {Szabados},
  {Szegedi-Elek}, {Taris}, {Taylor}, {Teixeira}, {Tolomei}, {Tonello}, {Torra},
  {Torra}, {Torralba Elipe}, {Trabucchi}, {Tsounis}, {Turon}, {Ulla}, {Unger},
  {Vaillant}, {van Dillen}, {van Reeven}, {Vanel}, {Vecchiato}, {Viala},
  {Vicente}, {Voutsinas}, {Weiler}, {Wevers}, {Wyrzykowski}, {Yoldas}, {Yvard},
  {Zhao}, {Zorec}, {Zucker}, \& {Zwitter}}]{Vallenari2023}
{Gaia Collaboration}, {Vallenari}, A., {Brown}, A.~G.~A., {et~al.} 2023, \aap,
  674, A1, \dodoi{10.1051/0004-6361/202243940}

\bibitem[{{Gould}(2000)}]{Gould2000}
{Gould}, A. 2000, \apj, 542, 785, \dodoi{10.1086/317037}

\bibitem[{{Gould}(2004)}]{Gould2004}
---. 2004, \apj, 606, 319, \dodoi{10.1086/382782}

\bibitem[{{Gould} {et~al.}(2006){Gould}, {Udalski}, {An}, {Bennett}, {Zhou},
  {Dong}, {Rattenbury}, {Gaudi}, {Yock}, {Bond}, {Christie}, {Horne},
  {Anderson}, {Stanek}, {DePoy}, {Han}, {McCormick}, {Park}, {Pogge},
  {Poindexter}, {Soszy{\'n}ski}, {Szyma{\'n}ski}, {Kubiak}, {Pietrzy{\'n}ski},
  {Szewczyk}, {Wyrzykowski}, {Ulaczyk}, {Paczy{\'n}ski}, {Bramich},
  {Snodgrass}, {Steele}, {Burgdorf}, {Bode}, {Botzler}, {Mao}, \&
  {Swaving}}]{Gould2006}
{Gould}, A., {Udalski}, A., {An}, D., {et~al.} 2006, \apjl, 644, L37,
  \dodoi{10.1086/505421}

\bibitem[{{Han} \& {Gould}(1995)}]{Han1995}
{Han}, C., \& {Gould}, A. 1995, \apj, 447, 53, \dodoi{10.1086/175856}

\bibitem[{{Han} \& {Gould}(2003)}]{Han2003}
---. 2003, \apj, 592, 172, \dodoi{10.1086/375706}

\bibitem[{{Hao} {et~al.}(2010){Hao}, {McKay}, {Koester}, {Rykoff}, {Rozo},
  {Annis}, {Wechsler}, {Evrard}, {Siegel}, {Becker}, {Busha}, {Gerdes},
  {Johnston}, \& {Sheldon}}]{Hao2010}
{Hao}, J., {McKay}, T.~A., {Koester}, B.~P., {et~al.} 2010, \apjs, 191, 254,
  \dodoi{10.1088/0067-0049/191/2/254}

\bibitem[{Jin {et~al.}(2016)Jin, Zhang, Balakrishnan, Wainwright, \&
  Jordan}]{Jin2016}
Jin, C., Zhang, Y., Balakrishnan, S., Wainwright, M.~J., \& Jordan, M.~I. 2016,
  in Proceedings of the 30th International Conference on Neural Information
  Processing Systems, NIPS'16 (Red Hook, NY, USA: Curran Associates Inc.),
  4123–4131

\bibitem[{{Johnson} {et~al.}(2020){Johnson}, {Penny}, {Gaudi}, {Kerins},
  {Rattenbury}, {Robin}, {Calchi Novati}, \& {Henderson}}]{Johnson2020}
{Johnson}, S.~A., {Penny}, M., {Gaudi}, B.~S., {et~al.} 2020, \aj, 160, 123,
  \dodoi{10.3847/1538-3881/aba75b}

\bibitem[{{Kelly}(2007)}]{Kelly2007}
{Kelly}, B.~C. 2007, \apj, 665, 1489, \dodoi{10.1086/519947}

\bibitem[{{Kim} {et~al.}(2016){Kim}, {Lee}, {Park}, {Kim}, {Cha}, {Lee}, {Han},
  {Chun}, \& {Yuk}}]{Kim2016}
{Kim}, S.-L., {Lee}, C.-U., {Park}, B.-G., {et~al.} 2016, Journal of Korean
  Astronomical Society, 49, 37, \dodoi{10.5303/JKAS.2016.49.1.37}

\bibitem[{{Koshimoto} {et~al.}(2021){Koshimoto}, {Baba}, \&
  {Bennett}}]{Koshimoto2021}
{Koshimoto}, N., {Baba}, J., \& {Bennett}, D.~P. 2021, \apj, 917, 78,
  \dodoi{10.3847/1538-4357/ac07a8}

\bibitem[{{Kruszy{\'n}ska} {et~al.}(2022){Kruszy{\'n}ska}, {Wyrzykowski},
  {Rybicki}, {Maskoli{\={u}}nas}, {Bachelet}, {Rattenbury}, {Mr{\'o}z},
  {Zieli{\'n}ski}, {Howil}, {Kaczmarek}, {Hodgkin}, {Ihanec}, {Gezer},
  {Gromadzki}, {Miko{\l}ajczyk}, {Stankevi{\v{c}}i{\={u}}t{\.{e}}},
  {{\v{C}}epas}, {Pak{\v{s}}tien{\.{e}}}, {{\v{S}}i{\v{s}}kauskait{\.{e}}},
  {Zdanavi{\v{c}}ius}, {Bozza}, {Dominik}, {Figuera Jaimes}, {Fukui},
  {Hundertmark}, {Narita}, {Street}, {Tsapras}, {Bronikowski},
  {Jab{\l}o{\'n}ska}, {Jab{\l}onowska}, \&
  {Zi{\'o}{\l}kowska}}]{Kruszynska2022}
{Kruszy{\'n}ska}, K., {Wyrzykowski}, {\L}., {Rybicki}, K.~A., {et~al.} 2022,
  \aap, 662, A59, \dodoi{10.1051/0004-6361/202142602}

\bibitem[{{Kurucz}(1993)}]{Kurucz1993}
{Kurucz}, R.~L. 1993, {SYNTHE spectrum synthesis programs and line data}

\bibitem[{{Lindegren}(1978)}]{Lindegren1978}
{Lindegren}, L. 1978, in IAU Colloq. 48: Modern Astrometry, 197

\bibitem[{Lucic {et~al.}(2018)Lucic, Faulkner, Krause, \& Feldman}]{Lucic2018}
Lucic, M., Faulkner, M., Krause, A., \& Feldman, D. 2018, Journal of Machine
  Learning Research (JMLR), 18, 1

\bibitem[{{Mann} {et~al.}(2019){Mann}, {Dupuy}, {Kraus}, {Gaidos}, {Ansdell},
  {Ireland}, {Rizzuto}, {Hung}, {Dittmann}, {Factor}, {Feiden}, {Martinez},
  {Ru{\'\i}z-Rodr{\'\i}guez}, \& {Thao}}]{Mann2019}
{Mann}, A.~W., {Dupuy}, T., {Kraus}, A.~L., {et~al.} 2019, \apj, 871, 63,
  \dodoi{10.3847/1538-4357/aaf3bc}

\bibitem[{{Mr{\'o}z} \& {Wyrzykowski}(2021)}]{Mroz2021}
{Mr{\'o}z}, P., \& {Wyrzykowski}, {\L}. 2021, \actaa, 71, 89,
  \dodoi{10.32023/0001-5237/71.2.1}

\bibitem[{{Mr{\'o}z} {et~al.}(2017){Mr{\'o}z}, {Udalski}, {Skowron}, {Poleski},
  {Koz{\l}owski}, {Szyma{\'n}ski}, {Soszy{\'n}ski}, {Wyrzykowski},
  {Pietrukowicz}, {Ulaczyk}, {Skowron}, \& {Pawlak}}]{Mroz2017}
{Mr{\'o}z}, P., {Udalski}, A., {Skowron}, J., {et~al.} 2017, \nat, 548, 183,
  \dodoi{10.1038/nature23276}

\bibitem[{{Mr{\'o}z} {et~al.}(2018){Mr{\'o}z}, {Ryu}, {Skowron}, {Udalski},
  {Gould}, {Szyma{\'n}ski}, {Soszy{\'n}ski}, {Poleski}, {Pietrukowicz},
  {Koz{\l}owski}, {Pawlak}, {Ulaczyk}, {OGLE Collaboration}, {Albrow}, {Chung},
  {Jung}, {Han}, {Hwang}, {Shin}, {Yee}, {Zhu}, {Cha}, {Kim}, {Kim}, {Kim},
  {Lee}, {Lee}, {Lee}, {Park}, {Pogge}, \& {KMTNet Collaboration}}]{Mroz2018}
{Mr{\'o}z}, P., {Ryu}, Y.~H., {Skowron}, J., {et~al.} 2018, \aj, 155, 121,
  \dodoi{10.3847/1538-3881/aaaae9}

\bibitem[{{Mr{\'o}z} {et~al.}(2019){Mr{\'o}z}, {Udalski}, {Skowron},
  {Szyma{\'n}ski}, {Soszy{\'n}ski}, {Wyrzykowski}, {Pietrukowicz},
  {Koz{\l}owski}, {Poleski}, {Ulaczyk}, {Rybicki}, \& {Iwanek}}]{Mroz2019}
{Mr{\'o}z}, P., {Udalski}, A., {Skowron}, J., {et~al.} 2019, \apjs, 244, 29,
  \dodoi{10.3847/1538-4365/ab426b}

\bibitem[{{Mr{\'o}z} {et~al.}(2020){Mr{\'o}z}, {Udalski}, {Szyma{\'n}ski},
  {Soszy{\'n}ski}, {Pietrukowicz}, {Koz{\l}owski}, {Skowron}, {Poleski},
  {Ulaczyk}, {Gromadzki}, {Rybicki}, {Iwanek}, \& {Wrona}}]{Mroz2020}
{Mr{\'o}z}, P., {Udalski}, A., {Szyma{\'n}ski}, M.~K., {et~al.} 2020, \apjs,
  249, 16, \dodoi{10.3847/1538-4365/ab9366}

\bibitem[{{Muraki} {et~al.}(2011){Muraki}, {Han}, {Bennett}, {Suzuki},
  {Monard}, {Street}, {Jorgensen}, {Kundurthy}, {Skowron}, {Becker}, {Albrow},
  {Fouqu{\'e}}, {Heyrovsk{\'y}}, {Barry}, {Beaulieu}, {Wellnitz}, {Bond},
  {Sumi}, {Dong}, {Gaudi}, {Bramich}, {Dominik}, {Abe}, {Botzler}, {Freeman},
  {Fukui}, {Furusawa}, {Hayashi}, {Hearnshaw}, {Hosaka}, {Itow}, {Kamiya},
  {Korpela}, {Kilmartin}, {Lin}, {Ling}, {Makita}, {Masuda}, {Matsubara},
  {Miyake}, {Nishimoto}, {Ohnishi}, {Perrott}, {Rattenbury}, {Saito},
  {Skuljan}, {Sullivan}, {Sweatman}, {Tristram}, {Wada}, {Yock}, {MOA
  Collaboration}, {Christie}, {DePoy}, {Gorbikov}, {Gould}, {Kaspi}, {Lee},
  {Mallia}, {Maoz}, {McCormick}, {Moorhouse}, {Natusch}, {Park}, {Pogge},
  {Polishook}, {Shporer}, {Thornley}, {Yee}, {{\ensuremath{\mu}}FUN
  Collaboration}, {Allan}, {Browne}, {Horne}, {Kains}, {Snodgrass}, {Steele},
  {Tsapras}, {RoboNet Collaboration}, {Batista}, {Bennett}, {Brillant},
  {Caldwell}, {Cassan}, {Cole}, {Corrales}, {Coutures}, {Dieters}, {Dominis
  Prester}, {Donatowicz}, {Greenhill}, {Kubas}, {Marquette}, {Martin},
  {Menzies}, {Sahu}, {Waldman}, {Williams}, {Zub}, {PLANET Collaboration},
  {Bourhrous}, {Matsuoka}, {Nagayama}, {Oi}, {Randriamanakoto}, {IRSF
  Observers}, {Bozza}, {Burgdorf}, {Calchi Novati}, {Dreizler}, {Finet},
  {Glitrup}, {Harps{\o}e}, {Hinse}, {Hundertmark}, {Liebig}, {Maier},
  {Mancini}, {Mathiasen}, {Rahvar}, {Ricci}, {Scarpetta}, {Skottfelt},
  {Surdej}, {Southworth}, {Wambsganss}, {Zimmer}, {MiNDSTEp Consortium},
  {Udalski}, {Poleski}, {Wyrzykowski}, {Ulaczyk}, {Szyma{\'n}ski}, {Kubiak},
  {Pietrzy{\'n}ski}, {Soszy{\'n}ski}, \& {OGLE Collaboration}}]{Muraki2011}
{Muraki}, Y., {Han}, C., {Bennett}, D.~P., {et~al.} 2011, \apj, 741, 22,
  \dodoi{10.1088/0004-637X/741/1/22}

\bibitem[{{Nataf} {et~al.}(2013){Nataf}, {Gould}, {Fouqu{\'e}}, {Gonzalez},
  {Johnson}, {Skowron}, {Udalski}, {Szyma{\'n}ski}, {Kubiak},
  {Pietrzy{\'n}ski}, {Soszy{\'n}ski}, {Ulaczyk}, {Wyrzykowski}, \&
  {Poleski}}]{Nataf2013}
{Nataf}, D.~M., {Gould}, A., {Fouqu{\'e}}, P., {et~al.} 2013, \apj, 769, 88,
  \dodoi{10.1088/0004-637X/769/2/88}

\bibitem[{{Paczynski}(1986)}]{Paczynski1986}
{Paczynski}, B. 1986, \apj, 304, 1, \dodoi{10.1086/164140}

\bibitem[{{Pasetto} {et~al.}(2018){Pasetto}, {Grebel}, {Chiosi},
  {Crnojevi{\'c}}, {Zeidler}, {Busso}, {Cassar{\`a}}, {Piovan}, {Tantalo}, \&
  {Brogliato}}]{Pasetto2018}
{Pasetto}, S., {Grebel}, E.~K., {Chiosi}, C., {et~al.} 2018, \apj, 860, 120,
  \dodoi{10.3847/1538-4357/aac1bb}

\bibitem[{{Pecaut} \& {Mamajek}(2013)}]{Pecaut2013}
{Pecaut}, M.~J., \& {Mamajek}, E.~E. 2013, \apjs, 208, 9,
  \dodoi{10.1088/0067-0049/208/1/9}

\bibitem[{Pedregosa {et~al.}(2011)Pedregosa, Varoquaux, Gramfort, Michel,
  Thirion, Grisel, Blondel, Prettenhofer, Weiss, Dubourg, Vanderplas, Passos,
  Cournapeau, Brucher, Perrot, \& Duchesnay}]{scikit-learn}
Pedregosa, F., Varoquaux, G., Gramfort, A., {et~al.} 2011, Journal of Machine
  Learning Research, 12, 2825

\bibitem[{{Penny} {et~al.}(2019){Penny}, {Gaudi}, {Kerins}, {Rattenbury},
  {Mao}, {Robin}, \& {Calchi Novati}}]{Penny2019}
{Penny}, M.~T., {Gaudi}, B.~S., {Kerins}, E., {et~al.} 2019, \apjs, 241, 3,
  \dodoi{10.3847/1538-4365/aafb69}

\bibitem[{{Poleski} \& {Yee}(2019)}]{Poleski2019}
{Poleski}, R., \& {Yee}, J.~C. 2019, Astronomy and Computing, 26, 35,
  \dodoi{10.1016/j.ascom.2018.11.001}

\bibitem[{{Rabus} {et~al.}(2019){Rabus}, {Lachaume}, {Jord{\'a}n}, {Brahm},
  {Boyajian}, {von Braun}, {Espinoza}, {Berger}, {Le Bouquin}, \&
  {Absil}}]{Rabus2019}
{Rabus}, M., {Lachaume}, R., {Jord{\'a}n}, A., {et~al.} 2019, \mnras, 484,
  2674, \dodoi{10.1093/mnras/sty3430}

\bibitem[{{Refsdal}(1966)}]{Refsdal1966}
{Refsdal}, S. 1966, \mnras, 134, 315, \dodoi{10.1093/mnras/134.3.315}

\bibitem[{{Robin} {et~al.}(2003){Robin}, {Reyl{\'e}}, {Derri{\`e}re}, \&
  {Picaud}}]{Robin2003}
{Robin}, A.~C., {Reyl{\'e}}, C., {Derri{\`e}re}, S., \& {Picaud}, S. 2003,
  \aap, 409, 523, \dodoi{10.1051/0004-6361:20031117}

\bibitem[{{Rota} {et~al.}(2021){Rota}, {Hirao}, {Bozza}, {Abe}, {Barry},
  {Bennett}, {Bhattacharya}, {Bond}, {Donachie}, {Fukui}, {Fujii}, {Silva},
  {Itow}, {Kirikawa}, {Koshimoto}, {Li}, {Matsubara}, {Miyazaki}, {Muraki},
  {Olmschenk}, {Ranc}, {Satoh}, {Sumi}, {Suzuki}, {Tristram}, \&
  {Yonehara}}]{Rota2021}
{Rota}, P., {Hirao}, Y., {Bozza}, V., {et~al.} 2021, \aj, 162, 59,
  \dodoi{10.3847/1538-3881/ac0155}

\bibitem[{{Sahu} {et~al.}(2022){Sahu}, {Anderson}, {Casertano}, {Bond},
  {Udalski}, {Dominik}, {Calamida}, {Bellini}, {Brown}, {Rejkuba}, {Bajaj},
  {Kains}, {Ferguson}, {Fryer}, {Yock}, {Mr{\'o}z}, {Koz{\l}owski},
  {Pietrukowicz}, {Poleski}, {Skowron}, {Soszy{\'n}ski}, {Szyma{\'n}ski},
  {Ulaczyk}, {Wyrzykowski}, {Barry}, {Bennett}, {Bond}, {Hirao}, {Silva},
  {Kondo}, {Koshimoto}, {Ranc}, {Rattenbury}, {Sumi}, {Suzuki}, {Tristram},
  {Vandorou}, {Beaulieu}, {Marquette}, {Cole}, {Fouqu{\'e}}, {Hill}, {Dieters},
  {Coutures}, {Dominis-Prester}, {Bennett}, {Bachelet}, {Menzies}, {Albrow},
  {Pollard}, {Gould}, {Yee}, {Allen}, {Almeida}, {Christie}, {Drummond},
  {Gal-Yam}, {Gorbikov}, {Jablonski}, {Lee}, {Maoz}, {Manulis}, {McCormick},
  {Natusch}, {Pogge}, {Shvartzvald}, {J{\o}rgensen}, {Alsubai}, {Andersen},
  {Bozza}, {Novati}, {Burgdorf}, {Hinse}, {Hundertmark}, {Husser}, {Kerins},
  {Longa-Pe{\~n}a}, {Mancini}, {Penny}, {Rahvar}, {Ricci}, {Sajadian},
  {Skottfelt}, {Snodgrass}, {Southworth}, {Tregloan-Reed}, {Wambsganss},
  {Wertz}, {Tsapras}, {Street}, {Bramich}, {Horne}, {Steele}, \& {RoboNet
  Collaboration}}]{Sahu2022}
{Sahu}, K.~C., {Anderson}, J., {Casertano}, S., {et~al.} 2022, \apj, 933, 83,
  \dodoi{10.3847/1538-4357/ac739e}

\bibitem[{Schwarz(1978)}]{Schwarz1978}
Schwarz, G. 1978, The Annals of Statistics, 6, 461 ,
  \dodoi{10.1214/aos/1176344136}

\bibitem[{{Shvartzvald} {et~al.}(2019){Shvartzvald}, {Yee}, {Skowron}, {Lee},
  {Udalski}, {Calchi Novati}, {Bozza}, {Beichman}, {Bryden}, {Carey}, {Gaudi},
  {Henderson}, {Zhu}, {Spitzer Team}, {Bachelet}, {Bolt}, {Christie}, {Maoz},
  {Natusch}, {Pogge}, {Street}, {Tan}, {Tsapras}, {LCO}, {{\ensuremath{\mu}}FUN
  Follow-up Teams}, {Pietrukowicz}, {Soszy{\'n}ski}, {Szyma{\'n}ski},
  {Mr{\'o}z}, {Poleski}, {Koz{\l}owski}, {Ulaczyk}, {Pawlak}, {Rybicki},
  {Iwanek}, {OGLE Collaboration}, {Albrow}, {Cha}, {Chung}, {Gould}, {Han},
  {Hwang}, {Jung}, {Kim}, {Kim}, {Kim}, {Lee}, {Lee}, {Park}, {Ryu}, {Shin},
  {Zang}, {KMTNet Collaboration}, {Dominik}, {Helling}, {Hundertmark},
  {J{\o}rgensen}, {Longa-Pe{\~n}a}, {Lowry}, {Sajadian}, {Burgdorf},
  {Campbell-White}, {Ciceri}, {Evans}, {Fujii}, {Hinse}, {Rahvar}, {Rabus},
  {Skottfelt}, {Snodgrass}, {Southworth}, \& {MiNDSTEp
  Collaboration}}]{Shvartzvald2019}
{Shvartzvald}, Y., {Yee}, J.~C., {Skowron}, J., {et~al.} 2019, \aj, 157, 106,
  \dodoi{10.3847/1538-3881/aafe12}

\bibitem[{{Skowron} {et~al.}(2011){Skowron}, {Udalski}, {Gould}, {Dong},
  {Monard}, {Han}, {Nelson}, {McCormick}, {Moorhouse}, {Thornley}, {Maury},
  {Bramich}, {Greenhill}, {Koz{\l}owski}, {Bond}, {Poleski}, {Wyrzykowski},
  {Ulaczyk}, {Kubiak}, {Szyma{\'n}ski}, {Pietrzy{\'n}ski}, {Soszy{\'n}ski},
  {OGLE Collaboration}, {Gaudi}, {Yee}, {Hung}, {Pogge}, {DePoy}, {Lee},
  {Park}, {Allen}, {Mallia}, {Drummond}, {Bolt}, {{\ensuremath{\mu}}FUN
  Collaboration}, {Allan}, {Browne}, {Clay}, {Dominik}, {Fraser}, {Horne},
  {Kains}, {Mottram}, {Snodgrass}, {Steele}, {Street}, {Tsapras}, {RoboNet
  Collaboration}, {Abe}, {Bennett}, {Botzler}, {Douchin}, {Freeman}, {Fukui},
  {Furusawa}, {Hayashi}, {Hearnshaw}, {Hosaka}, {Itow}, {Kamiya}, {Kilmartin},
  {Korpela}, {Lin}, {Ling}, {Makita}, {Masuda}, {Matsubara}, {Muraki},
  {Nagayama}, {Miyake}, {Nishimoto}, {Ohnishi}, {Perrott}, {Rattenbury},
  {Saito}, {Skuljan}, {Sullivan}, {Sumi}, {Suzuki}, {Sweatman}, {Tristram},
  {Wada}, {Yock}, {MOA Collaboration}, {Beaulieu}, {Fouqu{\'e}}, {Albrow},
  {Batista}, {Brillant}, {Caldwell}, {Cassan}, {Cole}, {Cook}, {Coutures},
  {Dieters}, {Dominis Prester}, {Donatowicz}, {Kane}, {Kubas}, {Marquette},
  {Martin}, {Menzies}, {Sahu}, {Wambsganss}, {Williams}, {Zub}, \& {PLANET
  Collaboration}}]{Skowron2011}
{Skowron}, J., {Udalski}, A., {Gould}, A., {et~al.} 2011, \apj, 738, 87,
  \dodoi{10.1088/0004-637X/738/1/87}

\bibitem[{{Sozzetti}(2005)}]{Sozzetti2005}
{Sozzetti}, A. 2005, \pasp, 117, 1021, \dodoi{10.1086/444487}

\bibitem[{{Specht} {et~al.}(2020){Specht}, {Kerins}, {Awiphan}, \&
  {Robin}}]{Specht2020}
{Specht}, D., {Kerins}, E., {Awiphan}, S., \& {Robin}, A.~C. 2020, \mnras, 498,
  2196, \dodoi{10.1093/mnras/staa2375}

\bibitem[{{Spergel} {et~al.}(2015){Spergel}, {Gehrels}, {Baltay}, {Bennett},
  {Breckinridge}, {Donahue}, {Dressler}, {Gaudi}, {Greene}, {Guyon}, {Hirata},
  {Kalirai}, {Kasdin}, {Macintosh}, {Moos}, {Perlmutter}, {Postman},
  {Rauscher}, {Rhodes}, {Wang}, {Weinberg}, {Benford}, {Hudson}, {Jeong},
  {Mellier}, {Traub}, {Yamada}, {Capak}, {Colbert}, {Masters}, {Penny},
  {Savransky}, {Stern}, {Zimmerman}, {Barry}, {Bartusek}, {Carpenter}, {Cheng},
  {Content}, {Dekens}, {Demers}, {Grady}, {Jackson}, {Kuan}, {Kruk}, {Melton},
  {Nemati}, {Parvin}, {Poberezhskiy}, {Peddie}, {Ruffa}, {Wallace}, {Whipple},
  {Wollack}, \& {Zhao}}]{Spergel2015}
{Spergel}, D., {Gehrels}, N., {Baltay}, C., {et~al.} 2015, arXiv e-prints,
  arXiv:1503.03757, \dodoi{10.48550/arXiv.1503.03757}

\bibitem[{Storn \& Price(1997)}]{Storn1997}
Storn, R., \& Price, K.~V. 1997, J. Glob. Optim., 11, 341.
\newblock \url{http://dblp.uni-trier.de/db/journals/jgo/jgo11.html#StornP97}

\bibitem[{{Street} {et~al.}(2019){Street}, {Bachelet}, {Tsapras},
  {Hundertmark}, {Bozza}, {Dominik}, {ROME/REA}, {MiNDSTEp Teams}, {ROME/REA
  Team}, {Bramich}, {Cassan}, {Horne}, {Mao}, {Saha}, {Wambsganss}, {Zang},
  {MiNDSTEp Team}, {J{\o}rgensen}, {Longa-Pe{\~n}a}, {Peixinho}, {Sajadian},
  {Burgdorf}, {Campbell-White}, {Dib}, {Evans}, {Fujii}, {Hinse}, {Khalouei},
  {Lowry}, {Rahvar}, {Rabus}, {Skottfelt}, {Snodgrass}, {Southworth}, \&
  {Tregloan-Reed}}]{Street2019}
{Street}, R.~A., {Bachelet}, E., {Tsapras}, Y., {et~al.} 2019, \aj, 157, 215,
  \dodoi{10.3847/1538-3881/ab1538}

\bibitem[{{Sumi} {et~al.}(2011){Sumi}, {Kamiya}, {Bennett}, {Bond}, {Abe},
  {Botzler}, {Fukui}, {Furusawa}, {Hearnshaw}, {Itow}, {Kilmartin}, {Korpela},
  {Lin}, {Ling}, {Masuda}, {Matsubara}, {Miyake}, {Motomura}, {Muraki},
  {Nagaya}, {Nakamura}, {Ohnishi}, {Okumura}, {Perrott}, {Rattenbury}, {Saito},
  {Sako}, {Sullivan}, {Sweatman}, {Tristram}, {Udalski}, {Szyma{\'n}ski},
  {Kubiak}, {Pietrzy{\'n}ski}, {Poleski}, {Soszy{\'n}ski}, {Wyrzykowski},
  {Ulaczyk}, \& {Microlensing Observations in Astrophysics (MOA)
  Collaboration}}]{Sumi2011}
{Sumi}, T., {Kamiya}, K., {Bennett}, D.~P., {et~al.} 2011, \nat, 473, 349,
  \dodoi{10.1038/nature10092}

\bibitem[{{Sumi} {et~al.}(2013){Sumi}, {Bennett}, {Bond}, {Abe}, {Botzler},
  {Fukui}, {Furusawa}, {Itow}, {Ling}, {Masuda}, {Matsubara}, {Muraki},
  {Ohnishi}, {Rattenbury}, {Saito}, {Sullivan}, {Suzuki}, {Sweatman},
  {Tristram}, {Wada}, {Yock}, \& {MOA Collaboratoin}}]{Sumi2013}
{Sumi}, T., {Bennett}, D.~P., {Bond}, I.~A., {et~al.} 2013, \apj, 778, 150,
  \dodoi{10.1088/0004-637X/778/2/150}

\bibitem[{{Sumi} {et~al.}(2023){Sumi}, {Koshimoto}, {Bennett}, {Rattenbury},
  {Abe}, {Barry}, {Bhattacharya}, {Bond}, {Fujii}, {Fukui}, {Hamada}, {Hirao},
  {Silva}, {Itow}, {Kirikawa}, {Kondo}, {Matsubara}, {Miyazaki}, {Muraki},
  {Olmschenk}, {Ranc}, {Satoh}, {Suzuki}, {Tomoyoshi}, {Tristram}, {Vandorou},
  {Yama}, \& {Yamashita}}]{Sumi2023}
{Sumi}, T., {Koshimoto}, N., {Bennett}, D.~P., {et~al.} 2023, \aj, 166, 108,
  \dodoi{10.3847/1538-3881/ace688}

\bibitem[{{Tak} {et~al.}(2018){Tak}, {Ghosh}, \& {Ellis}}]{Tak2018}
{Tak}, H., {Ghosh}, S.~K., \& {Ellis}, J.~A. 2018, \mnras, 481, 277,
  \dodoi{10.1093/mnras/sty2326}

\bibitem[{{Terry} {et~al.}(2021){Terry}, {Bhattacharya}, {Bennett}, {Beaulieu},
  {Koshimoto}, {Blackman}, {Bond}, {Cole}, {Henderson}, {Lu}, {Marquette},
  {Ranc}, \& {Vandorou}}]{Terry2021}
{Terry}, S.~K., {Bhattacharya}, A., {Bennett}, D.~P., {et~al.} 2021, \aj, 161,
  54, \dodoi{10.3847/1538-3881/abcc60}

\bibitem[{{Tisserand} {et~al.}(2007){Tisserand}, {Le Guillou}, {Afonso},
  {Albert}, {Andersen}, {Ansari}, {Aubourg}, {Bareyre}, {Beaulieu}, {Charlot},
  {Coutures}, {Ferlet}, {Fouqu{\'e}}, {Glicenstein}, {Goldman}, {Gould},
  {Graff}, {Gros}, {Haissinski}, {Hamadache}, {de Kat}, {Lasserre}, {Lesquoy},
  {Loup}, {Magneville}, {Marquette}, {Maurice}, {Maury}, {Milsztajn}, {Moniez},
  {Palanque-Delabrouille}, {Perdereau}, {Rahal}, {Rich}, {Spiro},
  {Vidal-Madjar}, {Vigroux}, {Zylberajch}, \& {EROS-2
  Collaboration}}]{Tisserand2007}
{Tisserand}, P., {Le Guillou}, L., {Afonso}, C., {et~al.} 2007, \aap, 469, 387,
  \dodoi{10.1051/0004-6361:20066017}

\bibitem[{{Tsapras} {et~al.}(2019){Tsapras}, {Cassan}, {Ranc}, {Bachelet},
  {Street}, {Udalski}, {Hundertmark}, {Bozza}, {Beaulieu}, {Marquette},
  {Euteneuer}, {Bramich}, {Dominik}, {Figuera Jaimes}, {Horne}, {Mao},
  {Menzies}, {Schmidt}, {Snodgrass}, {Steele}, {Wambsganss}, {Mr{\'o}z},
  {Szyma{\'n}ski}, {Soszy{\'n}ski}, {Skowron}, {Pietrukowicz}, {Koz{\l}owski},
  {Poleski}, {Ulaczyk}, {Pawlak}, {J{\o}rgensen}, {Skottfelt}, {Popovas},
  {Ciceri}, {Korhonen}, {Kuffmeier}, {Evans}, {Peixinho}, {Hinse}, {Burgdorf},
  {Southworth}, {Tronsgaard}, {Kerins}, {Andersen}, {Rahvar}, {Wang}, {Wertz},
  {Rabus}, {Calchi Novati}, {D'Ago}, {Scarpetta}, {Mancini}, {Abe}, {Asakura},
  {Bennett}, {Bhattacharya}, {Donachie}, {Evans}, {Fukui}, {Hirao}, {Itow},
  {Kawasaki}, {Koshimoto}, {Li}, {Ling}, {Masuda}, {Matsubara}, {Muraki},
  {Miyazaki}, {Nagakane}, {Ohnishi}, {Rattenbury}, {Saito}, {Sharan}, {Shibai},
  {Sullivan}, {Sumi}, {Suzuki}, {Tristram}, {Yamada}, {Yonehara}, {Robonet
  Team}, {Bramich}, {Dominik}, {Jaimes}, {Horne}, {Mao}, {Menzies}, {Schmidt},
  {Snodgrass}, {Steele}, {Wambsganss}, {Ogle Collaboration}, {Mr{\'o}z},
  {Szyma{\'n}ski}, {Soszy{\'n}ski}, {Skowron}, {Pietrukowicz}, {Koz{\l}owski},
  {Poleski}, {Ulaczyk}, {Pawlak}, {Mindstep Collaboration}, {J{\o}rgensen},
  {Skottfelt}, {Popovas}, {Ciceri}, {Korhonen}, {Kuffmeier}, {Evans},
  {Peixinho}, {Hinse}, {Burgdorf}, {Southworth}, {Tronsgaard}, {Kerins},
  {Andersen}, {Rahvar}, {Wang}, {Wertz}, {Rabus}, {Novati}, {D'Ago},
  {Scarpetta}, {Mancini}, {Moa Collaboration}, {Abe}, {Asakura}, {Bennett},
  {Bhattacharya}, {Donachie}, {Evans}, {Fukui}, {Hirao}, {Itow}, {Kawasaki},
  {Koshimoto}, {Li}, {Ling}, {Masuda}, {Matsubara}, {Muraki}, {Miyazaki},
  {Nagakane}, {Ohnishi}, {Rattenbury}, {Saito}, {Sharan}, {Shibai}, {Sullivan},
  {Sumi}, {Suzuki}, {Tristram}, {Yamada}, \& {Yonehara}}]{Tsapras2019}
{Tsapras}, Y., {Cassan}, A., {Ranc}, C., {et~al.} 2019, \mnras, 487, 4603,
  \dodoi{10.1093/mnras/stz1404}

\bibitem[{{Udalski}(2003)}]{Udalski2003}
{Udalski}, A. 2003, \actaa, 53, 291, \dodoi{10.48550/arXiv.astro-ph/0401123}

\bibitem[{{Udalski} {et~al.}(2018){Udalski}, {Han}, {Bozza}, {Gould}, {Bond},
  {and}, {Mr{\'o}z}, {Skowron}, {Wyrzykowski}, {Szyma{\'n}ski},
  {Soszy{\'n}ski}, {Ulaczyk}, {Poleski}, {Pietrukowicz}, {Koz{\l}owski}, {OGLE
  Collaboration}, {Abe}, {Barry}, {Bennett}, {Bhattacharya}, {Donachie},
  {Evans}, {Fukui}, {Hirao}, {Itow}, {Kawasaki}, {Koshimoto}, {Li}, {Ling},
  {Masuda}, {Matsubara}, {Miyazaki}, {Munakata}, {Muraki}, {Nagakane},
  {Ohnishi}, {Ranc}, {Rattenbury}, {Saito}, {Sharan}, {Sullivan}, {Sumi},
  {Suzuki}, {Tristram}, {Yamada}, {Yonehara}, {MOA Collaboration}, {Street},
  {Tsapras}, {Bachelet}, {Bramich}, {D{\'A}go}, {Dominik}, {Figuera Jaimes},
  {Horne}, {Hundertmark}, {Kains}, {Menzies}, {Schmidt}, {Snodgrass}, {Steele},
  {Wambsganss}, {Robonet Collaboration}, {Pogge}, {Jung}, {Shin}, {Yee}, {Kim},
  {{\ensuremath{\mu}}Fun Collaboration}, {Beichman}, {Carey}, {Calchi Novati},
  {Zhu}, \& {Spitzer Team}}]{Udalski2018}
{Udalski}, A., {Han}, C., {Bozza}, V., {et~al.} 2018, \apj, 853, 70,
  \dodoi{10.3847/1538-4357/aaa295}

\bibitem[{{Vandorou} {et~al.}(2020){Vandorou}, {Bennett}, {Beaulieu}, {Alard},
  {Blackman}, {Cole}, {Bhattacharya}, {Bond}, {Koshimoto}, \&
  {Marquette}}]{Vandorou2020}
{Vandorou}, A., {Bennett}, D.~P., {Beaulieu}, J.-P., {et~al.} 2020, \aj, 160,
  121, \dodoi{10.3847/1538-3881/aba2d3}

\bibitem[{Virtanen {et~al.}(2020)Virtanen, Gommers, Oliphant, Haberland, Reddy,
  Cournapeau, Burovski, Peterson, Weckesser, Bright, {van der Walt}, Brett,
  Wilson, Millman, Mayorov, Nelson, Jones, Kern, Larson, Carey, Polat, Feng,
  Moore, {VanderPlas}, Laxalde, Perktold, Cimrman, Henriksen, Quintero, Harris,
  Archibald, Ribeiro, Pedregosa, {van Mulbregt}, \& {SciPy 1.0
  Contributors}}]{Pauli2020}
Virtanen, P., Gommers, R., Oliphant, T.~E., {et~al.} 2020, Nature Methods, 17,
  261, \dodoi{10.1038/s41592-019-0686-2}

\bibitem[{{Wang} \& {Chen}(2019)}]{Wang2019}
{Wang}, S., \& {Chen}, X. 2019, \apj, 877, 116,
  \dodoi{10.3847/1538-4357/ab1c61}

\bibitem[{{Witt} \& {Mao}(1994)}]{Witt1994}
{Witt}, H.~J., \& {Mao}, S. 1994, \apj, 430, 505, \dodoi{10.1086/174426}

\bibitem[{{Worthey}(1994)}]{Worthey1994}
{Worthey}, G. 1994, \apjs, 95, 107, \dodoi{10.1086/192096}

\bibitem[{{Wyrzykowski} {et~al.}(2011){Wyrzykowski}, {Skowron}, {Koz{\l}owski},
  {Udalski}, {Szyma{\'n}ski}, {Kubiak}, {Pietrzy{\'n}ski}, {Soszy{\'n}ski},
  {Szewczyk}, {Ulaczyk}, {Poleski}, \& {Tisserand}}]{Wyrzykowski2011}
{Wyrzykowski}, L., {Skowron}, J., {Koz{\l}owski}, S., {et~al.} 2011, \mnras,
  416, 2949, \dodoi{10.1111/j.1365-2966.2011.19243.x}

\bibitem[{{Wyrzykowski} {et~al.}(2020){Wyrzykowski}, {Mr{\'o}z}, {Rybicki},
  {Gromadzki}, {Ko{\l}aczkowski}, {Zieli{\'n}ski}, {Zieli{\'n}ski},
  {Britavskiy}, {Gomboc}, {Sokolovsky}, {Hodgkin}, {Abe}, {Aldi}, {AlMannaei},
  {Altavilla}, {Al Qasim}, {Anupama}, {Awiphan}, {Bachelet}, {Bak{\i}{\c{s}}},
  {Baker}, {Bartlett}, {Bendjoya}, {Benson}, {Bikmaev}, {Birenbaum},
  {Blagorodnova}, {Blanco-Cuaresma}, {Boeva}, {Bonanos}, {Bozza}, {Bramich},
  {Bruni}, {Burenin}, {Burgaz}, {Butterley}, {Caines}, {Caton}, {Calchi
  Novati}, {Carrasco}, {Cassan}, {{\v{C}}epas}, {Cropper},
  {Chru{\'s}li{\'n}ska}, {Clementini}, {Clerici}, {Conti}, {Conti}, {Cross},
  {Cusano}, {Damljanovic}, {Dapergolas}, {D'Ago}, {de Bruijne}, {Dennefeld},
  {Dhillon}, {Dominik}, {Dziedzic}, {Erece}, {Eselevich}, {Esenoglu}, {Eyer},
  {Figuera Jaimes}, {Fossey}, {Galeev}, {Grebenev}, {Gupta}, {Gutaev},
  {Hallakoun}, {Hamanowicz}, {Han}, {Handzlik}, {Haislip}, {Hanlon}, {Hardy},
  {Harrison}, {van Heerden}, {Hoette}, {Horne}, {Hudec}, {Hundertmark},
  {Ihanec}, {Irtuganov}, {Itoh}, {Iwanek}, {Jovanovic}, {Janulis},
  {Jel{\'\i}nek}, {Jensen}, {Kaczmarek}, {Katz}, {Khamitov}, {Kilic},
  {Klencki}, {Kolb}, {Kopacki}, {Kouprianov}, {Kruszy{\'n}ska}, {Kurowski},
  {Latev}, {Lee}, {Leonini}, {Leto}, {Lewis}, {Li}, {Liakos}, {Littlefair},
  {Lu}, {Manser}, {Mao}, {Maoz}, {Martin-Carrillo}, {Marais},
  {Maskoli{\={u}}nas}, {Maund}, {Meintjes}, {Melnikov}, {Ment},
  {Miko{\l}ajczyk}, {Morrell}, {Mowlavi}, {Mo{\'z}dzierski}, {Murphy},
  {Nazarov}, {Netzel}, {Nesci}, {Ngeow}, {Norton}, {Ofek},
  {Pak{\v{s}}tien{\.{e}}}, {Palaversa}, {Pandey}, {Paraskeva}, {Pawlak},
  {Penny}, {Penprase}, {Piascik}, {Prieto}, {Qvam}, {Ranc},
  {Rebassa-Mansergas}, {Reichart}, {Reig}, {Rhodes}, {Rivet}, {Rixon},
  {Roberts}, {Rosi}, {Russell}, {Zanmar Sanchez}, {Scarpetta}, {Seabroke},
  {Shappee}, {Schmidt}, {Shvartzvald}, {Sitek}, {Skowron}, {{\'S}niegowska},
  {Snodgrass}, {Soares}, {van Soelen}, {Spetsieri},
  {Stankevi{\v{c}}i{\={u}}t{\.{e}}}, {Steele}, {Street}, {Strobl}, {Strubble},
  {Szegedi}, {Tinjaca Ramirez}, {Tomasella}, {Tsapras}, {Vernet}, {Villanueva},
  {Vince}, {Wambsganss}, {van der Westhuizen}, {Wiersema}, {Wium}, {Wilson},
  {Yoldas}, {Zhuchkov}, {Zhukov}, {Zdanavi{\v{c}}ius}, {Zo{\l}a}, \&
  {Zubareva}}]{Wyrzykowski2020}
{Wyrzykowski}, {\L}., {Mr{\'o}z}, P., {Rybicki}, K.~A., {et~al.} 2020, \aap,
  633, A98, \dodoi{10.1051/0004-6361/201935097}

\bibitem[{{Wyrzykowski} {et~al.}(2023){Wyrzykowski}, {Kruszy{\'n}ska},
  {Rybicki}, {Holl}, {Lec{\oe}ur-Ta{\"\i}bi}, {Mowlavi}, {Nienartowicz},
  {Jevardat de Fombelle}, {Rimoldini}, {Audard}, {Garcia-Lario}, {Gavras},
  {Evans}, {Hodgkin}, \& {Eyer}}]{Wyrzykowski2023}
{Wyrzykowski}, {\L}., {Kruszy{\'n}ska}, K., {Rybicki}, K.~A., {et~al.} 2023,
  \aap, 674, A23, \dodoi{10.1051/0004-6361/202243756}

\bibitem[{{Yee} {et~al.}(2014){Yee}, {Han}, {Gould}, {Skowron}, {Bond},
  {Udalski}, {Hundertmark}, {Monard}, {Porritt}, {Nelson}, {Bozza}, {Albrow},
  {Choi}, {Christie}, {DePoy}, {Gaudi}, {Hwang}, {Jung}, {Lee}, {McCormick},
  {Natusch}, {Ngan}, {Park}, {Pogge}, {Shin}, {Tan}, {{\ensuremath{\mu}}FUN
  Collaboration}, {Abe}, {Bennett}, {Botzler}, {Freeman}, {Fukui}, {Fukunaga},
  {Itow}, {Koshimoto}, {Larsen}, {Ling}, {Masuda}, {Matsubara}, {Muraki},
  {Namba}, {Ohnishi}, {Philpott}, {Rattenbury}, {Saito}, {Sullivan}, {Sumi},
  {Sweatman}, {Suzuki}, {Tristram}, {Tsurumi}, {Wada}, {Yamai}, {Yock},
  {Yonehara}, {MOA Collaboration}, {Szyma{\'n}ski}, {Ulaczyk}, {Koz{\l}owski},
  {Poleski}, {Wyrzykowski}, {Kubiak}, {Pietrukowicz}, {Pietrzy{\'n}ski},
  {Soszy{\'n}ski}, {OGLE Collaboration}, {Bramich}, {Browne}, {Figuera Jaimes},
  {Horne}, {Ipatov}, {Kains}, {Snodgrass}, {Steele}, {Street}, {Tsapras}, \&
  {RoboNet Collaboration}}]{Yee2014}
{Yee}, J.~C., {Han}, C., {Gould}, A., {et~al.} 2014, \apj, 790, 14,
  \dodoi{10.1088/0004-637X/790/1/14}

\bibitem[{{Yee} {et~al.}(2021){Yee}, {Zang}, {Udalski}, {Ryu}, {Green},
  {Hennerley}, {Marmont}, {Sumi}, {Mao}, {Gromadzki}, {Mr{\'o}z}, {Skowron},
  {Poleski}, {Szyma{\'n}ski}, {Soszy{\'n}ski}, {Pietrukowicz}, {Koz{\l}owski},
  {Ulaczyk}, {Rybicki}, {Iwanek}, {Wrona}, {Albrow}, {Chung}, {Gould}, {Han},
  {Hwang}, {Jung}, {Kim}, {Shin}, {Shvartzvald}, {Cha}, {Kim}, {Kim}, {Lee},
  {Lee}, {Lee}, {Park}, {Pogge}, {Bachelet}, {Christie}, {Hundertmark}, {Maoz},
  {McCormick}, {Natusch}, {Penny}, {Street}, {Tsapras}, {Beichman}, {Bryden},
  {Novati}, {Carey}, {Gaudi}, {Henderson}, {Johnson}, {Zhu}, {Bond}, {Abe},
  {Barry}, {Bennett}, {Bhattacharya}, {Donachie}, {Fujii}, {Fukui}, {Hirao},
  {Silva}, {Itow}, {Kirikawa}, {Kondo}, {Koshimoto}, {Alex Li}, {Matsubara},
  {Muraki}, {Miyazaki}, {Olmschenk}, {Ranc}, {Rattenbury}, {Satoh}, {Shoji},
  {Suzuki}, {Tanaka}, {Tristram}, {Yamawaki}, {Yonehara}, \& {MOA
  Collaboration}}]{Yee2021}
{Yee}, J.~C., {Zang}, W., {Udalski}, A., {et~al.} 2021, \aj, 162, 180,
  \dodoi{10.3847/1538-3881/ac1582}

\bibitem[{{Yoo} {et~al.}(2004){Yoo}, {DePoy}, {Gal-Yam}, {Gaudi}, {Gould},
  {Han}, {Lipkin}, {Maoz}, {Ofek}, {Park}, {Pogge}, {Mu-Fun Collaboration},
  {Udalski}, {Soszy{\'n}ski}, {Wyrzykowski}, {Kubiak}, {Szyma{\'n}ski},
  {Pietrzy{\'n}ski}, {Szewczyk}, {{\.Z}ebru{\'n}}, \& {OGLE
  Collaboration}}]{Yoo2004}
{Yoo}, J., {DePoy}, D.~L., {Gal-Yam}, A., {et~al.} 2004, \apj, 603, 139,
  \dodoi{10.1086/381241}

\bibitem[{{Zang} {et~al.}(2020){Zang}, {Shvartzvald}, {Wang}, {Udalski}, {Lee},
  {Sumi}, {Skottfelt}, {Li}, {Mao}, {Zhu}, {Yee}, {Calchi Novati}, {Beichman},
  {Bryden}, {Carey}, {Gaudi}, {Henderson}, {Spitzer Team}, {Mr{\'o}z},
  {Skowron}, {Poleski}, {Szyma{\'n}ski}, {Soszy{\'n}ski}, {Pietrukowicz},
  {Koz{\l}owski}, {Ulaczyk}, {Rybicki}, {Iwanek}, {OGLE Collaboration},
  {Bachelet}, {Christie}, {Green}, {Hennerley}, {Maoz}, {Natusch}, {Pogge},
  {Street}, {Tsapras}, {LCO Follow-Up Team}, {{\ensuremath{\mu}}FUN Follow-Up
  Team}, {Albrow}, {Chung}, {Gould}, {Han}, {Hwang}, {Jung}, {Ryu}, {Shin},
  {Cha}, {Kim}, {Kim}, {Kim}, {Lee}, {Lee}, {Park}, {KMTNet Collaboration},
  {Bond}, {Abe}, {Barry}, {Bennett}, {Bhattacharya}, {Donachie}, {Fukui},
  {Hirao}, {Itow}, {Kondo}, {Koshimoto}, {Alex Li}, {Matsubara}, {Muraki},
  {Miyazaki}, {Nagakane}, {Ranc}, {Rattenbury}, {Suematsu}, {Sullivan},
  {Suzuki}, {Tristram}, {Yonehara}, {MOA Collaboration}, {Dominik},
  {Hundertmark}, {J{\o}rgensen}, {Rahvar}, {Sajadian}, {Snodgrass}, {Bozza},
  {Burgdorf}, {Evans}, {Figuera Jaimes}, {Fujii}, {Mancini}, {Longa-Pe{\~n}a},
  {Helling}, {Peixinho}, {Rabus}, {Southworth}, {Unda-Sanzana}, {von Essen}, \&
  {MiNDSTEp Collaboration}}]{Zang2020}
{Zang}, W., {Shvartzvald}, Y., {Wang}, T., {et~al.} 2020, \apj, 891, 3,
  \dodoi{10.3847/1538-4357/ab6ff8}

\bibitem[{{Zhu} {et~al.}(2016){Zhu}, {Calchi Novati}, {Gould}, {Udalski},
  {Han}, {Shvartzvald}, {Ranc}, {J{\o}rgensen}, {Poleski}, {Bozza}, {Beichman},
  {Bryden}, {Carey}, {Gaudi}, {Henderson}, {Pogge}, {Porritt}, {Wibking},
  {Yee}, {SPITZER Team}, {Pawlak}, {Szyma{\'n}ski}, {Skowron}, {Mr{\'o}z},
  {Koz{\l}owski}, {Wyrzykowski}, {Pietrukowicz}, {Pietrzy{\'n}ski},
  {Soszy{\'n}ski}, {Ulaczyk}, {OGLE Group}, {Choi}, {Park}, {Jung}, {Shin},
  {Albrow}, {Park}, {Kim}, {Lee}, {Cha}, {Kim}, {Lee}, {KMTNET Group},
  {Friedmann}, {Kaspi}, {Maoz}, {WISE Group}, {Hundertmark}, {Street},
  {Tsapras}, {Bramich}, {Cassan}, {Dominik}, {Bachelet}, {Dong}, {Figuera
  Jaimes}, {Horne}, {Mao}, {Menzies}, {Schmidt}, {Snodgrass}, {Steele},
  {Wambsganss}, {RoboNeT Team}, {Skottfelt}, {Andersen}, {Burgdorf}, {Ciceri},
  {D'Ago}, {Evans}, {Gu}, {Hinse}, {Kerins}, {Korhonen}, {Kuffmeier},
  {Mancini}, {Peixinho}, {Popovas}, {Rabus}, {Rahvar}, {Tronsgaard},
  {Scarpetta}, {Southworth}, {Surdej}, {von Essen}, {Wang}, {Wertz}, \&
  {MiNDSTEP Group}}]{Zhu2016}
{Zhu}, W., {Calchi Novati}, S., {Gould}, A., {et~al.} 2016, \apj, 825, 60,
  \dodoi{10.3847/0004-637X/825/1/60}

\end{thebibliography}

\bibliographystyle{aasjournal}  

\end{document}